\documentclass[a4paper,11pt]{article}
\pdfoutput=1 

\usepackage{jheppub} 

\usepackage[T1]{fontenc} 
\usepackage{bm}
\usepackage{amsmath,amssymb,amsfonts,slashed}
\usepackage{cancel,color,xcolor}
\usepackage[normalem]{ulem}
\usepackage{xargs}
\usepackage[colorinlistoftodos]{todonotes}
\usepackage{tabularx}

\newcommandx{\JPL}[2][1=]{\todo[linecolor=red,backgroundcolor=red!25,bordercolor=red,#1]{\tiny JPL: #2}}

\newcommandx{\MF}[2][1=]{\todo[linecolor=blue,backgroundcolor=blue!25,bordercolor=blue,#1]{#2}}

\newcommandx{\MN}[2][1=]{\todo[linecolor=blue,backgroundcolor=magenta!25,bordercolor=blue,#1]{#2}}

\newcommandx{\SW}[2][1=]{\todo[linecolor=green,backgroundcolor=red!25,bordercolor=red,#1]{\tiny SW: #2}}

\title{\boldmath  
{Complete NLO BFKL  impact factors for quarkonium hadroproduction in NRQCD: the case of ${}^1S_0^{[1]}$, ${}^1S_0^{[8]}$, and ${}^3S_1^{[8]}$ states}}

\author[a]{Michael Fucilla,}
\author[b]{Jean-Philippe Lansberg,}
\author[c,1]{Maxim Nefedov\note{Corresponding author.}}
\author[a]{Lech Szymanowski,}
\author[b]{and Samuel Wallon}

\affiliation[a]{National Centre for Nuclear Research, Pasteura 7, Warsaw 02-093, Poland}
\affiliation[b]{Universit\'e Paris-Saclay, CNRS/IN2P3, IJCLab, F-91405, Orsay, France}
\affiliation[c]{Physics Department, Ben-Gurion University of the Negev, 
Beer Sheva 84105, Israel}

\emailAdd{Michael.Fucilla@ncbj.gov.pl}
\emailAdd{Jean-Philippe.Lansberg@in2p3.fr}
\emailAdd{nefedov@post.bgu.ac.il}
\emailAdd{Lech.Szymanowski@ncbj.gov.pl}
\emailAdd{Samuel.Wallon@ijclab.in2p3.fr}

\abstract{We present the first complete next-to-leading-order calculation of the impact factors for hadroproduction 
of the ${}^1S_0^{[1]}$, ${}^1S_0^{[8]}$, and ${}^3S_1^{[8]}$ NRQCD states
within the BFKL formalism. We complete the recent virtual-correction computation presented in \href{https://doi.org/10.1007/JHEP12(2024)129}{JHEP 12 (2024) 129} by that of the real-emission contributions. We observe the cancellation of the soft divergences between these real- and virtual-emission contributions and we note that the surviving collinear singularities are compatible with factorisation up to one loop for a novel class of processes where BFKL resummation can be applied. Our work indeed represents the first complete NLO quarkonium impact factor in the BFKL framework and paves the way to first next-to-leading-logarithmic-precision studies for hadroproduction of forward-backward quarkonium associated production at hadron colliders. 
}

\newcommand{\T}[1]{{\boldsymbol{#1}}_T}

\newcommand{\TT}[2]{{\boldsymbol{#1}}_{#2 T}}

\begin{document}
\maketitle
\flushbottom

\section{Introduction}
{The large centre-of-mass energies, $\sqrt{s}$, available at the CERN Large Hadron Collider (LHC) and at future circular colliders offer a unique opportunity to investigate QCD dynamics in previously unexplored regimes. In particular, they give access to the \textit{Regge-Gribov} (or \textit{semi-hard}) limit of QCD, characterised by the scale hierarchy $s \gg \{Q^2 \} \gg \Lambda_{{\rm{QCD}}}^2$, where $\{ Q \}$ is a set of hard scales characterising the process. In this limit, large logarithmic corrections of the ratio $s/Q^2$ can affect both parton densities and hard scattering cross sections. The Balitsky-Fadin-Kuraev-Lipatov (BFKL) framework~\cite{BFKL1,BFKL2,BFKL3} allows one to resum, to all orders, these contributions, both in the Leading Logarithmic (LL) approximation in which $\alpha_s^n \ln^n(s/Q^2)$ are resummed, and in the Next-to-Leading Logarithmic (NLL) approximation in which $\alpha_s^{n+1} \ln^n(s/Q^2)$ are resummed.}

In this framework, the cross section of inclusive processes can be expressed as the convolution of process-dependent impact factors (IFs) and universal Green's functions. The IFs are off-shell coefficient functions involving new degrees of freedom called reggeised gluons, which can be understood as off-shell $t$-channel gluons, emitted by fast-moving partons in the eikonal approximation. The Green's function describes strongly rapidity-ordered gluon emissions between two impact factors and is the solution of an integral equation, called the BFKL equation. 

The current frontier is the NLL precision\footnote{Some pieces of the next-to-NLO kernel have been recently found in $\mathcal{N}=4$ SYM~\cite{Byrne:2022wzk}, in pure-gauge QCD~\cite{DelDuca:2021vjq} and in full QCD~\cite{Caola:2021izf,Falcioni:2021dgr,Fadin:2023roz,Abreu:2024mpk,Buccioni:2024gzo,Abreu:2024xoh}.}, which requires the knowledge at Next-to-Leading Order (NLO) corrections to both the kernel of the BFKL equation ~\cite{Fadin:1998py,Ciafaloni:1998gs,Fadin:1998jv,Fadin:2000kx,Fadin:2000hu,Fadin:2004zq,Fadin:2005zj}
and to the impact factors (IFs)~\cite{Fadin:1999de,Fadin:1999df,Ciafaloni:1998kx,Ciafaloni:1998hu,Ciafaloni:2000sq,Bartels:2001ge,Bartels:2002yj,Caporale:2011cc,Ivanov:2012ms,Colferai:2015zfa,Chirilli:2012jd,Hentschinski:2014bra, Bartels:2000gt,Bartels:2001mv,Bartels:2002uz,Bartels:2003zi,Bartels:2004bi,Fadin:2001ap,Ivanov:2012iv,Nefedov:2024swu,Nefedov:2019mrg,Hentschinski:2020tbi,Celiberto:2022fgx,Fucilla:2024cpf,Celiberto:2024bbv}. 
Currently, there are only a few processes for which the full NLL BFKL cross section has been computed, such as the production of Mueller-Navelet~\cite{MN:87,Colferai:2010wu,Caporale:2012ih,Ducloue:2013hia,Ducloue:2013bva,Ducloue:2014koa,Caporale:2014gpa,Ducloue:2015jba,Mueller:2015ael} or Mueller-Tang~\cite{Mueller:1992pe,Colferai:2023hje} dijets. Those inclusive processes to be studied at the LHC all involve large transverse momenta, typically above tens of GeV, which reduces the size of BFKL effects, such as azimuthal de-correlation.
In the case of processes involving quarkonia, ${\cal Q}$, it is important to note that existing NLL cross sections are computed~\cite{Celiberto:2022dyf,Celiberto:2025euy,Celiberto:2023fzz} at leading-power in $P_T$, that is neglecting corrections scaling like to $M_{\cal Q}^2/P_T^2$. This is also known as the fragmentation approximation.

{In the present paper, we wish to demonstrate that it is possible to compute quarkonium-related cross sections up to NLL without this restriction when the quarkonium is produced at large forward (backward) rapidities in association with another particles produced at backward (forward) rapidities, in the exact analogy with the Mueller-Navelet dijets. To do so, we rely on the Non-Relativistic QCD (NRQCD) factorisation framework~\cite{Bodwin:1994jh} which has become a standard formalism for the description of hadronisation of heavy quark pair ($Q\bar{Q}$) into heavy quarkonium (${\cal Q}$)~\cite{Brambilla:2004wf,Brambilla:2010cs,Lansberg:2019adr,OniumEIC} and which is based on a double expansion on the strong coupling constant $\alpha_S$ and the relative velocity of the heavy quarks in the quarkonium rest frame, $v$. Within NRQCD,} the $Q\bar{Q}$-pair is produced at short distances in various states $Q\bar{Q}[m]$, denoted  $m={}^{2S+1}L^{[1,8]}_J$ using the spectroscopic notation with the total spin $S$, orbital momentum $L$, total angular momentum $J$ and with colour-singlet ({CS},${}^{[1]}$) or colour-octet ({CO},${}^{[8]}$) quantum numbers. The production of the state $m$ is described perturbatively, while the probability of hadronisation of the state $m$ into quarkonium ${\cal Q}$ is given by the Long-Distance Matrix Elements (LDMEs) of NRQCD: $\langle {\cal O}^{\cal Q}[m] \rangle$. The velocity-scaling rules for LDMEs contributing to production of various physical quarkonium states are summarised in tab.~\ref{tab:velocity-SR-LDMEs} together with the $\alpha_s$ orders of the corresponding LO IFs.

\begin{table}[]
    \centering \footnotesize
    \begin{tabularx}{\textwidth}{XX|XXXX|XXXXXXXX}
\hline
  $\left. {\cal Q} \middle\backslash m  \right.$ & & 
  $\bm{\mathit{{}^1S_0^{[1]}}}$
  & ${}^{3}S_1^{[1]}$ & 
  $\bm{\mathit{{}^1S_0^{[8]}}}$& 
  $\bm{\mathit{{}^3S_1^{[8]}}}$
  & ${}^1P_1^{[1]}$ & ${}^3P_0^{[1]}$ & ${}^3P_1^{[1]}$ & ${}^3P_2^{[1]}$ & ${}^1P_1^{[8]}$ & ${}^3P_0^{[8]}$ & ${}^3P_1^{[8]}$ & ${}^3P_2^{[8]}$ \\
\hline
\hspace*{-1.5mm}$\eta_c(nS),\eta_b(nS)$ &   & $v^0\alpha_s^2$             &                   & $v^4\alpha_s^2$           & $v^4\alpha_s^2$           &                 &                 &                 &                 & $v^4\alpha_s^2$           &                 &                 &                 \\
\hspace*{-1.5mm}$\psi(nS),\Upsilon(nS)$ &   &                 & $v^0\alpha_s^3$               & $v^4\alpha_s^2$           & $v^4\alpha_s^2$           &                 &                 &                 &                 &                 & $v^4\alpha_s^2$           & $v^4\alpha_s^2$           & $v^4\alpha_s^2$           \\
\hline
$h_c,h_b$  &    &                 &                   & $v^2\alpha_s^2$           &                 & $v^2\alpha_s^2$           &                 &                 &                 &                 &                 &                 &                 \\
$\chi_{c0},\chi_{b0}$ & &                 &                   &                 & $v^2\alpha_s^2$           &                 & $v^2\alpha_s^2$           &                 &                 &                 &                 &                 &                 \\
$\chi_{c1},\chi_{b1}$ & &                 &                   &                 & $v^2\alpha_s^2$           &                 &                 & $v^2\alpha_s^2$           &                 &                 &                 &                 &                 \\
$\chi_{c2},\chi_{b2}$ & &                 &                   &                 & $v^2\alpha_s^2$           &                 &                 &                 & $v^2\alpha_s^2$           &                 &                 &                 &                 \\
\hline
\end{tabularx}
    \caption{Velocity  scaling~\cite{Bodwin:1994jh,Bodwin:2005hm,Schuler:1997is} for LDMEs of different NRQCD states contributing to the production cross section of observable quarkonia up to $O(v^4)$ and the $\alpha_s$ order of the corresponding LO IFs. The states for which we are computing the IF up to $\alpha_s^3$ are highlighted in bold.}
    \label{tab:velocity-SR-LDMEs}
\end{table}

The current state of the art in the NRQCD computations for heavy-quarkonium production cross sections is the Next-to-Leading order (NLO) in $\alpha_s$ for the short-distance part~\cite{Kuhn:1992qw,Kramer:1995nb,Petrelli:1997ge,Maltoni:1997pt,Campbell:2007ws,Gong:2008sn,Artoisenet:2008fc,Artoisenet:2009xh,Gong:2008ft,Li:2008ym,Brodsky:2009cf,Lansberg:2010vq,Butenschoen:2010rq,Ma:2010yw,Ma:2010vd,Butenschoen:2011yh,Gong:2012ah,Gong:2012ug,Butenschoen:2012px,Butenschoen:2012qr,Chao:2012iv,Lansberg:2013qka,Gong:2013qka,Feng:2015cba,Shao:2014fca,Sun:2014gca,Lansberg:2017ozx,Feng:2018ukp,Sun:2017wxk,Feng:2019zmn,Feng:2020cvm,ColpaniSerri:2021bla,Feng:2024heh}.

{Existing quarkonium BFKL computations in the forward-backward limit, and yet valid at any $P_T$, are limited to LL and to the production of quarkonium+jet \cite{Boussarie:2017oae} and quarkonium pairs \cite{He:di-Jpsi}. We wish to provide here the missing pieces to upgrade them to NLL.} Quarkonium production also plays a key role in the search for gluon saturation \cite{Morreale:2021pnn}. However, within the saturation formalism, most calculations are currently limited to leading logarithmic accuracy \cite{Kang:2013hta,Ma:2014mri,Salazar:2021mpv,Kang:2023doo,Cheung:2024qvw,Gimeno-Estivill:2024gbu}, with the only exception of exclusive diffractive quarkonium photoproduction \cite{Mantysaari:2022kdm}. 

{Recently, the one-loop {$\alpha_s^3$} corrections to $2 \to 1$ IFs for the production of 3 $S$-wave NRQCD states: $m={}^1S_0^{[1]}$, ${}^1S_0^{[8]}$ and ${}^3S_1^{[8]}$ have been computed~\cite{Nefedov:2024swu} with the help of the Gauge-Invariant Effective Field Theory (EFT)~\cite{Lipatov95} for Multi-Regge processes in QCD.  In the present paper we complete the computation started in~\cite{Nefedov:2024swu} by adding the real-emission corrections to the IFs, by verifying that the infrared divergences of the real- and virtual-emission corrections cancel and thereby consolidating the BFKL factorisation framework up to NLL for the forward-backward quarkonium-production processes.}

The paper is organised as follows. First, we present in a self-contained manner the results for the NLO impact factors, which can be used directly in numerical computations. Second, we explain in detail their derivation. In Sect.~\ref{Sec:BFKLcross}, we present the general expressions for the BFKL NLL cross section. Then in Sec.~\ref{Sec:ColorSinglet} and Sec.~\ref{Sec:ColorOctet} we collect and discuss the final expressions for the NLO corrections to the IFs for the 3 considered states ${}^1S_0^{[1]}$, ${}^1S_0^{[8]}$ and ${}^3S_1^{[8]}$. Sec.~\ref{Sec:ColorSingletDeriv} and Sec.~\ref{Sec:ColorOctetDeriv} are dedicated to the details of the derivation of the results resp. 
for the ${}^1S_0^{[1]}$ state and the ${}^1S_0^{[8]}$ and ${}^3S_1^{[8]}$ states. Finally, in Sec.~\ref{Sec:Conclusion} we give our conclusions. In the appendix~\ref{append:real-LLA-NLLA} some details of the derivation of the LL and NLL-resummed partonic cross section are discussed.  Appendix~\ref{append:exact-MEs} provides the explicit expressions for the squared matrix elements, which are too lengthy to be presented in the main text.

\section{Cross section for high-energy backward-forward heavy quarkonium associated production in the BFKL framework}
\label{Sec:BFKLcross}

{As discussed in the introduction, we focus here on the specific regime  where two particles are produced with a large rapidity separation, like the Mueller-Navelet dijets, where BFKL resummation needs to be taken into account. In the case of quarkonia, it applies to  forward-backward inclusive production of a pair of quarkonia or to the forward-backward associated inclusive production with an identified hard particle, like a jet, a heavy quark, or even an electroweak boson. 

In what follows, we focus on quarkonium-pair production in hadron--hadron collisions, with the initial-state hadrons denoted by $h$.
Although we restrict ourselves to hadronic initial states, the formalism can be straightforwardly generalised to collisions involving nuclei\footnote{If the nuclear PDF approximation can be employed.} or resolved photons.
The reaction is written as
\begin{equation}
    h(P_1) + h(P_2) \to {\cal Q}(p_1) + X + {\cal Q}(p_2)\,.
\end{equation}
} To emphasise that  ${\cal Q}(p_1)$ and ${\cal Q}(p_2)$
are highly separated in rapidity, we have put the symbol $X$ between them.

In the frame in which the incoming hadron $h(P_1)$ ($h(P_2)$)  defines the $+$ (resp. $-$) direction (neglecting the hadron masses), and denoting the squared hadronic centre-of-mass energy, $S=(P_1 + P_2)^2$, the momenta of the quarkonia expanded in this Sudakov basis\footnote{The components along $P_1$ and $P_2$ thus define the $+$ and $-$ component for any 4-vector, presently assumed to be in the c.m.s so that $P_1^+=P_2^-.$} read
\begin{eqnarray}
\label{momenta-quarkonia}
p_1^\mu &=& z_1 x_1 P_1^\mu + \frac{M_{T1}^2}{z_1 x_1 S} P_2^\mu + p_{T1}^\mu\,, \,\\
p_2^\mu &=&  \frac{M_{T2}^2}{z_2 x_2 S} P_1^\mu + z_2 x_2 P_2^\mu + p_{T2}^\mu \,,
\end{eqnarray}
where
$M_{T1,2}=\sqrt{M_{{\cal Q}_{1,2}}^2+\TT{p}{1,2}^2}$ are the transverse mass of the produced quarkonia.\footnote{Throughout this article, we use bold symbols for Euclidean vectors, so that generically $p_T^2= - \T{p}^2.$}
The quarkonium rapidities 
\begin{equation}
\label{yi}
y_i = \frac{1}2 \ln \frac{p_i^+}{p_i^-}
\end{equation}
then read
\begin{equation}
\label{y12}
y_1 = \ln \frac{z_1 x_1 \sqrt{S}}{M_{T1}}\quad \hbox{and} \quad
y_2 = \ln \frac{M_{T2}}{z_2 x_2 \sqrt{S}}\,
\end{equation}
$\phi_{1,2}$ are the azimuthal angles of the quarkonia in the transverse plane with respect to the colliding-hadron direction.

The multi-differential\footnote{in the modulus of the transverse momenta $|\TT{p}{1,2}|$, azimuthal angles $\phi_{1,2}$ and rapidities $y_{1,2}$ for each of these quarkonia. } forward-backward BFKL cross section is similar to that in collinear factorisation and reads like
\begin{eqnarray}
    &&\frac{ d \sigma}{dy_1 d|\TT{p}{1}|d\phi_1  dy_{2} d|\TT{p}{2}|d\phi_{2}}  \nonumber \\
    &&=\sum\limits_{a,b=q,\bar{q},g} \int\limits_0^1 dx_1 dx_2 f_a(x_1,\mu_F) f_b(x_2,\mu_F) \frac{ d \hat{\sigma}_{ab}}{dy_1 d|\TT{p}{1}|d\phi_1  dy_{2} d|\TT{p}{2}|d\phi_{2}}, \label{eq:coll-fact}
\end{eqnarray}
where $f_{a,b}(x,\mu_F)$ denote the usual DGLAP-evolved Parton Distribution Functions (PDFs) of parton species $a$ carrying a hadron-momentum fraction $x$ and where $\hat{\sigma}_{ab}$ is the BFKL-resummed partonic cross-section.

\begin{figure}
\hspace{.4cm}\includegraphics[width=0.90\linewidth, clip]{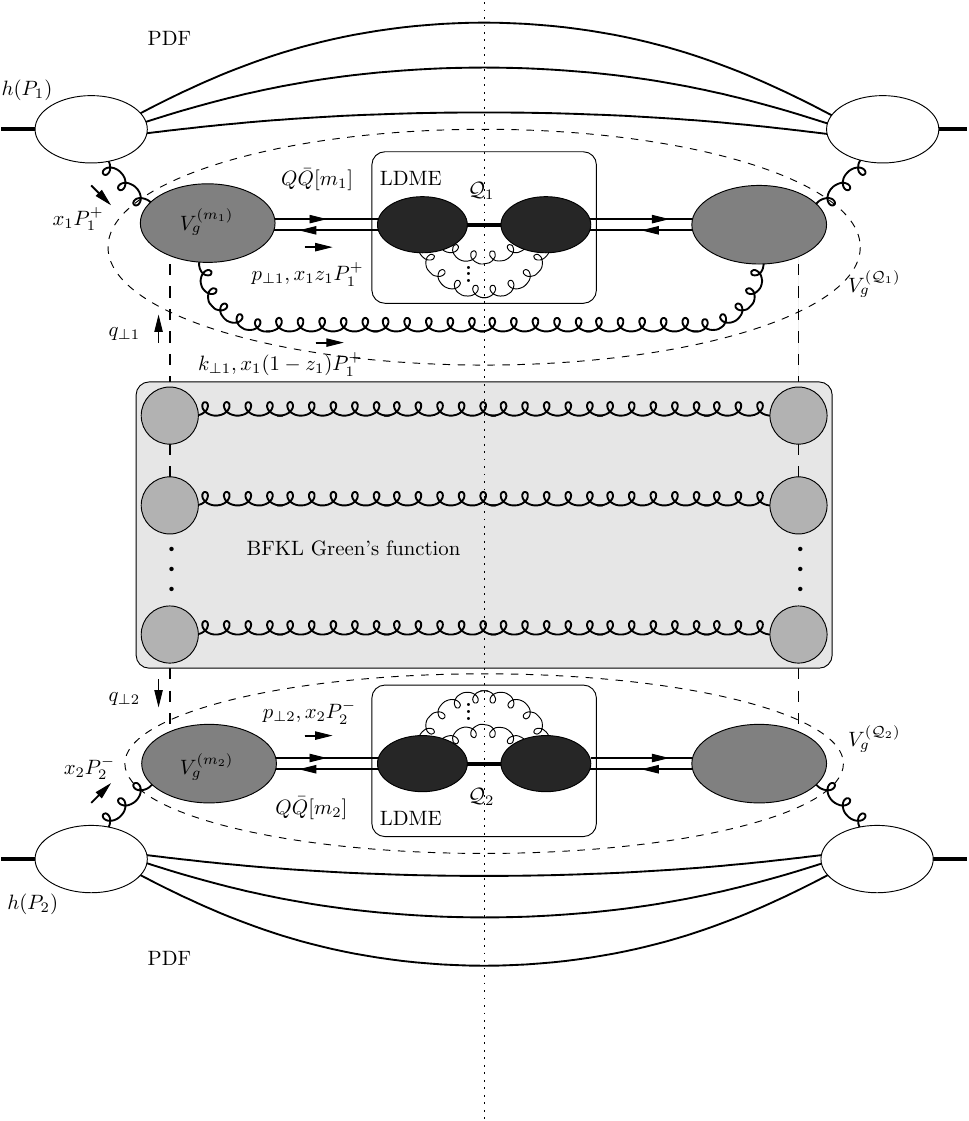}
    \caption{
    Inclusive production of a quarkonium pair with a large rapidity separation in hadron-hadron collision
    in the BFKL framework, at cross-section level. The figure shows the production mechanism in the case of two incoming on-shell gluons of momentum $x_1 P_1^+$ and $x_2 P_2^-$, out of the two scattered hadrons of momentum $P_1$ and $P_2$, whose distributions are described by two PDFs. 
    The cross section involves the convolution of the two 
 IFs $V_g^{({\cal Q}_1)}$ and $V_g^{({\cal Q}_2)}$, drawn as dashed ellipses, with a BFKL Green's function, drawn as rectangular box, which takes into account the multiple gluonic emissions treated inclusively, through the exchange of reggeons (drawn as dashed lines).
    In each IFs $V_{a_1}^{({\cal Q}_1)}$ and $V_{a_2}^{({\cal Q}_2)}$ (here with $a_1=a_2=g$), 
    a $Q \bar{Q}$ pair, of quantum numbers $m_1$ and $m_2$, are produced perturbatively, as described by the impact factors $V_{a_1}^{(m_1)}$ and $V_{a_2}^{(m_1)}$, each of them being depicted as the two upper grey blobs and the two lower blobs respectively. The hadronisation of each state $Q \bar{Q}[m_1]$ or $Q \bar{Q}[m_2]$ into each quarkonium ${\cal Q}_1$ and ${\cal Q}_2$ is encoded in the corresponding LDME, drawn as a rectangular box. This hadronisation may involve an arbitrary number of soft gluonic emissions, whose minimal number depends on the $Q \bar{Q}[m_1]$ state, 
    as illustrated by the two drawn gluons in the LDME boxes through dark blobs. As illustrated in the upper part of the diagram, in the case of NLO real emissions, a hard gluon with transverse momentum $\TT{k}{1}$ and momentum fraction $1-z_1$ may be involved in the short-distance production of the $Q \bar{Q}[m_1]$ state. }
    \label{fig:double-quarkonia-BFKL}
\end{figure}

The BFKL framework is valid in the limit of 
large rapidity separation $Y=y_1-y_2$ between quarkonia, related to the partonic Mandelstam variable $s$ in this production channel through
$s=(p_1+p_2)^2 \approx M_{T1}M_{T2}e^{Y},$ with $s \approx  z_1 z_2 x_1 x_2 S$ for $Y\gg 1$. In this framework, as illustrated by fig.~\ref{fig:double-quarkonia-BFKL},
the partonic cross section reads
\begin{eqnarray}
\label{sigma-BFKL}
    & \displaystyle\frac{  d\hat{\sigma}_{ab}}{dy_1 d|\TT{p}{1}| d\phi_1   dy_{2} d |\TT{p}{2}| d\phi_2 } =    \label{eq:sighat-res}  \\
    & \displaystyle  \!
    \int d^2\TT{q}{1} d^2\TT{q}{2} V^{({\cal Q})}_a(\TT{q}{1},\TT{p}{1},z_1;s_0) G(\ln({s}/{s_0});\TT{q}{1},\TT{q}{2})  V^{({\cal Q})}_b(\TT{q}{2},\TT{p}{2},z_2 ; s_0) , \nonumber
\end{eqnarray}
where $s_0$ is the BKFL scale. In eqn.~(\ref{sigma-BFKL}), $\TT{q}{1},\TT{q}{2}$ are the tranverse momenta of the $t-$channel reggeons oriented from the BFKL pomeron to the IFs $V^{({\cal Q})}_a$ as shown in fig.~\ref{fig:double-quarkonia-BFKL}. In the LLA the BFKL Green's function $G(Y;\TT{q}{1},\TT{q}{2})$ admits a spectral representation in terms of the eigenfunctions of the LO BFKL kernel and is therefore given by~(\cite{BFKL1,BFKL2,BFKL3}, see also~\cite{kovchegov_levin_2012}):
\begin{eqnarray}
    && G^{\text{(LLA)}}(Y;\TT{q}{1},\TT{q}{2})= \sum\limits_{n=-\infty}^{\infty}  \int \frac{d\gamma}{ 2 \pi^2 i} \frac{e^{in (\phi_1-\phi_2)}}{\TT{q}{1}^2}  \left[ \frac{\TT{q}{1}^2}{ \TT{q}{2}^2} \right]^\gamma  \exp \left[ \frac{\alpha_s C_A}{\pi} \chi_{0}(n,\gamma) Y \right],\label{eq:G-BFKL-LLA}
\end{eqnarray}
 where $s_0$ has been fixed to be $M_{T1} M_{T2}$ so that $Y=\ln(s/s_0)$, $\chi_0(n,\gamma) = 2\psi(1) - \psi(|n|/2+\gamma) - \psi(|n|/2+1-\gamma)$ with $\psi(z)=\Gamma'(z)/\Gamma(z)$ and the contour in the $\gamma$ complex plane is going parallel to the imaginary axis, passing through the point $\gamma=1/2$.

Relying on NRQCD factorisation, the IFs $V^{({\cal Q})}_a$ can be expanded in a series in $v^2$. Taking into account contributions up to $O(v^4)$ (tab.~\ref{tab:velocity-SR-LDMEs}): 
\begin{eqnarray}
    V^{({\cal Q})}_a(\T{q},\T{p},z; s_0)&=&\sum\limits_{m={}^{2S+1}S_J^{[1,8]}} V^{(m)}_a(\T{q},\T{p},z; s_0) \frac{\langle {\cal O}^{\cal Q}[m] \rangle}{M_{\cal Q}^3} \nonumber \\
    &&+ \sum\limits_{m={}^{2S+1}P_J^{[1,8]}} V^{(m)}_a(\T{q},\T{p},z; s_0) \frac{\langle {\cal O}^{\cal Q}[m] \rangle}{M_{\cal Q}^5}. \label{eq:NRQCD-exp}
\end{eqnarray}
In the present paper we will consider only the contribution of $S$-wave intermediate states ${}^{2S+1}S_J^{[1,8]}$ to the IF, leaving the $P$-wave states for future work. The IFs of production of a state $Q\bar{Q}[m]$ at short distances can be expanded perturbatively as:
\begin{equation}
    V^{(m)}_a(\T{q},\T{p},z; s_0) = V^{(m,0)}_a(\T{q},\T{p},z) + \frac{\alpha_s(\mu_R)}{2\pi} V^{(m,1)}_a(\T{q},\T{p},z ; s_0) + O(\alpha_s^4), \label{eq:IF-as-expansion}
\end{equation}
with the LO IFs being
\begin{eqnarray}
     V^{(m,0)}_g(\T{q},\T{p},z) = |\T{p}| \delta^{(2)}(\T{p}-\T{q})\delta(1-z)  h_m^{(0)}(\T{p}^2), \label{eq:LO-IF}
\end{eqnarray}
with 
\begin{eqnarray}
h_{{}^3S_1^{[1]}}^{(0)}(\T{p}^2) &=& 0, \\
    h_{{}^1S_0^{[1,8]}}^{(0)}(\T{p}^2) &=& \alpha_s^2(\mu_R) \frac{F_{0}({}^1S_0^{[1,8]})}{8\pi^2} \frac{M^2 \sqrt{N_c^2-1}}{(M^2+\T{p}^2)^2} , \label{eq:LO-IF-1S0} \\
    h_{{}^3S_1^{[8]}}^{(0)}(\T{p}^2) &=& \alpha_s^2(\mu_R) \frac{F_{0}({}^3S_1^{[8]})}{8\pi^2} \frac{\T{p}^2 \sqrt{N_c^2-1}}{(M^2+\T{p}^2)^2} ,   \label{eq:LO-IF-3S18}
\end{eqnarray}
with $F_{0}({}^1S_0^{[1]})=64\pi^3/(N_c^2(N_c^2-1))$, $F_{0}({}^1S_0^{[8]})=64\pi^3(N_c^2-4)/(N_c(N_c^2-1)^2)$ and $F_0({}^3S_1^{[8]})=128\pi^3/(N_c^2-1)^2$. To lighten the notation, here and below, we put $M=M_{\cal Q}$.  Note that the IF for production of ${}^3S_1^{[8]}$-state vanishes for $\T{p}^2\to 0$, consistently with the results of the collinear computation in Ref.~\cite{Cho:1995vh} (see eqn.~(2.13) there), while being non-zero at finite $\T{p}$. For the CS ${}^3S_1^{[1]}$-state, the IF is equal to zero at $\T{p}=0$ due to Landau-Yang theorem and it stays zero even at finite $\T{p}$.  The LO Feynman diagrams corresponding to the expressions (\ref{eq:LO-IF-1S0}) and (\ref{eq:LO-IF-3S18}) are shown in the fig.~\ref{fig:Rg-QQbar-LO} with the details of their derivation being presented in the Appendix~\ref{append:MRK}. 

\begin{figure}
    \centering
    \includegraphics[width=0.7\linewidth]{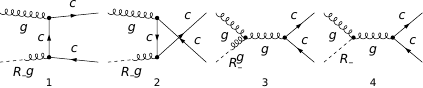}
    \caption{Feynman diagrams for the process $Rg\to Q\bar{Q}$ in the EFT~\cite{Lipatov95} \, contributing to the LO IFs~(\ref{eq:LO-IF}) -- (\ref{eq:LO-IF-3S18}). Dashed line denotes the reggeised gluon ($R$) and the lines labelled by $c$ denote a heavy quark $Q$.}
    \label{fig:Rg-QQbar-LO}
\end{figure}

The LDME for the colour-singlet state can be related with the radial part of the potential-model wavefunction at the origin $R(0)$ as $\left\langle{\cal O}\left[^1S_0^{[1]} \right] \right\rangle=2N_c |R(0)|^2/(4\pi)$ up to $\mathcal{O}(v^4)$ corrections. The CO LDMEs $\left\langle{\cal O}\left[^{2S+1}S_J^{[8]} \right] \right\rangle$ should be fitted to data.

The goal of the present paper is to compute the NLO correction to the IF $V^{(m,1)}_a$ for $a=g,q$ and $m={}^1S_0^{[1]}$, ${}^1S_0^{[8]}$ and ${}^3S_1^{[8]}$. It turns out that the key technical differences in the computation concern the distinction between {colour-singlet(CS, ${}^{[1]}$)} and {colour-octet (CO, ${}^{[8]}$)} cases, while the difference between $S=0$ and $S=1$ cases is marginal. Therefore in the next two sections we present respectively the results for the NLO IFs of production of the CS and CO $Q\bar{Q}$-state and then proceed with explaining the details of their derivation.

\section{Result for the NLO corrections to the BFKL IFs for the ${}^1S_0^{[1]}$ state}
\label{Sec:ColorSinglet}

\subsection{The subtraction approach for the real-emission part}

The IF is a function of $z$, $\T{q}$ and $\T{p}$, which should be integrated over $\T{q}$ with the Green's function in the eqn.~(\ref{eq:sighat-res}) {and over $x_{1,2}$ with the PDFs in eqn.~(\ref{eq:coll-fact}). We recall (eqn.~(\ref{y12})) that the variables $x_{1,2}$ are related to the variables $z_{1,2}=M_{T1,2}e^{y_{1,2}}/(x_{1,2}\sqrt{S})$}. These integrations will contain collinear ($\T{p}-\T{q}\to 0$), soft ($\T{p}-\T{q}\to 0$ and $z\to 1$) and rapidity ($z\to 1$) divergences which should be isolated. These will either cancel against the corresponding $1/\epsilon$ and $1/\epsilon^2$ poles in the virtual corrections to the IF, computed in Ref.~\cite{Nefedov:2024swu}, or should be properly subtracted according to the chosen factorisation scheme. The real-emission contributions to the IF are too complicated for this purpose. Therefore in the present paper we use the subtractive approach, similar to one used in Ref.~\cite{Hentschinski:2020tbi} to isolate the divergences. We add and subtract a suitable function, \({\cal J}\), which removes all singularities from the reduced squared matrix element, $ \tilde{H} $, and for which the divergences can be computed in closed form:
\begin{eqnarray}
    V_{a}^{(m,1,\text{ fin.)}} (\T{q},\T{p},z)&=& \frac{z |\T{p}|}{\pi} h_m^{(0)}(\T{p}^2) \nonumber \\
    &&\times\bigg[ \Tilde{H}^{(m)}_{Ra}(\T{q},\T{p}, z) - \left.{\cal J}^{(m)}_{Ra}(\T{q}-\T{p},\T{p}, z)\right\vert_{\begin{tiny}\begin{array}{c}r=0,\\\epsilon=0\end{array}\end{tiny}} \bigg],  \label{eq:IF-NLO-num}
\end{eqnarray}
where the subscript \(a\) labels a generic collinear parton (\(a=q\) for a quark and \(a=g\) for a gluon) interacting with a reggeised gluon, denoted by \(R\). See the Appendix~\ref{append:QMRK} for the details of the definition of $ \Tilde{H}^{(m)}_{Ra}$.  The subtraction term ${\cal J}$ in general depends on a rapidity regulator $r$ and on the dimensional regularisation parameter $\epsilon$, which becomes important for the computations in Sec.~\ref{Sec:ColorSingletDeriv} below. However, in eqn.~(\ref{eq:IF-NLO-num}), the regularisation parameters should be set to zero, since this equation by construction does not contain IR, collinear or rapidity divergences. The expressions for $\Tilde{H}^{(m)}_{Ra}$ are too long to be shown in the main text, so we have moved them to the Appendix~\ref{append:exact-MEs}, see eqns.~(\ref{eq:Htil-1S01}) -- (\ref{eq:Hq-3S18}) there. The subtraction terms in eqn.~(\ref{eq:IF-NLO-num}) read
\begin{eqnarray}
  {\cal J}^{(m)}_{Rg}(\T{k},\T{p},z) &=& \frac{2C_A}{\T{k}^2} \bigg[ \frac{1-z}{(1-z)^2 + r\frac{\T{k}^2}{q_+^2}} + \frac{1}{z} +z(1-z) -2 + \Delta {j}^{(m,\text{pol.})}_{Rg} (\T{k},\T{p},z) \bigg]   \nonumber \\
  &&  + \Delta {\cal J}^{(m,\text{CO})}_{Rg} (\T{k},\T{p},z), \label{eq:sub-term-Rg} \\
 {\cal J}^{(m)}_{Rq}(\T{k},\T{p},z) &=& \frac{C_F}{\T{k}^2} \bigg[ \frac{1+(1-z)^2}{z}-\epsilon z  + \Delta {j}^{(m,\text{pol.})}_{Rq} (\T{k},\T{p},z) \bigg] , \label{eq:sub-term-Rq}  
\end{eqnarray}
where $q_+$ is the large light-cone component of the momentum of the incoming on-shell gluon, which is equal to $x_1P_1^+$ ($x_2P_2^-$) for the IF for the production of a quarkonium with positive (negative) rapidity. Please note that $\T{k}$ in eqns.~(\ref{eq:sub-term-Rg}) and (\ref{eq:sub-term-Rq}) is not necessarily the transverse momentum of an emitted gluon $\T{q}-\T{p}$ but should rather be understood as the argument of the function ${\cal J}^{(m)}_{Ra}$. The polarisation corrections $\Delta {j}^{(m,\text{pol.})}_{Rg}$ and $\Delta {j}^{(m,\text{pol.})}_{Rq}$ contain the dependence of the subtraction term on the azimuthal angle of $\T{k}$, which turns out to be different for the $Q\bar{Q}$ states with different total angular momentum. For the $J=0$ states, ${}^1S_0^{[1]}$ and ${}^1S_0^{(8)}$, the following correction terms arise in the gluon and quark channels:
\begin{eqnarray}
  \Delta {j}^{({}^1S_0^{[1]},\text{pol.})}_{Rg}= \Delta {j}^{({}^1S_0^{[8]},\text{pol.})}_{Rg} = \frac{1-z}{z} \bigg( 1-\frac{2(\T{k}\cdot \T{p})^2}{\T{k}^2\T{p}^2}+\epsilon \bigg),  \label{eq:DeltaJ-pol-g}\\
  \Delta {j}^{({}^1S_0^{[1]},\text{pol.})}_{Rq}= \Delta {j}^{({}^1S_0^{[8]},\text{pol.})}_{Rq} = \frac{2(1-z)}{z}\bigg( 1-\frac{2(\T{k}\cdot \T{p})^2}{\T{k}^2\T{p}^2}+\epsilon \bigg). \label{eq:DeltaJ-pol-q}
\end{eqnarray}
While for the $J=1$ state ${}^3S_1^{[8]}$, these corrections are absent: 
\begin{eqnarray}
     \Delta {j}^{({}^3S_1^{[8]},\text{pol.})}_{Rg} = \Delta {j}^{({}^3S_1^{[8]},\text{pol.})}_{Rq} =0.  
\end{eqnarray}
The eqns.~(\ref{eq:DeltaJ-pol-g}) and (\ref{eq:DeltaJ-pol-q}) have a useful property that they vanish upon integration over the azimuthal angle of $\T{k}$ up to $O(\epsilon^2)$:
\begin{eqnarray}
    \int d^{2-2\epsilon}\Omega_{\T{k}} \Delta {j}^{({}^1S_0^{[1]},\text{pol.})}_{Ra} (\T{k},\T{p},z) = \int d^{2-2\epsilon}\Omega_{\T{k}} \Delta {j}^{({}^1S_0^{[8]},\text{pol.})}_{Ra} (\T{k},\T{p},z) =O(\epsilon^2),  \label{eq:phi-dep-vanishing-property}
\end{eqnarray}
which is important for the derivation of the analytic part of the IF in Sec.~\ref{Sec:SubtractionTerm}.

The term $\Delta {\cal J}^{(m,\text{CO})}_{Rg}$ in eqn.~(\ref{eq:sub-term-Rg}) is nonzero only for colour-octet states ${}^1S_0^{[8]}$ and ${}^3S_1^{[8]}$ and will be discussed in the Sec.~\ref{sec:Sub-term-CO-soft} below.

The subtraction terms (\ref{eq:sub-term-Rg}) and (\ref{eq:sub-term-Rq}) are chosen in such a way as to subtract the singular behaviour of the exact matrix element in the first term of eqn.~(\ref{eq:IF-NLO-num}).  In eqns.~(\ref{eq:sub-term-Rg}) and (\ref{eq:sub-term-Rq}), we show the $D=4-2\epsilon$ dimensional  versions of the subtraction terms, which also depend on the rapidity regulator variable $r$ and will be used in the derivations of the Sec.~\ref{Sec:SubtractionTerm}. However, in eqn.~(\ref{eq:IF-NLO-num}), one should put $\epsilon=0$ and $r=0$ such that the subtraction term cancels the singular behaviour from the exact four-dimensional matrix element in the first term. We describe the details of the derivation of the subtraction terms in the beginning of Sec.~\ref{Sec:SubtractionTerm}. 

\subsection{The analytic contribution for the $g+R$ channel}
Besides the subtracted real-emission correction~(\ref{eq:IF-NLO-num}), the NLO correction to the IF for $a=g$ includes the following ``analytic'' contribution which contains the virtual correction and all the distributions, such as $\delta$-functions and $(+)$-distributions in $\T{q}$ and $z$, so that it can be safely integrated over these variables when substituted to eqns.~(\ref{eq:coll-fact}) and (\ref{eq:sighat-res}):
\begin{eqnarray}
 &&  V_{g}^{\text{($m$, 1, analyt.)}} (\T{q},\T{p},z; s_0) = z |\T{p}| h_m^{(0)}(\T{p}^2) \Bigg\{ 2C_A \int \frac{d^2 \T{k}}{\pi} K_{\text{BFKL}}(\T{q},\T{k},\T{p})   \nonumber \\
 &&\times\bigg[ \frac{1}{(1-z)_+}  + \frac{1}{z} +z(1-z) -2 + \left.\Delta {j}^{(m,\text{pol.})}_{Rg} (\T{k},\T{p},z)\right\vert_{\epsilon=0}  + \delta(1-z)\ln\left(\frac{\sqrt{s_0}}{|\T{k}|}\right) \bigg] \nonumber \\
 &&   + \delta^{(2)}(\T{q}-\T{p}) \bigg\{ -\ln\frac{\mu_F^2}{\T{p}^2} P_{gg}(z)   +\delta(1-z) \bigg[ -\frac{\pi^2}{6}C_A +\frac{4}{3}C_A - \frac{5}{6}\beta_0 - 2C_F \left( 2 + \frac{3}{2} \ln \frac{\T{p}^2}{m_Q^2} \right) \nonumber \\
 && +\beta_0 \ln \frac{\mu_R^2}{\T{p}^2} + F^{\text{(virt.)}}_{m}(\T{p}^2/M^2) \bigg] \bigg\} \Bigg\} , \label{eq:Vg-NLO-analyt-CS}
\end{eqnarray}
where $m_Q=M/2$, $P_{gg}(z)=2C_A\left[ \frac{z}{(1-z)_+} + \frac{1-z}{z} + z(1-z) \right] + \frac{\beta_0}{2}\delta(1-z)$. The dependence of eqn.~(\ref{eq:Vg-NLO-analyt-CS}) on the energy scale $s_0$ cancels the $s_0$-dependence of the Green's function in eqn.~(\ref{eq:sighat-res}) up to next-to-NL accuracy. The LO BFKL kernel in eqn.~(\ref{eq:Vg-NLO-analyt-CS}) is
\begin{equation}
    K_{\text{BFKL}}(\T{q},\T{k},\T{p}) = \frac{1}{\T{k}^2}\left[ \delta^{(2)}(\T{q}-\T{k}-\T{p}) - \frac{\T{p}^2}{\T{p}^2 + \T{k}^2} \delta^{(2)}(\T{q}-\T{p}) \right].\label{eq:K-BFKL}
\end{equation}

Note that, when integrated with a smooth function of $\T{q}$, such as the BFKL Green's function in eqn.~(\ref{eq:sighat-res}), the BKFL kernel (\ref{eq:K-BFKL}) gives:
\begin{equation}
    \int d^2\T{q}\, G(\T{q}) K_{\text{BFKL}}(\T{q},\T{k},\T{p}) = \frac{1}{\T{k}^2}\bigg[ G(\T{p}+\T{k}) - \frac{\T{p}^2}{\T{p}^2+\T{k}^2} G(\T{p}) \bigg],\label{eq:K-BFKL-IR-safety}
\end{equation}
which does not have a singularity at $\T{k}\to 0$.  The result (\ref{eq:Vg-NLO-analyt-CS}) is threfore IR-safe. This form of the LO BFKL kernel is somewhat different from other representations  which can be found in the literature. We prefer to use this form because eqn.~(\ref{eq:K-BFKL-IR-safety}) is a smooth function of $\T{p}$, which is advantageous for numerics. The denominators in (\ref{eq:K-BFKL}) depend only on $\T{k}^2$ and $\T{p}^2$, and not on $(\T{k}+\T{p})^2$, which would complicate the analytic integration in Sec.~\ref{Sec:SubtractionTerm}.  We demonstrate the equivalence of eqn.~(\ref{eq:K-BFKL}) to the more standard forms of the kernel in Appendix~\ref{append:LO-kernel}.

 The function $F^{\text{(virt.)}}_{m}(\tau)$ in eqn.~(\ref{eq:Vg-NLO-analyt-CS}) is the finite part of the one-loop correction. For $m={}^1S_0^{[1]}$ it is given by:
\begin{equation}
F^{\text{(virt.)}}_{{}^1S_0^{[1]}}(\tau)=-\frac{10}{9} n_F + C_F C[gR\to {}^1S_0^{[1]},C_F] + C_A C[gR\to {}^1S_0^{[1]},C_A],\label{eq:F-virt-CS}
\end{equation}
where the coefficients in front of $C_F$ and $C_A$ are given by Eqns. (5.4), (5.14) -- (5.17) in Ref.~\cite{Nefedov:2024swu}. 

To summarise, the complete NLO correction to the IF for $m={}^1S_0^{[1]}$, pertaining to the eqn.~(\ref{eq:IF-as-expansion}), is:
\begin{equation}
    V_{a}^{({}^1S_0^{[1]},1)} (\T{q},\T{p},z; s_0) =  V_{a}^{({}^1S_0^{[1]},1,\text{ fin.)}} (\T{q},\T{p},z) + V_{a}^{({}^1S_0^{[1]},1,\text{ analyt.)}} (\T{q},\T{p},z; s_0), \label{eq:VNLOa-CS-decomp}
\end{equation}
where the first term is given by  eqn.~(\ref{eq:IF-NLO-num}), while the second term for $a=g$ and $a=q$ is given by  eqn.~(\ref{eq:Vg-NLO-analyt-CS}) or eqn.~(\ref{eq:Vq-NLO-result}) below. 

\subsection{The analytic contribution for the $q+R$ channel}

The quark channel ($a=q$) also opens at NLO, and in this case the following analytic part of the NLO correction should be added to the subtracted real-emission correction (\ref{eq:IF-NLO-num}):
\begin{eqnarray}
 && \hspace{-1cm} V_{q}^{\text{($m,1,$ analyt.)}} (\T{q},\T{p},z)  =  z |\T{p}| h_m^{(0)}(\T{p}^2) 
 \nonumber \\
 \times
 &\Bigg\{&C_F \int \frac{d^2 \T{k}}{\pi} K_{\text{BFKL}}(\T{q},\T{k},\T{p})  
 \left( \frac{1+(1-z)^2}{z} + \left.\Delta {j}^{(m,\text{pol.})}_{Rq} (\T{k},\T{p},z)\right\vert_{\epsilon=0}  \right) \nonumber \\
 &+&   \delta^{(2)}(\T{q}-\T{p}) \left( -\ln\frac{\mu_F^2}{\T{p}^2} P_{gq}(z) +C_F z  \right)\Bigg\} , \label{eq:Vq-NLO-result}
\end{eqnarray}
where $P_{gq}(z)=C_F(1+(1-z)^2)/z$.
\section{Results for the NLO corrections to the BFKL IFs for the ${}^1S_0^{[8]}$ and ${}^3S_1^{[8]}$ states}
\label{Sec:ColorOctet}

\subsection{Subtraction term for the CO cases}\label{sec:Sub-term-CO-soft}

In the case of the gluon channel for $S$-wave colour-octet states ($a=g$), besides the initial-state collinear and soft divergences, one encounters an additional soft divergence due to possibility of radiation of soft gluons by the total colour charge\footnote{In the case of $P$-wave states, in addition to those, one also has an additional IR-divergent contribution due to chromoelectric-dipole $S\to P$ transitions, see e.g. eqns.~(26) -- (30) in Ref.~\cite{Butenschoen:2019lef}. $P$-wave states are however beyond the scope of the present paper.} of the final-state $Q\bar{Q}$ pair. This final-state soft divergence is taken care of by the term $\Delta {\cal J}^{(m,\text{CO})}_{Rg}$ in eqn.~(\ref{eq:sub-term-Rg}) and it is the same for both states $m={}^1S_0^{[8]}$ and ${}^3S_1^{[8]}$:
\begin{eqnarray}
 && \Delta {\cal J}^{({}^1S_0^{[8]},\text{CO})}_{Rg}=\Delta {\cal J}^{({}^3S_1^{[8]},\text{CO})}_{Rg}= \Delta {\cal J}^{({\bf 8})}_{Rg}(\T{k},\T{p},z), \label{eq:sub-term-CO} \\
 && \Delta {\cal J}^{({\bf 8})}_{Rg}= \frac{2C_A}{z\T{k}^2}  \frac{(\boldsymbol{k}_T \cdot \boldsymbol{p}_T) [(\boldsymbol{k}_T- (1-z)\boldsymbol{p}_T)^2+ (1-z)^2 M^2] - (1-z) M^2 \boldsymbol{k}_T^2}{  [(\boldsymbol{k}_T - (1-z) \boldsymbol{p}_T)^2 + (1-z)^2 M^2]^2}. \label{eq:delta-sub-term-CO}
\end{eqnarray}
We have checked both analytically and numerically that eqn.~(\ref{eq:sub-term-Rg}) with the CO correction term (\ref{eq:delta-sub-term-CO}) indeed reproduces the collinear, Regge and soft limits of the exact matrix elements for the CO NRQCD states, collected in Appendix~\ref{append:exact-MEs}, where the soft-gluon limit can be defined by rescaling: $\T{k}\to \lambda\T{k}$ with $\lambda=1-z\ll 1$. The soft limit of the squared matrix element (\ref{eq:delta-sub-term-CO}) can also be derived from the standard Bremsstrahlung approximation for QCD real-emission amplitudes, see eqn.~(\ref{eq:Asoft-eik}) in Sec.~\ref{sec:Octet-soft} below. 

The subtraction term for the quark channel ($a=q$) is given by eqn.~(\ref{eq:sub-term-Rq}) both for CS and CO states.  The corresponding analytic contribution to the NLO IF  is given by  eqn.~(\ref{eq:Vq-NLO-result}) also for the CO case because the only initial-state collinear singularity introduced by quark emission
does not depend on the colour of the final state. 

\subsection{The analytic contribution for the $g+R$ channel in the CO cases}

For the case of CO-states $m={}^1S_0^{[8]}$ and ${}^3S_1^{[8]}$ one has to substitute to the eqn.~(\ref{eq:Vg-NLO-analyt-CS}) the corresponding finite part of the one-loop correction.  For the ${}^1S_0^{[8]}$-state it is given by:  
\begin{equation}
  F^{\text{(virt.)}}_{{}^1S_0^{[8]}}(\tau) =  -\frac{10n_F}{9}+C_F C[gR\to {}^1S_0^{[8]},C_F] + C_A C[gR\to {}^1S_0^{[8]},C_A], \label{eq:Fvirt_1S08}
\end{equation} 
with the coefficients in front of $C_F$ and $C_A$ given by eqns. (5.21), (5.22) in Ref.~\cite{Nefedov:2024swu}.  For the ${}^3S_1^{[8]}$-state the finite part of the virtual contribution is given by:
\begin{eqnarray}
   F^{\text{(virt.)}}_{{}^3S_1^{[8]}}(\tau) &=&  n_F C[gR\to {}^3S_1^{[8]},\text{LO},n_F] \nonumber \\
   &&+C_F C[gR\to {}^3S_1^{[8]},\text{LO},C_F] + C_A C[gR\to {}^3S_1^{[8]},\text{LO},C_A], \label{eq:Fvirt_3S18}
\end{eqnarray}
with coefficients in front of $n_F$, $C_F$ and $C_A$ given by eqns. (5.33), (5.34) and (5.35) in Ref.~\cite{Nefedov:2024swu}. All the expressions for the coefficients are also available in the computer-readable format from the appendix to Ref.~\cite{Nefedov:2024swu}.

Since the structure of IR divergences and thus the subtraction term~(\ref{eq:sub-term-Rg}) has changed in the case of CO NRQCD states $m={}^1S_0^{[8]}$ and ${}^3S_1^{[8]}$ in comparison with the CS case, the result for the analytic part of the NLO correction to the IF in eqn.~(\ref{eq:Vg-NLO-analyt-CS}), which was derived assuming that $\Delta {\cal J}^{(m,\text{CO})}_{Rg}=0$, is incomplete. The following correction term, corresponding to the additional IR-divergence in eqn.~(\ref{eq:delta-sub-term-CO}) should be {\it added} to  eqn.~(\ref{eq:Vg-NLO-analyt-CS}):
\begin{eqnarray}
     &&\Delta V_{g}^{(m,1,{\bf 8})} (\T{q},\T{p},z) =   |\T{p}| h_{m}^{(0)}(\T{p}^2) 
     \nonumber \\
     \!\!&\!\!\Bigg\{&\!\!\! 2C_A\!\!\int \frac{d^{2} \T{k}}{\pi} K_{\text{BFKL}}(\T{q},\T{k},\T{p}) 
      \frac{(\T{k} \cdot \T{p}) [(\T{k}- (1-z)\T{p})^2+ (1-z)^2 M^2] - (1-z) M^2 \T{k}^2}{  [(\T{k} - (1-z) \T{p})^2 + (1-z)^2 M^2]^2}  \nonumber \\ 
     && + C_A \delta^{(2)}(\T{q}-\T{p}) \bigg\{\delta(1-z)  \bigg[\text{Li}_2\left(-\frac{\T{p}^2}{M^2} \right) -\ln\frac{M^2+\T{p}^2}{\T{p}^2}  \ln\frac{M^2+\T{p}^2}{M^2} \bigg]  \\
     && \hspace{3.cm}- \frac{2}{(1-z)_+} \bigg[ 1 - \ln\frac{M^2+\T{p}^2}{M^2}\bigg] -\frac{2}{1-z} g\bigg(\T{p},M,\frac{|\T{p}|}{(1-z)}\bigg) \bigg\} \Bigg\}, \label{eq:DeltaVg-8-result} \nonumber
\end{eqnarray}
where the function $g$ is:
\begin{eqnarray}
   && g(\T{p},M,M_1)= \coth^{-1} \bigg( \frac{M^2 +M_1^2 + \T{p}^2}{\sqrt{M^4 + 2M^2(\T{p}^2-M_1^2) + (\T{p}^2+M_1^2)^2}} \bigg) \nonumber \\ 
&& \times\frac{\big( M^2-M_1^2+\T{p}^2 \big)\big( M^4 +2M^2 \T{p}^2 + (M_1^2+\T{p}^2)^2 \big)}{\big( (M-M_1)^2 + \T{p}^2 \big) \big( (M+M_1)^2 + \T{p}^2\big) \sqrt{M^4 + 2M^2(\T{p}^2-M_1^2) + (\T{p}^2+M_1^2)^2} } \nonumber \\
&&  + \frac{1}{2}\ln \frac{M_1^2}{M^2} - \frac{M^4 - M^2 (M_1^2-2\T{p}^2) + \T{p}^2(M_1^2+\T{p}^2)}{\big( (M-M_1)^2 + \T{p}^2 \big) \big( (M+M_1)^2 + \T{p}^2\big)},
\end{eqnarray}
with $\coth^{-1}(x)=\frac{1}{2}\ln\big(\frac{x+1}{x-1} \big)$. The function $g\bigg(\T{p},M,\frac{|\T{p}|}{(1-z)}\bigg)\to 0$ for $z\to 1$, so  its contribution to eqn.~(\ref{eq:DeltaVg-8-result}) can be safely integrated over $z$ with the gluon PDF.

To summarise, one of the key results of the present paper is the complete NLO correction to the IF for $m={}^1S_0^{[8]}$ and ${}^3S_1^{[8]}$, pertaining to  eqn.~(\ref{eq:IF-as-expansion}) in the gluon channel:
\begin{equation}
    V_{g}^{(m,1)} (\T{q},\T{p},z; s_0) =  V_{g}^{(m,1,\text{ fin.)}}  + V_{g}^{(m,1,\text{ analyt.)}} +\Delta V_{g}^{(m,1,{\bf 8})} , \label{eq:Vg-NLO-CO} 
\end{equation}
where the first term is given by  eqn.~(\ref{eq:IF-NLO-num}) with the subtraction term (\ref{eq:sub-term-Rg}) taking into account the $\Delta {\cal J}^{({\bf 8})}_{Rg}$ correction (\ref{eq:delta-sub-term-CO}), the second term  is given by  eqn.~(\ref{eq:Vg-NLO-analyt-CS}) and the third term  by  eqn.~(\ref{eq:DeltaVg-8-result}). The NLO correction to the quark-channel IF ($a=q$) is given by eqn.~(\ref{eq:VNLOa-CS-decomp}) for the corresponding NRQCD state $m$.

\section{Derivation of the analytic part of the ${}^1S_0^{[1]}$ NLO IFs}
\label{Sec:ColorSingletDeriv}

In this section, eqns.~(\ref{eq:Vg-NLO-analyt-CS}) and (\ref{eq:Vq-NLO-result}) for the analytic part of the IFs for the production of a $Q\bar{Q}[{}^1S_0^{[1]}]$ NRQCD state will be derived.

\subsection{Extracting divergences from the CS subtraction term}
\label{Sec:SubtractionTerm}

In this section, we focus on the IR-divergence structure of the real-emission NLO corrections, all IR singularities of which are captured by the subtraction terms  (\ref{eq:sub-term-Rg}) and (\ref{eq:sub-term-Rq}).

 Let us first come back to the expressions (\ref{eq:sub-term-Rg}) and (\ref{eq:sub-term-Rq}) and explain their derivation. For the case of $m={}^1S_0^{[1]}$, we have started with the $D=4-2\epsilon$ dimensional version of the corresponding reduced matrix element (eqns.~(\ref{eq:Htil-1S01}) and (\ref{eq:HRq-1S01}) in Appendix~\ref{append:exact-MEs}), expressed in terms of $\T{p}$, $\T{k}=\T{q}-\T{p}$ and $z$ (with the help of eqns.~(\ref{eq:s-def}) -- (\ref{eq:y-expr})) and taken the collinear limit $\T{k}\to 0$. In this limit, we immediately obtain eqn.~(\ref{eq:sub-term-Rg}) with $r=0$ and $\Delta {\cal J}^{(m,\text{CO})}_{Rg}=0$ in the gluon channel, and eqn.~(\ref{eq:sub-term-Rq}) in the quark channel. 
 
 For the quark channel, this completes the derivation because there are no any additional soft or rapidity divergences in this channel. For the gluon channel, we check that the first term of eqn.~(\ref{eq:sub-term-Rg}) at $r=0$ which is proportional to $1/(1-z)$ also reproduces the Regge limit $z\to 1$ of the exact squared amplitude, independently on $\T{k}$. This allows us to restore its dependence on the rapidity regulator $r$. In the ``tilted Wilson line'' regularisation for rapidity divergences in Lipatov's EFT, used in Ref.~\cite{Nefedov:2024swu} and proposed in Refs.~\cite{Chachamis:2012cc,Chachamis:2013hma}, the $1-z=k^+/q^+$ is replaced by 
\begin{equation}
\frac{\tilde{k}_+}{q_+} = \frac{k_+ + r k_-}{q_+} = (1-z) + \frac{r \T{k}^2}{q_+^2(1-z)},    
\end{equation}
which explains the structure of the $r$-dependent term in eqn.~(\ref{eq:sub-term-Rg}). Finally, we have checked that, in the CS case, the subtraction term (\ref{eq:sub-term-Rg}) with $r=0$ and $\Delta {\cal J}^{(m,\text{CO})}_{Rg}=0$ reproduces the soft limit $\T{k}\to \lambda \T{k}$ with $\lambda \sim 1-z\ll 1$ of the exact reduced matrix element $\tilde{H}_{Rg}^{({}^1S_0^{[1]})}$, so the subtraction term (\ref{eq:sub-term-Rg}) is applicable in all potentially singular limits in the CS case.  For the CO cases, an additional soft singularity in $\Delta {\cal J}^{(m,\text{CO})}_{Rg}$ appears, which is discussed in Sec.~\ref{sec:Octet-soft} below.

Now we pass to the explanation on how to convert the phase-space singularities present in the subtraction terms (\ref{eq:sub-term-Rg}) and (\ref{eq:sub-term-Rq}) into explicit $1/\epsilon$ and $1/\epsilon^2$-poles in the dimensional regularisation parameter $\epsilon$, such that the remaining finite part is given in terms of well-defined distributions in $z$ and $\T{q}$. First, we rewrite the contribution we have subtracted in Eq. (\ref{eq:IF-NLO-num}) in $D=4-2\epsilon$ dimensions: 
\begin{eqnarray}
     V_{a\text{, ST}}^{(m,1)} (\T{q},\T{p},z)= \frac{2z|\T{p}|}{(2\pi)^{1-2\epsilon}} h_m^{(0)}(\T{p}^2)  {\cal J}^{(m)}_{Ra}(\T{q}-\T{p},\T{p}, z), \label{eq:V-NLO-sub-def}
\end{eqnarray}
where the subtraction terms (ST) up to $O(\epsilon)$ and for non-zero $r \ll 1$ are given in eqns.~(\ref{eq:sub-term-Rg}) and (\ref{eq:sub-term-Rq}).

To extract the divergences from the subtraction term, we rewrite Eq. (\ref{eq:V-NLO-sub-def}) as:
\begin{eqnarray}
    V_{a\text{, ST}}^{(m,1)}  &=&  \frac{2z|\T{p}| {\mu^{2\epsilon}}}{(2\pi)^{1-2\epsilon}}  h_m^{(0)}(\T{p}^2) \int d^{2-2\epsilon} \T{k} \, {\cal J}^{(m)}_{Ra}(\T{k},\T{p}, z) \delta^{(2{-2\epsilon})}(\T{q}-\T{k}-\T{p}). \label{eq:V-NLO-sub-01}
\end{eqnarray}
Then we split Eq. (\ref{eq:V-NLO-sub-01}) into a sum of two contributions:
\begin{equation}
    V_{a\text{, ST}}^{(m,1)}(\T{q},\T{p},z) = V_{a\text{, ST-1}}^{(m,1)}(\T{q},\T{p},z) + V_{a\text{, ST-2}}^{(m,1)}(\T{q},\T{p},z),
\end{equation}
where the first one contains $K_{\text{BFKL}}$ (eqn.~(\ref{eq:K-BFKL})) and is IR-finite, as explained after eqn.~(\ref{eq:K-BFKL-IR-safety}), and therefore can be written for $\epsilon=0$:
\begin{eqnarray}
    && V_{a\text{, ST-1}}^{(m,1)} = \frac{z|\T{p}|}{\pi} h_{m}^{(0)}(\T{p}^2) \int d^{2} \T{k} \, \T{k}^2 {\cal J}^{(m)}_{Ra}(\T{k},\T{p}, z) K_{\text{BFKL}}(\T{q},\T{k},\T{p}). \label{eq:V-NLO-sub-cont1} 
\end{eqnarray}
The second contribution simply contains the term which we subtracted to obtain eqn.~(\ref{eq:V-NLO-sub-cont1}) and it encapsulates all the IR-divergences regularised by dimensional regularisation:
\begin{eqnarray}
 V_{a\text{, ST-2}}^{(m,1)} &=&   \frac{2z|\T{p}| {\mu^{2\epsilon}}}{(2\pi)^{1-2\epsilon}} h_{m}^{(0)}(\T{p}^2) \delta^{(2{-2\epsilon})}(\T{q}-\T{p}) \int d^{2-2\epsilon} \T{k} \frac{\T{p}^2  {\cal J}^{(m)}_{Ra}(\T{k},\T{p}, z)}{\T{p}^2 + \T{k}^2}.  \label{eq:V-NLO-sub-cont2}
\end{eqnarray}
We note that, in the present section, we deal with the case of the CS state, so the term $\Delta {\cal J}^{(m,\text{CO})}_{Rg}$ is absent in eqn.~(\ref{eq:sub-term-Rg}).
 Both contributions (\ref{eq:V-NLO-sub-cont1}) and (\ref{eq:V-NLO-sub-cont2}) contain a rapidity divergence, which is regularised in eqn.~(\ref{eq:sub-term-Rg}) by the parameter $r\ll 1$. The rapidity logarithm at $r\ll 1$ can be made explicit using the following distributional expansion for the rapidity-divergent term in eqn.~(\ref{eq:sub-term-Rg}):
\begin{equation}
   \frac{1-z}{(1-z)^2 + r\frac{\T{k}^2}{q_+^2}} = \frac{1}{(1-z)_+} - \delta(1-z) \frac{1}{2}\ln \frac{r \T{k}^2}{q_+^2} + O(r), \label{eq:RD-expansion} 
\end{equation}
which was derived by considering the $r$-expansion of the integral of this term with an arbitrary smooth test function $f(z)$.

With the help of eqn.~(\ref{eq:RD-expansion}) the first part for the gluon case takes the form:
\begin{eqnarray}
&&\hspace{-1cm}  V_{g\text{, ST-1}}^{(m,1)} =  2C_A z|\T{p}| h_{m}^{(0)}(\T{p}^2) 
\int \frac{d^{2} \T{k}}{\pi} \left[ \frac{1}{(1-z)_+}  + \frac{1}{z} +z(1-z) -2   \right. \nonumber \\
&&+ \left.\Delta {j}^{(m,\text{pol.})}_{Rg} (\T{k},\T{p},z)\right\vert_{\epsilon=0}  - \dashuline{\delta(1-z) \frac{1}{2}\ln \frac{r \T{k}^2}{q_+^2}}  \bigg] K_{\text{BFKL}}(\T{q},\T{k},\T{p}), \label{eq:VNLO-sub1-result} 
\end{eqnarray}
and where the divergence at $z\to 1$ is regulated by the $(+)$-prescription. The appearance of $K_{\text{BFKL}}$ multiplying $\ln r$ is expected, since $\ln r$ in the highlighted term in eqn.~(\ref{eq:VNLO-sub1-result}) is a proxy for a high-energy/rapidity logarithm which we resum by solving the LL BFKL equation for the Green's function. This rapidity-divergent term should be properly subtracted according to the chosen rapidity-factorisation scheme, and this question is discussed in Sec.~\ref{sec:BFKL-CT-real}.   

For the quark case, there is no rapidity divergence and we simply have:
\begin{eqnarray}
&&\hspace{-1cm}  V_{q\text{, ST-1}}^{(m,1)}  =  C_F z|\T{p}| h_{m}^{(0)}(\T{p}^2) 
\nonumber \\
&& \times 
\int \frac{d^{2} \T{k}}{\pi} \left[ \frac{1+(1-z)^2}{z} + \left.\Delta {j}^{(m,\text{pol.})}_{Rq} (\T{k},\T{p},z)\right\vert_{\epsilon=0}\right] K_{\text{BFKL}}(\T{q},\T{k},\T{p}). \label{eq:Vq-NLO-sub1} 
\end{eqnarray}
{ The appearance of $K_{\text{BFKL}}$ in this equation may be surprising since $K_{\text{BFKL}}$ is the coefficient of the rapidity divergence which is absent in the quark case. However, in our case it appears just as an artifact of our procedure of isolating the collinear singularity at $\boldsymbol{k}_T\to 0$.} 

Coming back to the second contribution, eqn.~(\ref{eq:V-NLO-sub-cont2}), one can integrate-out the $2-2\epsilon$-dimensional azimuthal angle, taking into account eqns.~(\ref{eq:phi-dep-vanishing-property}): 
\begin{eqnarray}
 && \hspace{-10mm}  \int \frac{d^{2-2\epsilon}\Omega_{\T{k}}}{\Omega_{2-2\epsilon}} {\cal J}^{(m)}_{Rg}(\T{k},\T{p},z) = \frac{2C_A}{\T{k}^2}\bigg[ \frac{1-z}{(1-z)^2 + r\frac{\T{k}^2}{q_+^2}} + z(1-z) + \frac{1}{z}  -2 +O(\epsilon^2) \bigg], \label{eq:Av-JRg} \\
 && \hspace{-10mm}   \int \frac{d^{2-2\epsilon}\Omega_{\T{k}}}{\Omega_{2-2\epsilon}} {\cal J}^{(m)}_{Rq} (\T{k},\T{p},z) = \frac{P_{gq}(z)-\epsilon C_F z}{\T{k}^2}+O(\epsilon^2), \label{eq:Av-JRq}
\end{eqnarray}
where $\Omega_{2-2\epsilon}=2\pi^{1-\epsilon}/\Gamma(1-\epsilon)$. The $O(\epsilon)$ parts of the splitting functions in these results agree with eqns.(2.37) -- (2.44) in Ref.~\cite{Harris:2001sx}. Thus, one obtains:
\begin{eqnarray}
  \hspace{-10mm}V_{g\text{, ST-2}}^{(m,1)} &=& C_{\overline{MS}}  z|\T{p}|  \ h_m^{(0)} (\T{p}^2) \delta^{(2{-2\epsilon})}(\T{q}-\T{p})  \label{eq:V-NLO-sub2-g-01} \\
  &&  \times 2C_A\int\limits_0^\infty \frac{d\T{k}^2}{\T{k}^2} \bigg(\frac{\T{k}^2}{\mu^2}\bigg)^{-\epsilon} \frac{\T{p}^2}{\T{p}^2 + \T{k}^2}\bigg[ \frac{1-z}{(1-z)^2 + r\frac{\T{k}^2}{q_+^2}} + z(1-z) + \frac{1}{z}  -2 \bigg], \nonumber \\
   \hspace{-10mm} V_{q\text{, ST-2}}^{(m,1)} &=& C_{\overline{MS}}  z|\T{p}| h_m^{(0)}(\T{p}^2) \delta^{(2{-2\epsilon})}(\T{q}-\T{p}) \label{eq:V-NLO-sub2-q-01}  
  \int\limits_0^\infty \frac{d\T{k}^2}{\T{k}^2} \bigg(\frac{\T{k}^2}{\mu^2}\bigg)^{-\epsilon} \frac{\T{p}^2}{\T{p}^2 + \T{k}^2}\left[ P_{gq}(z)-\epsilon C_F z \right], \,
\end{eqnarray}
where $C_{\overline{MS}} =(4\pi)^\epsilon/\Gamma(1-\epsilon)$. Expanding at $r\ll 1$ using eqn.~(\ref{eq:RD-expansion}) and taking the integrals over $\T{k}^2$ leads to:
  \begin{eqnarray}
  &&  V_{g\text{, ST-2}}^{(m,1)} = C_{\overline{MS}}  z|\T{p}|  h_{m}^{(0)}(\T{p}^2) \delta^{(2{-2\epsilon})}(\T{q}-\T{p}) \label{eq:VNLO-sub-2-result} \, \\
  &&  \times \left(\frac{\mu^{2}}{\T{p}^2} \right)^\epsilon \left\{ -\frac{1}{\epsilon}P_{gg}(z) + \delta(1-z)\left[ \underline{\frac{C_A}{\epsilon^2}} + \underline{\frac{C_A}{\epsilon} \ln\frac{r \T{p}^2}{q_+^2}}  + \underline{\underline{\frac{\beta_0}{2}\frac{1}{\epsilon} }}   - \frac{\pi^2}{6} C_A  \right] + O(\epsilon^2) \right\} , \nonumber
\end{eqnarray}
for the gluon case and
\begin{eqnarray}
   && \hspace{-1cm} V_{q\text{, ST-2}}^{(m,1)} = C_{\overline{MS}} z|\T{p}|  h_{m}^{(0)}(\T{p}^2) \delta^{(2{-2\epsilon})}(\T{q}-\T{p}) \label{eq:Vq-NLO-sub2} 
   \\ &&  \times 
  \bigg( \frac{\mu^2}{\T{p}^2}\bigg)^{\epsilon} \bigg( -\frac{1}{\epsilon} + O(\epsilon)\bigg)\left[ P_{gq}(z)-\epsilon C_F z \right],
  \nonumber
\end{eqnarray}
for the quark case. 

The initial-state collinear divergences proportional to the splitting functions $P_{ab}(z)$ in eqns.~(\ref{eq:VNLO-sub-2-result}) and (\ref{eq:Vq-NLO-sub2}) are removed by the standard DGLAP counterterms, corresponding to the renormalisation of PDFs of initial-state partons in eqn.~(\ref{eq:coll-fact}) in the $\overline{MS}$ scheme, which boils down to the replacement:
\begin{equation}
    -\frac{1}{\epsilon}P_{ab}(z) \to P_{ab}(z) \ln\frac{\T{p}^2}{\mu_F^2}. \label{eq:DGLAP-CT}
\end{equation}
For the quark case, this is the only divergence to be taken care of and the sum of $V_{q\text{, ST-1}}^{(m,1)}+V_{q\text{, ST-2}}^{(m,1)}$ (Eq. (\ref{eq:Vq-NLO-sub1}) + Eq. (\ref{eq:Vq-NLO-sub2})) with the replacement (\ref{eq:DGLAP-CT}) reproduces Eq.~(\ref{eq:Vq-NLO-result}). For the gluon case, one has to take into account the loop correction and the transformation to the BKFL rapidity-factorisation scheme to cancel all the divergences remaining in eqn.~(\ref{eq:VNLO-sub-2-result}).

\subsection{Loop correction and renormalisation}\label{sec:renorm}

The one-loop correction to the IF of the process $gR\to Q\bar{Q}[{}^1S_0^{[1]}]$ with the subtracted on-shell-quark-mass-renormalisation counterterm is given in Eq.~(5.13) of Ref.~\cite{Nefedov:2024swu}. In the notation of the present paper,\footnote{The one-loop corrections are written in Ref.~\cite{Nefedov:2024swu} at the level of the amplitude, so they should be multiplied by a factor of 2 to get the interference term contributing at NLO.} it reads:
\begin{eqnarray}
      V_{g\text{, V}}^{({}^1S_0^{[1]},1)} &&= C_{\overline{MS}} \delta^{(2{-2\epsilon})}(\T{q}-\T{p})\delta(1-z) |\T{p}|  h_{{}^1S_0^{[1]}}^{(0)}(\T{p}^2)  \left(\frac{\mu^2}{\T{p}^2} \right)^{\epsilon}  \\ && \times \bigg\{ -\frac{C_A}{\epsilon^2}  
       + \frac{1}{\epsilon}\left[ C_A\ln\frac{q_+^2}{r \T{p}^2} + \beta_0 + 3C_F - C_A  \right]  + F^{\text{(virt.)}}_{{}^1S_0^{[1]}}(\T{p}^2/M^2) + O(r,\epsilon) \bigg\}, \label{eq:VNLO-1L-non-ren} \nonumber
\end{eqnarray}
where $F_{{}^1S_0^{[1]}}^{\text{(virt.)}}(\tau)$ is the finite part of the one-loop correction, see eqn.~(\ref{eq:F-virt-CS}). In Ref.~\cite{Nefedov:2024swu}, the one-loop amplitude was properly LSZ-amputated with respect to the on-shell particles, so one needs to add to it just two UV counterterms:
\begin{enumerate}
    \item The $\overline{\rm MS}$ renormalisation of $\alpha_s$ which comes with a factor 2 because $h_{m}^{(0)}\propto \alpha_s^2$:
    \[
    2\left(-\frac{\beta_0}{2\epsilon}\right),
    \]
    \item Wave-function renormalisation for two external heavy quarks in the on-shell scheme (see e.g.~\cite{Grozin:2005yg}, Chapt. 6), which is the standard renormalisation scheme for heavy-quark mass and wave-function in quarkonium physics, in which all the NLO computations are done, see e.g.~\cite{Kramer:1995nb,Gong:2008sn,Gong:2008ft,Ma:2010yw}:
    \[
    2\left[-\frac{3 C_F}{2\epsilon} - C_F \left( 2+\frac{3}{2} \ln \frac{\mu^2}{m_Q^2} \right)\right].
    \]
\end{enumerate}
The treatment of the Reggeon is different and is discussed in Sec.~\ref{sec:BFKL-CT-virt}. In total we should add to  Eq. (\ref{eq:VNLO-1L-non-ren}):
\begin{equation}
C_{\overline{MS}} \delta^{(2{-2\epsilon})}(\T{q}-\T{p})\delta(1-z) |\T{p}|  h_{m}^{(0)}(\T{p}^2)  \left\{-\frac{1}{\epsilon}\left[ \beta_0 + 3C_F \right] - 2C_F\left( 2+\frac{3}{2} \ln \frac{\mu_R^2}{m_Q^2} \right)\right\}, \label{eq:UV-CTs} 
\end{equation}
yielding:
\begin{eqnarray}
  && \hspace{-0.5cm} V_{g\text{, VR}}^{({}^1S_0^{[1]},1)} =  C_{\overline{MS}}\delta^{(2{-2\epsilon})}(\T{q}-\T{p})\delta(1-z) |\T{p}|  h_{{}^1S_0^{[1]}}^{(0)}(\T{p}^2) \left(\frac{\mu^2}{\T{p}^2} \right)^{\epsilon}  \label{eq:VNLO-virt-ren}
  \\
  && \hspace{-0.5cm} \times \!\bigg\{ \underline{ -\frac{C_A}{\epsilon^2}} - \frac{C_A}{\epsilon} \left( \underline{ \ln\frac{r\T{p}^2}{q_+^2}}\underline{\underline{+ 1}} \right)     \!+\!\beta_0 \ln \frac{\mu^2}{\T{p}^2}  \!- \!2C_F\left( 2+\frac{3}{2} \ln \frac{\T{p}^2}{m_Q^2} \right)  \!+\! F^{\text{(virt.)}}_{{}^1S_0^{[1]}}(\T{p}^2/M^2) + O(r,\epsilon)  \bigg\}.    \nonumber
\end{eqnarray}
Note that the terms underlined by a single line cancel against the corresponding terms in eqn.~(\ref{eq:VNLO-sub-2-result}). The remaining divergences, underlined by two lines in eqns.~(\ref{eq:VNLO-sub-2-result}) and (\ref{eq:VNLO-virt-ren}), will be cancelled by the counterterm corresponding to the transition to the BFKL rapidity-factorisation scheme, discussed in the next subsection.

\subsection{BFKL counterterm for the virtual part}\label{sec:BFKL-CT-virt}

The virtual part of the BFKL counterterm is derived in Sec. 6.1 of Ref.~\cite{Nefedov:2024swu}, eqn. (6.8), by matching of the result of the high-energy EFT~\cite{Lipatov95} for the Regge limit of the real part of the one-loop QCD $2\to 2$ scattering amplitude with gluon quantum numbers in the  $t$-channel on the standard one-Reggeon-exchange Ansatz:
\begin{equation}
    s\gg -t:\; \text{Re\,} {\cal M}^{\text{(QCD)}}_{2\to 2} = \frac{s}{t} \Gamma_{\text{proj.}} \bigg(\frac{s}{s_0} \bigg)^{\omega_g(t)} \Gamma_{\text{targ.}}+O(s^{0}), 
\end{equation}
where $\Gamma_{\text{proj.}}$ and $\Gamma_{\text{targ.}}$ are (amplitude-level) IFs of the projectile and target and $\omega_g(t)$ is the gluon Regge trajectory (see eqn.~(\ref{eq:1-loop-omega}) in Appendix~\ref{append:LO-kernel}). We will not repeat this derivation here and just state the result for the virtual part of the scheme-transformation term:
\begin{eqnarray}
     &&  V_{g\text{, BFKL-V}}^{(m,1)}  =  C_{\overline{MS}}\delta^{(2{-2\epsilon})}(\T{q}-\T{p})\delta(1-z) |\T{p}|  h_{m}^{(0)}(\T{p}^2)  \left(\frac{\mu^2}{\T{p}^2} \right)^{\epsilon} \nonumber \\
     && \times \bigg\{ \underline{\underline{-\frac{1}{2\epsilon}\big( \beta_0- 2C_A \big)}} \uwave{+ \frac{C_A}{\epsilon} \ln \frac{r s_0}{(q_+)^2}} +\frac{4}{3}C_A - \frac{5}{6}\beta_0   \bigg\}. \label{eq:VgBFKL-virt}
\end{eqnarray}
As anticipated above, the doubly-underlined divergent terms in eqn.~(\ref{eq:VgBFKL-virt}) cancel the corresponding terms coming from eqns.~(\ref{eq:VNLO-sub-2-result}) and (\ref{eq:VNLO-virt-ren}). These terms in eqn.~(\ref{eq:VgBFKL-virt}) have their origin from the non-rapidity-divergent part of the one-loop correction to the reggeised gluon propagator in the EFT~\cite{Chachamis:2012cc,Nefedov:2019mrg}. The $\ln r/\epsilon$ term, highlighted with the wavy line in eqn.~(\ref{eq:VgBFKL-virt}) will cancel the IR divergence of the scheme-transformation term for the real correction, derived in the next subsection (see eqn.~(\ref{eq:Vg_BFKL_R2})). 

\subsection{BFKL counterterm for the real-emission part}\label{sec:BFKL-CT-real}

The definition of a rapidity-factorisation scheme  at NLO is just the prescription on how the contribution when the emitted gluon with momentum $k$ is highly separated in rapidity from the quarkonium: $y_p-y_k\gg 1$ is subtracted. In this {\it Multi-Regge} kinematics, the probability of emission of an additional gluon factorises from the impact-factor (see also the discussion in Appendix~\ref{append:MRK}):
\begin{eqnarray}
    V_{g\text{, MRK}}^{(g,1)} =  \frac{2z|\T{p}|  {\mu^{2\epsilon}}  }{(2\pi)^{1-2\epsilon}} h_m^{(0)}(\T{p}^2) \int d^{2-2\epsilon} \T{k}\,  {\cal J}_{RR}(k)  \delta^{(2{-2\epsilon})}(\T{q}-\T{p}-\T{k}).\label{eq:V-NLO-Regge-00}
\end{eqnarray}
The emission probability, also known as the square of the Lipatov's vertex (eqn.~(\ref{eq:G-BFKL-as})), is:
\begin{equation}
    {\cal J}_{RR}(k) = \frac{1}{1-z}\frac{2C_A}{k^+ k^-},
\end{equation}
where the factor $1/(1-z)$ came from the phase-space measure of $k$. Since $k^+k^-=\T{k}^2$, the rest of the integrand, besides $1/(1-z)$-factor is $z$-independent and the integral over $z$ is divergent at $z\to 1$. 

To regularise this divergence, different regularisation schemes are employed in different approaches. In the EFT formalism~\cite{Lipatov95}, the regularisation with tilted Wilson line prescription~\cite{Chachamis:2012cc,Chachamis:2013hma,Nefedov:2019mrg,Nefedov:2024swu} ($k^+\to \tilde{k}^+= k^+ + rk^-$) leads to, as discussed in the beginning of Sec.~\ref{Sec:ColorSingletDeriv}:
\begin{equation}
    {\cal J}_{RR}^{(r)}(k)=\frac{2C_A}{\T{k}^2} \frac{1-z}{(1-z)^2+\frac{r\T{k}^2}{q_+^2}}.\label{eq:JRR_r-scheme}
\end{equation}

On the other hand, in the BFKL framework, the regularisation with the help of a cut on the invariant mass of the final state produced in the IF is customary: $s=(k+p)^2<s_\Lambda$ with $s=(k+p)^2\simeq k^-q^+ = \T{k}^2/(1-z)$. So in this approach one obtains:
\begin{equation}
    {\cal J}_{RR}^{(s_\Lambda)}(k)= \frac{2C_A}{\T{k}^2}\frac{1}{1-z}\theta\big(1-z> \frac{\T{k}^2}{s_\Lambda}\big).\label{eq:JRR_sL-scheme}
\end{equation}

To derive the scheme-transformation term, one takes the difference of eqn.~(\ref{eq:JRR_sL-scheme}) and (\ref{eq:JRR_r-scheme}):
\begin{eqnarray}
 &&   \Delta {\cal J}_{RR}(k) = {\cal J}_{RR}^{(s_\Lambda)} -{\cal J}_{RR}^{(r)} =  \frac{2C_A}{\T{k}^2}\bigg[ \frac{1}{1-z}\theta\big(1-z> \frac{\T{k}^2}{s_\Lambda}\big)-\frac{1-z}{(1-z)^2+\frac{r\T{k}^2}{q_+^2}} \bigg]\nonumber \\
 && =  \frac{2C_A}{\T{k}^2}\bigg[ \theta\big(1-z> \frac{\T{k}^2}{s_\Lambda}\big) \bigg( \frac{1}{1-z} -\frac{1-z}{(1-z)^2+\frac{r\T{k}^2}{q_+^2}} \bigg)-\frac{1-z}{(1-z)^2+\frac{r\T{k}^2}{q_+^2}  } \theta\big(1-z< \frac{\T{k}^2}{s_\Lambda}\big) \bigg]\nonumber \\
&&= \frac{2C_A}{\T{k}^2}\bigg[ O(r) - \frac{1-z}{(1-z)^2+\frac{r\T{k}^2}{q_+^2}}\theta\big(1-z< \frac{\T{k}^2}{s_\Lambda}\big) \bigg]. 
\end{eqnarray}

The obtained distribution in $z$ should be expanded in the limit $r\to 0$ and $s_\Lambda\to\infty$, using eqn. (\ref{eq:RD-expansion}) and the fact that for $s_\Lambda\to\infty$:
\begin{eqnarray}
  &&  \int\limits_0^1\frac{dz f(z) }{(1-z)_+} \theta\big(1-z<\frac{\T{k}^2}{s_\Lambda}\big) =   -f(1) \ln \frac{s_\Lambda}{\T{k}^2} + O(\T{k}^2/s_\Lambda),
\end{eqnarray}
thus one has:
\begin{equation}
    \Delta {\cal J}_{RR}(k) = -\frac{2C_A}{\T{k}^2} \delta(1-z)\bigg[ - \frac{1}{2}\ln\frac{r\T{k}^2}{q_+^2} - \ln \frac{s_\Lambda}{\T{k}^2}\bigg]+ O(r,\T{k}^2/s_\Lambda).
\end{equation}

In the BFKL formalism, the dependence on $s_\Lambda$ is cancelled by the following counterterm (see e.g. Eq. (5.19) in Ref.~\cite{Celiberto:2022fgx}):
\begin{equation}
    \Delta {\cal J}^{(s_\Lambda-\text{BFKL})}_{RR}(k) = -\frac{2C_A}{\T{k}^2} \ln\frac{s_\Lambda}{\sqrt{s_0 \T{k}^2 }},
\end{equation}
which finally leads to the the following expression for the emission probability:
\begin{equation}
    \Delta {\cal J}^{\text{(BFKL)}}_{RR}(k) = \Delta {\cal J}_{RR} + \Delta {\cal J}^{(s_\Lambda-\text{BFKL})}_{RR} = \frac{2C_A}{\T{k}^2} \delta(1-z)\bigg[\frac{1}{2} \ln\frac{r\T{k}^2}{q_+^2} + \ln \frac{\sqrt{s_0}}{|\T{k}|}\bigg].
\end{equation}
Plugging it into  Eq. (\ref{eq:V-NLO-Regge-00}) and extracting the IR divergences in the same way as in the sec.~\ref{Sec:SubtractionTerm}, one obtains:
\begin{equation}
    V_{g,\text{BFKL-R}}^{(m,1)} = V_{g,\text{BFKL-R-1}}^{(m,1)} + V_{g,\text{BFKL-R-2}}^{(m,1)}\,. \label{eq:Vg_BFKL-R_exp}
\end{equation}

Where the first contribution is:
\begin{equation}
    V_{g,\text{BFKL-R-1}}^{(m,1)} = 2C_A |\T{p}| \delta(1-z) h_m^{(0)}(\T{p}^2) \int \frac{d^2\T{k}}{\pi} K_{\text{BFKL}}(\T{q},\T{k},\T{p}) \bigg[ \dashuline{\frac{1}{2} \ln\frac{r\T{k}^2}{q_+^2}} + \ln \frac{\sqrt{s_0}}{|\T{k}|}\bigg],\label{eq:Vg_BFKL_R1}
\end{equation}
where we note that the underlined $r$-dependent logarithm cancels the corresponding logarithm in eqn.~(\ref{eq:VNLO-sub1-result}).

The second contribution in eqn.~(\ref{eq:Vg_BFKL-R_exp}) is:
  \begin{eqnarray}
    &&\hspace{-0.5cm} V_{g,\text{BFKL-R-2}}^{(m,1)} = \frac{4C_A|\T{p}| {\mu^{2\epsilon}} }{(2\pi)^{1-2\epsilon}} h_m^{(0)}(\T{p}^2)\delta^{(2{-2\epsilon})}(\T{q}-\T{p}) \delta(1-z) \int \frac{d^{2-2\epsilon}\T{k}\, \T{p}^2}{\T{k}^2 (\T{k}^2 + \T{p}^2) }    \frac{1}{2}\ln\frac{rs_0}{q_+^2} \label{eq:Vg_BFKL_R2} \\
    &&=  C_{\overline{MS}} |\T{p}| h_m^{(0)}(\T{p}^2) \delta^{(2{-2\epsilon})}(\T{q}-\T{p}) \delta(1-z) \left( \frac{\T{p}^2}{\mu^2}\right)^{-\epsilon} \bigg( \uwave{-\frac{C_A}{\epsilon} \ln\frac{rs_0}{q_+^2}} + O(\epsilon)  \bigg) \delta(1-z). \nonumber
\end{eqnarray}
 The underlined term cancels the corresponding term in Eq. (\ref{eq:VgBFKL-virt}). 

The results of eqns. (\ref{eq:VgBFKL-virt}), (\ref{eq:Vg_BFKL_R1}) and (\ref{eq:Vg_BFKL_R2}) can be combined into a single transition term to the BFKL scheme:
 \begin{eqnarray}
      &&V_{g,\text{BFKL}}^{(m,1)} =   V_{g,\text{BFKL-V}}^{(m,1)} + V_{g,\text{BFKL-R-1}}^{(m,1)} + V_{g,\text{BFKL-R-2}}^{(m,1)}  \nonumber \\
      && =  |\T{p}| h_m^{(0)}(\T{p}^2) \delta(1-z) \bigg\{ C_{\overline{MS}}\left(\frac{\mu^2}{\T{p}^2} \right)^{\epsilon} \bigg[ -\frac{1}{2\epsilon}\big(\beta_0- 2C_A \big) +\frac{4}{3}C_A - \frac{5}{6}\beta_0 \bigg] \delta^{(2)}(\T{q}-\T{p})\nonumber \\
     && +  2C_A\int \frac{d^2\T{k}}{\pi} K_{\text{BFKL}}(\T{q},\T{k},\T{p})\bigg[  \dashuline{\frac{1}{2}\ln\frac{r\T{k}^2}{q_+^2}} + \ln \frac{\sqrt{s_0}}{|\T{k}|}\bigg] \bigg\}. \label{eq:Vg-BFKL-sch}
 \end{eqnarray}
 This scheme-transformation contribution is  universal and process-independent.

 \subsection{Collecting the contributions together for the $Q\bar{Q}[{}^1S_0^{[1]}]$ state}

 The final result for the impact-factor in the gluon case is a sum of the contribution of the subtraction term (\ref{eq:VNLO-sub1-result})+(\ref{eq:VNLO-sub-2-result}), renormalised virtual  contribution (\ref{eq:VNLO-virt-ren}) and the rapidity-factorisation scheme-transition contribution for the BFKL scheme (\ref{eq:Vg-BFKL-sch}). Below we write it in a form, which emphasises the origin and cancellation of various terms:
 \begin{eqnarray}
    && V_g^{({}^1S_0^{[1]},1{,\text{analyt.}})} (\T{q},\T{p},z) =   V_{g\text{, ST-1}}^{({}^1S_0^{[1]},1)} +   V_{g\text{, ST-2}}^{({}^1S_0^{[1]},1)} + V_{g\text{, VR}}^{({}^1S_0^{[1]},1)}   + V_{g,\text{BFKL}}^{({}^1S_0^{[1]},1)} \nonumber \\
    && =   z|\T{p}| h_{{}^1S_0^{[1]}}^{(0)}(\T{p}^2) \bigg\{ 2 C_A \int \frac{d^{2} \T{k}}{\pi} \bigg[ \frac{1}{(1-z)_+} + \frac{1}{z} +z(1-z) -2  \nonumber \\
&&  + \left.\Delta {j}^{(m,\text{pol.})}_{Rg} (\T{k},\T{p},z)\right\vert_{\epsilon=0}  + \delta(1-z)\bigg( \dashuline{-\frac{1}{2}\ln \frac{r \T{k}^2}{q_+^2} }  \underbrace{+ \dashuline{ \frac{1}{2}\ln\frac{r\T{k}^2}{q_+^2}} + \ln \frac{\sqrt{s_0}}{|\T{k}|}}_{\text{(BFKL)}}\bigg) \bigg] \nonumber \\
&& \times K_{\text{BFKL}}(\T{q},\T{k},\T{p}) +  \delta^{(2{-2\epsilon})}(\T{q}-\T{p}) C_{\overline{MS}}\left(\frac{\mu^2}{\T{p}^2} \right)^{\epsilon} \bigg[ \underbrace{-\ln\frac{\mu_F^2}{\T{p}^2}P_{gg}(z)}_{\text{(ST-2)+(DGLAP-CT)}} \nonumber \\
&&  +  \delta(1-z)  \bigg(  \underbrace{\underline{\frac{C_A}{\epsilon^2}+ \frac{C_A}{\epsilon} \ln\frac{r \T{p}^2}{q_+^2}} \underline{\underline{ + \frac{\beta_0}{2}\frac{1}{\epsilon}}} - \frac{\pi^2}{6} C_A }_{\text{(ST-2)}} \underbrace{ \underline{ -\frac{C_A}{\epsilon^2}} - \frac{C_A}{\epsilon} \left( \underline{ \ln\frac{r\T{p}^2}{q_+^2}}\underline{\underline{+ 1}} \right) }_{\text{(VR)}} \nonumber \\
&& \underbrace{+\beta_0 \ln \frac{\mu^2}{\T{p}^2}  - 2C_F\left( 2+\frac{3}{2} \ln \frac{\T{p}^2}{m_Q^2} \right)  + F^{\text{(virt.)}}_{{}^1S_0^{[1]}}(\T{p}^2/M^2)}_{\text{(VR)}}  \underbrace{-\frac{1}{2\epsilon}\big(\underline{\underline{\beta_0-2C_A}} \big) +\frac{4}{3}C_A - \frac{5}{6}\beta_0}_{\text{(BFKL)}} \bigg) \bigg] \bigg\}, \nonumber \\
&&\,\label{eq:V-NLO-1S01-labels}
 \end{eqnarray}
which leads to the Eq. (\ref{eq:Vg-NLO-analyt-CS}) for $m={}^1S_0^{[1]}$. The non-labelled terms in eqn.~(\ref{eq:V-NLO-1S01-labels}) come from the contribution $ V_{g\text{, ST-1}}^{({}^1S_0^{[1]},1)}$, eqn.~(\ref{eq:VNLO-sub1-result}). 

\section{Derivation of the analytic part of CO IFs (${}^1S_0^{[8]}$, ${}^3S_1^{[8]}$)}
\label{Sec:ColorOctetDeriv}

\subsection{Additional soft divergence for $S$-wave octet\label{sec:Octet-soft}}

Considering the collinear ($\T{k}^2\to 0$) and Regge ($z\to 1$) limits of the exact real-emission matrix element $\Tilde{H}^{(m)}_{Rg}$ in eqn. (\ref{eq:IF-NLO-num}) for the CO NRQCD states $m={}^1S_0^{[8]}$ and ${}^3S_1^{[8]}$ we find that the corresponding leading-power terms proportional to $1/\T{k}^2$ and $1/(1-z)$ respectively, are reproduced by the eqn.~(\ref{eq:sub-term-Rg}) without $\Delta {\cal J}^{(m,\text{CO})}_{Rg}$-term. However we expect an additional non-integrable singularity to appear in the soft limit $k^\mu\sim (\lambda,\lambda,\lambda,\lambda)^\mu$ with $\lambda\ll 1$, due to the possibility of radiation of soft gluons by the colour charge of the CO $Q\bar{Q}$-pair. Indeed, making the following substitutions in the exact matrix element $\Tilde{H}^{(m)}_{Rg}(\T{p}+\T{k},\T{p},z)$:
\begin{equation}
    \T{k}\to \lambda \T{k}, \, (1-z)\to \lambda(1-z),
\end{equation}
and taking the limit $\lambda\ll 1$ we find:
\begin{eqnarray}
    \Tilde{H}^{(m)}_{Rg}(\T{p}+\lambda\T{k},\T{p},1-\lambda(1-z)) = \frac{1}{\lambda^3} \frac{|A_{\text{soft}}|^2}{2(1-z)N_c^2(N_c^2-1)}+O(1/\lambda^2),\label{eq:H-CO-soft-limit}
\end{eqnarray}
where
\begin{eqnarray}
   |A_{\text{soft}}|^2 &=& 2N_c^2(N_c^2-1)(1-z)\frac{2C_A}{\T{k}^2}\bigg[\frac{1}{(1-z)} +  \frac{(\boldsymbol{k}_T \cdot \boldsymbol{p}_T) }{  (\boldsymbol{k}_T - (1-z) \boldsymbol{p}_T)^2 + (1-z)^2 M^2} \nonumber \\
   &&-\frac{(1-z) M^2 \boldsymbol{k}_T^2}{  [(\boldsymbol{k}_T - (1-z) \boldsymbol{p}_T)^2 + (1-z)^2 M^2]^2}\bigg].\label{eq:Asoft-squared} 
\end{eqnarray}

In fact, the squared soft-emission amplitude (\ref{eq:Asoft-squared}) can be straightforwardly reproduced from the following  bremsstrahlung amplitude:
\begin{eqnarray}
    A_{\text{soft}}^{abcd}= f^{cae}f^{edb}\frac{n_+\cdot\varepsilon(k)}{n_+\cdot k} - f^{cde}f^{eab}\frac{n_-\cdot\varepsilon(k)}{n_-\cdot k} + f^{ceb}f^{eda}\frac{p\cdot\varepsilon(k)}{p\cdot k},\label{eq:Asoft-eik}
\end{eqnarray}
where the first term corresponds to the emission of the soft gluon (with the colour index $d$) from the Reggeon ($b$), the second -- from the incoming on-shell gluon ($c$), the third -- from the CO $Q\bar{Q}$-pair ($a$) and we parametrise the momenta of quarkonium $p$ and of the gluon $k$ in the soft limit (implying that $1-z\ll 1$) as:
\begin{eqnarray}
  &&  p^\mu = \frac{1}{2}\big( q_+ n_-^\mu + \frac{M^2+\T{p}^2}{q_+} n_+^\mu \big) + p_T^\mu, \\ 
  &&  k^\mu= \frac{1}{2}\big( q_+(1-z) n_-^\mu + \frac{\T{k}^2}{q_+(1-z)} n_+^\mu \big) + k_T^\mu.
\end{eqnarray}
The factor $1/(1-z)$ in eqn.~(\ref{eq:H-CO-soft-limit}) comes from the phase-space measure of the soft gluon: $d^2\T{k} dz/(1-z)$ (see eqn.~(\ref{eq:Htil-M2-gen})). This is yet another instance of the phenomenon observed in Refs.~\cite{Nefedov:2020ecb,vanHameren:2022mtk,vanHameren:2025hyo}, that in the soft limit the initial-state reggeised gluon gives the same contribution as if it is an ordinary on-shell gluon.

The first term in the square bracket in eqn.~(\ref{eq:Asoft-squared}) double-counts the singularity $\sim 2C_A/\big((1-z) \T{k}^2\big)$, which is already taken into account in the subtraction term (\ref{eq:sub-term-Rg}) without $\Delta {\cal J}^{(m,\text{CO})}_{Rg}$-term. Therefore we include all but this term from the square bracket of the eqn.~(\ref{eq:Asoft-squared}) into the correction for the subtraction term for the CO case~(\ref{eq:delta-sub-term-CO}).

\subsection{Extracting the additional soft divergence from the IFs in the CO case}

For the term (\ref{eq:delta-sub-term-CO}), one can again use the procedure described in section \ref{Sec:SubtractionTerm}, obtaining

\begin{eqnarray}
&& \Delta V_{g\text{, ST-1}}^{(m,1,{\bf 8})} (\T{q},\T{p},z) =  2C_A |\T{p}| h_{g}^{\text{(${\cal Q}$, LO)}}(\T{p}^2) \int \frac{d^{2} \T{k}}{\pi} K_{\text{BFKL}}(\T{q},\T{k},\T{p}) \nonumber \\
 &&\times \frac{(\boldsymbol{k}_T \cdot \boldsymbol{p}_T) [(\boldsymbol{k}_T- (1-z)\boldsymbol{p}_T)^2+ (1-z)^2 M^2] - (1-z) M^2 \boldsymbol{k}_T^2}{  [(\boldsymbol{k}_T - (1-z) \boldsymbol{p}_T)^2 + (1-z)^2 M^2]^2} ,  \label{eq:DeltaVg_octet_ST-1}
\end{eqnarray}
and
\begin{eqnarray}
&& \Delta V_{g\text{, ST-2}}^{(m,1,{\bf 8})}  =  \frac{2|\T{p}| \mu^{2\epsilon}}{(2\pi)^{1-2\epsilon}} h_{m}^{(0)}(\T{p}^2) \delta^{(2{-2\epsilon})}(\T{q}-\T{p}) \label{eq:DeltaVg_octet_ST-2} \\
&&\times  2C_A \int  \frac{d^{2-2\epsilon} \T{k}}{\T{k}^2} \frac{\T{p}^2 }{\T{p}^2 + \T{k}^2} \frac{(\boldsymbol{k}_T \cdot \boldsymbol{p}_T) [(\boldsymbol{k}_T- (1-z)\boldsymbol{p}_T)^2+ (1-z)^2 M^2] - (1-z) M^2 \boldsymbol{k}_T^2}{  [(\boldsymbol{k}_T - (1-z) \boldsymbol{p}_T)^2 + (1-z)^2 M^2]^2} \; . \nonumber
\end{eqnarray}
The contribution (\ref{eq:DeltaVg_octet_ST-1}) becomes a part of the final result, eqn.~(\ref{eq:DeltaVg-8-result}), as it stands, so the task that remains is to integrate-out $\T{k}$ in eqn. (\ref{eq:DeltaVg_octet_ST-2}).  By rescaling $\boldsymbol{k}_T \rightarrow (1-z) \boldsymbol{k}_{T}$ we ensure that the limit of the soft emitted gluon corresponds to simply taking $z\to 1$.  Then, after partial-fractioning of the denominators $\T{k}^2$ and $\T{p}^2+(1-z)^2\T{k}^2$ one obtains:
\begin{eqnarray}
&& \Delta V_{g\text{, ST-2}}^{(m,1,{\bf 8})}(\T{q},\T{p},z) =   \frac{2 |\T{p}| \mu^{2\epsilon}}{(2\pi)^{1-2\epsilon}} h_{m}^{(0)}(\T{p}^2) \delta^{(2{-2\epsilon})}(\T{q}-\T{p})  \label{eq:DeltaVg_octet_ST-2_v2} \\
&& \times 2C_A \int  d^{2-2\epsilon} \T{k} \frac{(\boldsymbol{k}_T \cdot \boldsymbol{p}_T) [(\boldsymbol{k}_T- \boldsymbol{p}_T)^2+  M^2] -  M^2 \boldsymbol{k}_T^2}{  [(\boldsymbol{k}_T - \boldsymbol{p}_T)^2 + M^2]^2} \nonumber \\
&&\times \bigg[  \frac{(1-z)^{-1-2 \epsilon}}{\T{k}^2} -   \frac{(1-z)^{1-2 \epsilon} }{\T{p}^2 + (1-z)^2 \T{k}^2}   \bigg] \; . \nonumber
\end{eqnarray}
In the first term in the square brackets, one can expand the $(1-z)^{-1-2\epsilon}$ in $\epsilon$ in a distributional sense:
\begin{equation}
    (1-z)^{-2 \epsilon -1} = - \frac{\delta (1-z)}{2 \epsilon} + \frac{1}{(1-z)_{+}} + \mathcal{O} (\epsilon) \; ,
\end{equation}
thus exposing the IR-divergence. The $\T{k}$ integral in front of it is finite, but we have to compute it up to  $O(\epsilon)$ to obtain the complete finite part of the IF. The $\T{k}$ integral of the second term in the square brackets in eqn.~(\ref{eq:DeltaVg_octet_ST-2_v2}) is also finite, and its $z$ dependence requires no special treatment due to the $(1-z)$ factor in front of the integral, so one can put $\epsilon=0$ in this term right away. Given these observations, we split eqn.~(\ref{eq:DeltaVg_octet_ST-2_v2}) into a sum of two contributions: 
\begin{eqnarray}
    \Delta V_{g\text{, ST-2}}^{(m,1,{\bf 8})} = \Delta V_{g\text{, ST-2-D}}^{(m,1,{\bf 8})} + \Delta V_{g\text{, ST-2-F}}^{(m,1,{\bf 8})}, \label{eq:VNLO-sub2-oct-exp1} 
\end{eqnarray}
where:
\begin{eqnarray}
&&  \Delta V_{g\text{, ST-2-D}}^{(m,1,{\bf 8})} =  2 C_A    \frac{2 |\T{p}| \mu^{2\epsilon}}{(2\pi)^{1-2\epsilon}} h_{m}^{(0)}(\T{p}^2) \delta^{(2{-2\epsilon})}(\T{q}-\T{p}) \bigg( -\frac{\delta(1-z)}{2 \epsilon} + \frac{1}{(1-z)_+} \bigg)  \nonumber   \\
&& \hspace{-0.5 cm} \times  \int  \frac{d^{2-2\epsilon} \T{k}}{\T{k}^2}  \frac{(\boldsymbol{k}_T \cdot \boldsymbol{p}_T) [(\boldsymbol{k}_T- \boldsymbol{p}_T)^2+  M^2] -  M^2 \boldsymbol{k}_T^2}{  [(\boldsymbol{k}_T - \boldsymbol{p}_T)^2 + M^2]^2}, \label{eq:DeltaVg_octet_ST-2-D}\\
&& \Delta V_{g\text{, ST-2-F}}^{(m,1,{\bf 8})} =  - |\T{p}|  h_{m}^{(0)}(\T{p}^2) \delta^{(2)}(\T{q}-\T{p}) \nonumber \\
&&\times  \frac{2C_A}{1-z} \int \frac{d^{2} \T{k}}{\pi }  \frac{(\boldsymbol{k}_T \cdot \boldsymbol{p}_T) [(\boldsymbol{k}_T- \boldsymbol{p}_T)^2+  M^2] -  M^2 \boldsymbol{k}_T^2}{ [ \T{k}^2 + \T{p}^2/(1-z)^2 ]  [(\boldsymbol{k}_T - \boldsymbol{p}_T)^2 + M^2]^2} \; . \label{eq:DeltaVg_octet_ST-2-F}
\end{eqnarray}

The overall factor $1/(1-z)$ in the last expression is harmless because, as we will eventually show, the $\T{k}$ integral behaves as $(1-z)^2$ for $z\to 1$. The next  subsection is devoted to the computation of the $\T{k}$ integrals above, where that in eqn.~(\ref{eq:DeltaVg_octet_ST-2-D}) should be computed up to $O(\epsilon^1)$ and that in eqn.~(\ref{eq:DeltaVg_octet_ST-2-F}) only up to $O(\epsilon^0)$.

\subsection{Computation of $\Delta V_{g\text{, ST-2-D}}^{(m,1,{\bf 8})}$ and $\Delta V_{g\text{, ST-2-F}}^{(m,1,{\bf 8})}$ }

The computation of the $\T{k}$ integrals in both eqns. (\ref{eq:DeltaVg_octet_ST-2-D}) and (\ref{eq:DeltaVg_octet_ST-2-F}) can be conveniently organised in terms of the following family of two-dimensional one-loop integrals:
\begin{equation}
    j_{M_1}(a,b)=\int\frac{2^{2\epsilon}\mu^{2\epsilon} d^{2-2\epsilon}\T{k}}{\pi^{1-2\epsilon} [\T{k}^2+M_1^2]^a [(\T{k}-\T{p})^2+M^2]^b} = \int\frac{2^{2\epsilon}\mu^{2\epsilon} d^{2-2\epsilon}\T{k}}{\pi^{1-2\epsilon} D_1^a D_2^b}. \label{eq:JM1-family}
\end{equation}
For eqn.~(\ref{eq:DeltaVg_octet_ST-2-D}) only the case $M_1=0$ is relevant while, in eqn. (\ref{eq:DeltaVg_octet_ST-2-F}), one has $M_1^2=\T{p}^2/(1-z)^2$. By expressing the scalar products of the ``loop'' momentum $\T{k}$ in terms of denominators $D_1$ and $D_2$, one rewrites eqns.~(\ref{eq:DeltaVg_octet_ST-2-D}) and (\ref{eq:DeltaVg_octet_ST-2-F}) in terms of integrals $j_{M_1}(a,b)$ as:
\begin{eqnarray}
    && \Delta V_{g\text{, ST-2-D}}^{(m,1,{\bf 8})} =  2C_A  |\T{p}|  h_{m}^{(0)}(\T{p}^2) \delta^{(2{-2\epsilon})}(\T{q}-\T{p}) \bigg( -\frac{\delta(1-z)}{2 \epsilon} + \frac{1}{(1-z)_+} \bigg)  \nonumber \\
&&  \times    \bigg[ \frac{1}{2}\bigg( (M^2+\T{p}^2)j_0(1,1) + j_0(0,1) \bigg) - M^2j_0(0,2) \bigg], \label{eq:DeltaVg_8_ST2-D_00} \\  
&& \Delta V_{g\text{, ST-2-F}}^{(m,1,{\bf 8})} =  - |\T{p}|  h_{m}^{(0)}(\T{p}^2) \delta^{(2)}(\T{q}-\T{p}) \frac{2C_A}{1-z} \nonumber \\
&& \times    \bigg[ -\frac{1}{2}\bigg( j_{M_1}(1,0) -  j_{M_1}(0,1) - \big(M^2-M_1^2+\T{p}^2 \big) j_{M_1}(1,1) \bigg) \nonumber \\
&& -M^2 j_{M_1}(0,2) + M_1^2M^2 j_{M_1}(1,2) \bigg].  \label{eq:DeltaVg_8_ST2-F_00}  
\end{eqnarray}
In the expressions above, we encounter integrals $j_{M_1}(1,2)$ and $j_{M_1}(0,2)$ and we would like to express them in terms of ``master'' integrals $j_{M_1}(0,1)$, $j_{M_1}(1,0)$ and $j_{M_1}(1,1)$. To achieve this, we note that the integrals of the family (\ref{eq:JM1-family}) satisfy the following two independent sets of integration-by-parts (IBP) identities~\cite{Chetyrkin:1981qh} for any values of $a$ and $b$:
\begin{eqnarray}
   && \int d^{2-2\epsilon}\T{k} (\T{p}\cdot\nabla_{\T{k}}) \frac{1}{D_1^a D_2^b} =0 \Rightarrow \nonumber \\
   &&  \bigg[(a-b) + a\big(  - {\bf 1}^+ {\bf 2}^- + (M^2-M_1^2+\T{p}^2){\bf 1}^+ \big) \nonumber \\
   && + b \big( {\bf 1}^-{\bf 2}^+  + (M^2-M_1^2-\T{p}^2){\bf 2}^+   \big) \bigg]j_{M_1}(a,b) =0, \label{eq:IBP1}\\
  &&  \int d^{2-2\epsilon}\T{k} \nabla_{\T{k}}^i \frac{(\T{k})_i}{D_1^a D_2^b} =0 \Rightarrow \nonumber \\
 && \bigg[ (2-2\epsilon-2a-b) +2a M_1^2 {\bf 1}^+ \nonumber \\
 && -b\big(  {\bf 1}^- {\bf 2}^+ -(M^2+M_1^2+\T{p}^2){\bf 2}^+\big)  \bigg]j_{M_1}(a,b)=0, \label{eq:IBP2}
\end{eqnarray}
where the operators ${\bf 1}^{\pm}$ and ${\bf 2}^{\pm}$ act on integrals in the family as: ${\bf 1}^{\pm}j_{M_1}(a,b)=j_{M_1}(a\pm 1,b)$ and ${\bf 2}^{\pm}j_{M_1}(a,b)=j_{M_1}(a,b\pm 1)$. 

Following a poor man's version of Laporta algorithm~\cite{Laporta:2000dsw}, we take the relations (\ref{eq:IBP1}) and (\ref{eq:IBP2}) at $a=b=1$,  the relation (\ref{eq:IBP1}) at $a=0$, $b=1$ and the relation (\ref{eq:IBP2}) for $a=1$, $b=0$ as well as for $a=0$ and $b=1$:
\begin{eqnarray}
&& -j_{M_1}(2,0)+(M^2-M_1^2+\T{p}^2)j_{M_1}(2,1) \nonumber \\
&& + j_{M_1}(0,2)+(M^2-M_1^2-\T{p}^2)j_{M_1}(1,2)=0, \\   
&&  (-1-2\epsilon)j_{M_1}(1,1) +2M_1^2j_{M_1}(2,1) - j_{M_1}(0,2)\nonumber \\
&& +(M^2+M_1^2+\T{p}^2)j_{M_1}(1,2)  =0, \\
&& -j_{M_1}(0,1)+j_{M_1}(-1,2) +(M^2-M_1^2-\T{p}^2)j_{M_1}(0,2)=0, \\
&& -2\epsilon j_{M_1}(1,0) + 2 M_1^2j_{M_1}(2,0)=0, \\
&& (1-2\epsilon) j_{M_1}(0,1)-j_{M_1}(-1,2)+(M^2+M_1^2+\T{p}^2)j(0,2)=0.
\end{eqnarray}
We solve this system of linear equations with respect to the integrals $j_{M_1}(1,2)$ and $j_{M_1}(0,2)$ which we need, as well as with respect to auxiliary integrals $j_{M_1}(2,1)$, $j_{M_1}(2,0)$, $j_{M_1}(-1,2)$ as unknowns. The solutions for $j_{M_1}(1,2)$ and $j_{M_1}(0,2)$ look as follows:
\begin{eqnarray}
    && j_{M_1}(0,2) = \frac{\epsilon}{M^2} j_{M_1}(0,1), \label{eq:tadpole-IBP} \\
    && j_{M_1}(1,2) = \frac{1}{{M^2 \big( (M-M_1)^2 + \T{p}^2 \big) \big( (M+M_1)^2 + \T{p}^2\big)}} \bigg[-2M^2 \epsilon j_{M_1}(1,0)  \nonumber \\
    && + \big( M^2 + M_1^2 + \T{p}^2 \big)\epsilon j_{M_1}(0,1) + M^2\big( M^2-M_1^2+\T{p}^2 \big) (1+2\epsilon)j_{M_1}(1,1) \bigg], \label{eq:bubble-IBP}
\end{eqnarray}

For $M_1=0$, the required master integrals are:
\begin{eqnarray}
    && j_0(0,1)=C_{\overline{MS}} \left(\frac{\mu^2}{M^2} \right)^{\epsilon} \bigg[ \frac{1}{\epsilon} + \frac{\pi^2}{6}\epsilon + O(\epsilon^2) \bigg], \label{eq:2D-tadpole-res}\\
    && j_0(1,1)=C_{\overline{MS}} \left( \frac{\mu^2}{\T{p}^2}\right)^{\epsilon} \frac{1}{M^2+\T{p}^2} \bigg[ -\frac{1}{\epsilon} + \bigg( 2\ln\frac{M^2+\T{p}^2}{M^2} - \ln\frac{\T{p}^2}{M^2} \bigg) \nonumber \\
    && +\epsilon \bigg( -2\text{Li}_2\left(-\frac{\T{p}^2}{M^2} \right) -\frac{\pi^2}{6} -\frac{1}{2}\ln^2\frac{\T{p}^2}{M^2} \nonumber \\
    && + 2\ln\frac{\T{p}^2}{M^2} \ln\frac{M^2+\T{p}^2}{M^2} - 2\ln^2\frac{M^2+\T{p}^2}{M^2} \bigg) + O(\epsilon^2)\bigg],
\end{eqnarray}
where the result for $j_0(1,1)$ was obtained with the help of Feynman's parametrisation and expanded in $\epsilon$ with the help of \texttt{HypExp} package~\cite{Huber:2005yg} with the cross check against an independent computation with the help of Mellin-Barnes method~\cite{Smirnov}.

For $M_1\neq 0$, we obtain the following results for master integrals:
\begin{eqnarray}
    && j_{M_1}(0,1)=j_0(0,1), \\
    && j_{M_1}(1,0)=C_{\overline{MS}} \left(\frac{\mu^2}{M_1^2} \right)^{\epsilon} \bigg[ \frac{1}{\epsilon} + O(\epsilon) \bigg], \\
    &&  j_{M_1}(1,1) = \frac{2}{\sqrt{M^4 + 2M^2(\T{p}^2-M_1^2) + (\T{p}^2+M_1^2)^2}} \nonumber \\
    && \times \coth^{-1} \bigg( \frac{M^2 +M_1^2 + \T{p}^2}{\sqrt{M^4 + 2M^2(\T{p}^2-M_1^2) + (\T{p}^2+M_1^2)^2}} \bigg) + O(\epsilon), \label{eq:M1-bubble-res}
\end{eqnarray}
where the result for $j_{M_1}(1,1)$ is straightforwardly obtained by integrating the corresponding Feynman's parametrisation for $\epsilon=0$.

Substituting the integrals $j_0(0,2)$, $j_{M_1}(0,2)$ and $j_{M_1}(1,2)$ from eqns.~(\ref{eq:DeltaVg_8_ST2-D_00}) and (\ref{eq:DeltaVg_8_ST2-F_00}) with the help of IBP relations (\ref{eq:tadpole-IBP}) -- (\ref{eq:bubble-IBP}) and using the results for master integrals (\ref{eq:2D-tadpole-res}) -- (\ref{eq:M1-bubble-res}) we obtain:
\begin{eqnarray}
    && \Delta V_{g\text{, ST-2-D}}^{(m,1,{\bf 8})} = |\T{p}| h_{m}^{(1)}(\T{p}^2) \delta^{(2{-2\epsilon})}(\T{q}-\T{p}) C_{\overline{MS}} \left(\frac{\mu^2}{\T{p}^2} \right)^{\epsilon}  \bigg\{ \delta(1-z)  \bigg[  \uwave{\uwave{\frac{C_A}{\epsilon}\bigg(1 - \ln\frac{M^2+\T{p}^2}{M^2}\bigg)}}   \nonumber \\
&&  +C_A \bigg( \text{Li}_2\left(-\frac{\T{p}^2}{M^2} \right) -\ln\frac{M^2+\T{p}^2}{\T{p}^2}  \ln\frac{M^2+\T{p}^2}{M^2} \bigg) + O(\epsilon)  \bigg] \nonumber \\
&& - \frac{2C_A}{(1-z)_+} \bigg[ 1 - \ln\frac{M^2+\T{p}^2}{M^2}\bigg]\bigg\},\label{eq:VNLO-sub-oct-div-result} 
\end{eqnarray}
for eqn.~(\ref{eq:DeltaVg_8_ST2-D_00}) while, for the eqn.~(\ref{eq:DeltaVg_8_ST2-F_00}), we get:
\begin{eqnarray}
&& \Delta V_{g\text{, ST-2-F}}^{(m,1,{\bf 8})}  =  -|\T{p}|  h_{m}^{(1)}(\T{p}^2) \delta^{(2)}(\T{q}-\T{p})  \\
&& \times   \frac{2C_A}{1-z}  \bigg[ \frac{\big( M^2-M_1^2+\T{p}^2 \big)\big( M^4 +2M^2 \T{p}^2 + (M_1^2+\T{p}^2)^2 \big)}{\big( (M-M_1)^2 + \T{p}^2 \big) \big( (M+M_1)^2 + \T{p}^2\big) \sqrt{M^4 + 2M^2(\T{p}^2-M_1^2) + (\T{p}^2+M_1^2)^2} } \nonumber \\ 
&& \times \coth^{-1} \bigg( \frac{M^2 +M_1^2 + \T{p}^2}{\sqrt{M^4 + 2M^2(\T{p}^2-M_1^2) + (\T{p}^2+M_1^2)^2}} \bigg) + \frac{1}{2}\ln \frac{M_1^2}{M^2} \nonumber \\
&& - \frac{M^4 - M^2 (M_1^2-2\T{p}^2) + \T{p}^2(M_1^2+\T{p}^2)}{\big( (M-M_1)^2 + \T{p}^2 \big) \big( (M+M_1)^2 + \T{p}^2\big)} + O(\epsilon) \bigg], \label{eq:VgNLO-sub-oct-fin-result} \nonumber
\end{eqnarray}
where $M_1^2=\T{p}^2/(1-z)^2$. One can check that the last expression leads to a smooth function of $\T{p}^2>0$ and goes to zero at $z\to 1$ as it should.

\subsection{Virtual part in the CO case and the full result for the CO IF}

Adding the UV counterterms, discussed in Sec.~\ref{sec:renorm} to the results for the virtual corrections to the IFs of production of $^1S_0^{[8]}$ and $^3S_1^{[8]}$ states, described in Ref.~\cite{Nefedov:2024swu} (Eqns. (5.20) and (5.30)\footnote{In Ref.~\cite{Nefedov:2024swu} the new Lorentz structures (5.28) and (5.29), for the ${}^3S_1^{[8]}$ state do not contribute to the interference with the Born IF, because they were chosen to be orthogonal to the latter one. For this reason,  we take only eqns.~(5.20) and (5.30).}) one obtains the following result for the renormalised virtual correction:
\begin{eqnarray}
  &&  V_{g\text{, VR}}^{(m, {\bf 8})} =  \delta^{(2{-2\epsilon})}(\T{q}-\T{p})\delta(1-z) |\T{p}|  h_{m}^{(0)}(\T{p}^2) C_{\overline{MS}} \left(\frac{\mu^2}{\T{p}^2} \right)^{\epsilon}   \label{eq:VNLO-virt-ren-CO}  \bigg\{ -\frac{C_A}{\epsilon^2} - \frac{C_A}{\epsilon} \bigg( { \ln\frac{r\T{p}^2}{q_+^2}} +1 \nonumber \\
  && \hspace{-6mm} +\uwave{\uwave{ 1 - \ln \frac{M^2+\T{p}^2}{M^2}}} \bigg) +\beta_0 \ln \frac{\mu^2}{\T{p}^2}  - 2C_F\left( 2+\frac{3}{2} \ln \frac{\T{p}^2}{m_Q^2} \right)  + F_m^{\text{(virt.)}}(\T{p}^2/M^2) + O(r,\epsilon)  \bigg\}, \label{eq:Vg_virt-ren-octet}
\end{eqnarray}
where the finite parts of the one-loop correction $F_{m}^{\text{(virt.)}}(\tau)$ are given in eqns.~(\ref{eq:Fvirt_1S08}) and (\ref{eq:Fvirt_3S18}) for the CO $S$-wave states.  The divergent contribution, highlighted in the second line of eqn.~(\ref{eq:Vg_virt-ren-octet}), is cancelled by the divergence which appears in eqn.~(\ref{eq:VNLO-sub-oct-div-result}) due to the radiation of soft gluons by the CO final state. The structure of other divergences in eqn.~(\ref{eq:Vg_virt-ren-octet}) is exactly the same as in eqn.~(\ref{eq:VNLO-virt-ren}) for the CS case, and therefore the rest of the derivation of the eqn.~(\ref{eq:Vg-NLO-analyt-CS}) proceeds in exactly the same way. Hence the sum of eqns.~(\ref{eq:DeltaVg_octet_ST-1}), (\ref{eq:VNLO-sub-oct-div-result}) and (\ref{eq:VgNLO-sub-oct-fin-result}) with the additional divergence cancelled constitutes the ``correction'' term to the IF (\ref{eq:DeltaVg-8-result}), which appears in the CO case:
\begin{eqnarray}
&& \Delta V_{g}^{(m,1,{\bf 8})} =  \Delta V_{g\text{, ST-1}}^{(m,1,{\bf 8})}  + \Delta V_{g\text{, ST-2-D}}^{(m,1,{\bf 8})}  +  \Delta V_{g\text{, ST-2-F}}^{(m,1,{\bf 8})}  \nonumber \\
     && +  \delta^{(2{-2\epsilon})}(\T{q}-\T{p})\delta(1-z) |\T{p}|  h_{m}^{(0)}(\T{p}^2)  \left(\frac{\mu^2}{\T{p}^2} \right)^{\epsilon} \bigg\{ - \frac{C_A}{\epsilon} \bigg(  1 - \ln \frac{M^2+\T{p}^2}{M^2} \bigg) \bigg\}. 
\end{eqnarray}

In this way, we derive eqn.~(\ref{eq:Vg-NLO-CO}).

\section{Conclusion}
\label{Sec:Conclusion}

 The results presented here constitute the first next-to-leading-order (NLO) calculation of impact factors (IFs) within the BFKL framework for inclusive production of quarkonia, taking into account the heavy-quark-mass effects exactly. { Considering inclusive quarkonium hadroproduction (i.e., gluon- or quark-initiated channels), this is also the first NLO result obtained in any small-$x$ formalism.} 
 
The main results of this paper are as follows:
\begin{enumerate}\setlength\itemsep{-1mm}
    \item the gluon-induced NLO (${\cal O}(\alpha_s^3$) IFs for the production of the
\( Q\bar{Q}[{}^1S_0^{[1]}] \) state, given in eqn.~(\ref{eq:VNLOa-CS-decomp}), and of the
\( Q\bar{Q}[{}^1S_0^{[8]}] \) and \( Q\bar{Q}[{}^3S_1^{[8]}] \) states, given in
eqn.~(\ref{eq:Vg-NLO-CO});
\item the corresponding quark-induced NLO (${\cal O}(\alpha_s^3$) IFs, given by the sum of eqn.~(\ref{eq:Vq-NLO-result}) and eqn.~(\ref{eq:IF-NLO-num}) supplemented by the subtraction term~(\ref{eq:sub-term-Rq}).
\end{enumerate}

 This achievement lays the groundwork for precision studies of quarkonium production in proton--proton collisions at the LHC in the semi-hard QCD regime and opens the door to a broad range of phenomenological applications. A first  application is the study of inclusive associated quarkonium--jet or quarkonium--pair production with a large rapidity separation in hadron-hadron collisions, as discussed in Sec.~\ref{Sec:BFKLcross}. 
 
  A second important application is the computation of cross sections for single {\it forward} quarkonium production\footnote{At high energies and in the forward region, fixed-order collinear quarkonium results feature perturbative instabilities at NLO~\cite{Schuler:1994hy,Mangano:1996kg,Feng:2015cba,Lansberg:2020ejc}, which are resolved by high-energy resummation~\cite{Lansberg:2021vie,Lansberg:2023kzf}.} at full NLL accuracy, either using unintegrated gluon distributions (see e.g.~\cite{Anikin:2011sa,Bolognino:2019pba,Bolognino:2021niq,Celiberto:2018muu}) or using High-Energy Factorisation~\cite{Catani:1990xk,Catani:1990eg,Catani:1992rn,Catani:1994sq,Collins:1991ty}, possibly matched to NLO collinear factorisation computations~\cite{Petrelli:1997ge} using the technique similar to that used in Refs.~\cite{Lansberg:2021vie,Lansberg:2023kzf}.  For this purpose the obtained results should be transferred to the appropriate rapidity-factorisation scheme by the known scheme-transformation terms~\cite{Nefedov:2024swu}.  
 
 Our results can be applied to compute cross sections for both pseudoscalar $\eta_c$ ($\eta_b$) and vector $J/\psi$ ($\Upsilon$) states, for which abundant data is available at the LHC. To obtain the complete NRQCD prediction accurate up to ${\cal O}(\alpha_s^3)$, the NLO IFs of production of $P$-wave states should be computed, which is in our future plans. This will also allow for studies of $\chi_Q$ production.

\acknowledgments

The work of M.N. is supported by binational Science Foundation grants \#2022132 and \#2021789, and by the ISF grant \#910/23 and by the Marie Sk{\l}odowska-Curie action ``RadCor4HEF'', under grant
agreement No.~101065263. The work of M.F. is supported by the ULAM fellowship program of NAWA No. BNI/ ULM/2024/1/00065 “Color glass condensate effective theory beyond the eikonal approximation”. M.F., M.N., and L.S. gratefully acknowledge the warm hospitality and financial support of IJCLab, where part of this work was carried out. This work was supported by the "P2I - Graduate School of Physics", in the framework ``Investissements d’Avenir'' (ANR-11-IDEX-0003-01) managed by the Agence Nationale de la Recherche (ANR), France. This work was also supported in part by the European Union’s Horizon 2020 research and innovation program under Grant Agreement No. 824093 (Strong2020). This project has also received funding from the French Agence Nationale de la Recherche (ANR) via the grant ANR-20-CE31-0015 (``PrecisOnium'')  and was also partly supported by the French CNRS via the COPIN-IN2P3 bilateral agreement and via the IN2P3 project “QCDFactorisation@NLO”. L. S. was supported by the Grant No. 2024/53/B/ST2/00968 of the National Science Center in Poland.

\appendix

\section{Details of the derivation of the LLA and NLLA cross sections}\label{append:real-LLA-NLLA}

\subsection{LLA and Multi-Regge Kinematics}\label{append:MRK}
The goal of this appendix is to derive the real-emission LLA and NLLA contributions to the factorisation formula for the partonic cross section (\ref{eq:sighat-res}), by considering the process:
\begin{equation}
    a(q_1)+b(q_2)\to {\cal Q}_1(p_1) + g(k) + {\cal Q}_2(p_2), \label{eq:proc-MRK}
\end{equation}
with $q_{1,2}^\mu= x_{1,2} P_{1,2}^\mu$. In the LLA case, the initial partons $a=b=g$ and final-state partons are in the Multi-Regge Kinematics (MRK) limit:
\begin{eqnarray}
    p_1^+\sim q_1^+ \gg k^+ \gg p_2^+, \label{eq:MRK-1} \\
    p_1^- \ll k^- \ll p_2^-\sim q_2^-, \label{eq:MRK-2}
\end{eqnarray}
which is the kinematic region which generates the leading-logarithmic contribution to the cross section. We start with the standard expression for the partonic cross section in QCD:
\begin{eqnarray}
    && d\hat{\sigma} = \frac{|{\cal A}_{\text{MRK}}|^2}{I_{ab} N_{a}^{\text{(col.)}} N_b^{\text{(col.)}} N_a^{\text{(pol.)}} N_b^{\text{(pol.)}} } (2\pi)^D \delta^{(D)}(q_1+q_2 - p_1-k-p_2) \nonumber \\
    &&\times \frac{d^D p_1 \delta_+(p_1^2-M^2)}{(2\pi)^{D-1}} \frac{d^D k \delta_+(k^2)}{(2\pi)^{D-1}} \frac{d^D p_2 \delta_+(p_2^2-M^2)}{(2\pi)^{D-1}}, \label{eq:sigma-MRK-der-01}
\end{eqnarray}
where the ${\cal A}_{\text{MRK}}$ is the ordinary tree-level QCD amplitude of the process (\ref{eq:proc-MRK}) in the MRK limit (\ref{eq:MRK-1})+(\ref{eq:MRK-2}), the factors $N_a^{\text{(pol.)}}$ stand for the averaging over helicities of the incoming on-shell parton ($N_g^{\text{(pol.)}}=D-2=2-2\epsilon$ and $N_q^{\text{(pol.)}}=2$), the $N_a^{\text{(col.)}}$ average over colours of the parton ($N_q^{\text{(col.)}}=N_c$ and $N_g^{\text{(col.)}}=N_c^2-1$), the partonic flux factor is $I_{ab}=2(q_1+q_2)^2 = 2q_1^+ q_2^-$ and $\delta_+(p^2-m^2) = \delta(p^2-m^2)\theta(p^0)$.  

Due to conditions (\ref{eq:MRK-1}) and (\ref{eq:MRK-2}), the momentum-conservation $\delta$ function can be approximated as:
\begin{eqnarray}
    && \delta^{(D)}(q_1+q_2 - p_1-k-p_2) = 2\delta(q_1^+ - p_1^+ -\underbrace{k^+}_{\simeq 0} -\underbrace{p_2^+}_{\simeq 0}) \delta(q_2^- -\underbrace{p_1^-}_{\simeq 0} - \underbrace{k^-}_{\simeq 0} - p_2^-) \nonumber \\
    && \times \delta^{(2)}(\TT{p}{1}+\T{k}+\TT{p}{2}) \simeq 2 \delta(q_1^+ - p_1^+) \delta(q_2^- - p_2^-) \delta^{(2)}(\TT{p}{1}+\T{k}+\TT{p}{2}),  \label{eq:delta-MC-appr}
\end{eqnarray}
where the factor of $2$ comes from our definition of Sudakov components $k^{\pm} = k_{\pm} = n_{\pm}\cdot k$ with $n_{\pm}^\mu=(n^{\pm})^\mu=(1,0,0,\mp 1)^\mu$ in the $pp$ centre-of-momentum frame.

For the final-state particles, one transitions to the transverse momentum and rapidity variables in a standard way, e.g.:
\begin{eqnarray}
    \frac{d^D p_1 \delta_+(p_1^2-M^2)}{(2\pi)^{D-1}} \to \frac{d^{2-2\epsilon}\TT{p}{1} dy_1 }{2(2\pi)^{D-1}}.
\end{eqnarray}
and the MRK conditions (\ref{eq:MRK-1}) and (\ref{eq:MRK-2}) are equivalent to:
\begin{equation}
    y_1\gg y_k \gg y_2, \label{eq:MRK-3}
\end{equation}
with transverse momenta being fixed. 

\begin{figure}
    \centering
    \begin{tabular}{cc}
        \parbox{0.7\linewidth}{\includegraphics[width=\linewidth]{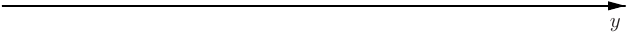}} &  \\
        \parbox{0.7\linewidth}{\includegraphics[width=\linewidth]{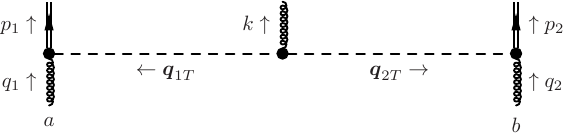}} & (a)  \\
        \parbox{0.7\linewidth}{\includegraphics[width=\linewidth]{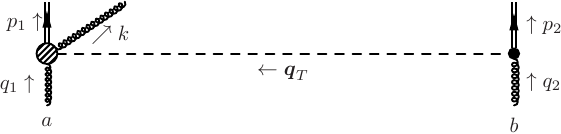}} & (b) \\
    \end{tabular}
    \caption{Diagrammatic representation for the factorisation of the amplitude (\ref{eq:proc-MRK}) in the MRK (\ref{eq:A-MRK}) and QMRK (\ref{eq:AQMRK^2}). The dashed lines denote reggeised-gluon propagators.}
    \label{fig:MRK-QMRK}
\end{figure}

The crucial step of course is the factorisation for ${\cal A}_{\text{MRK}}$ in the kinematic regime (\ref{eq:MRK-3}), which is illustrated diagrammatically by fig.~\ref{fig:MRK-QMRK}(a). The corresponding factorised form holds up to corrections $O(e^{y_k-y_1})$ and $O(e^{y_2-y_k})$ and can be written in terms of reggeised-gluon exchanges and vertices with the help of Lipatov gauge-invariant EFT for multi-Regge processes in QCD~\cite{Lipatov95}:
\begin{eqnarray}
    {\cal A}_{\text{MRK}} &=& {\cal M}(a+R^{c_1}_- (\TT{q}{1})\to {\cal Q}_1) \frac{i}{2{\TT{q}{1}^2}} \big( -g_s f^{c_1 c_k c_2} L^\mu (\TT{q}{1},\TT{q}{2},k) \varepsilon^*_\mu(k) \big) \nonumber \\
    && \times \frac{i}{2{\TT{q}{2}^2}} {\cal M}(a+R^{c_2}_+ (\TT{q}{2})\to {\cal Q}_2),\label{eq:A-MRK}
\end{eqnarray}
where $\TT{q}{1,2}=-\TT{p}{1,2}$ are transverse momenta of reggeised gluons in the $t$ channels, the factors $i/(2\TT{q}{1,2}^2)$ are the reggeised gluon propagators, the LO EFT particle-particle-Reggeon amplitudes ${\cal M}$ (``impact-factors'') for the quarkonium production processes considered in the present paper are given by diagrams in  fig.~\ref{fig:Rg-QQbar-LO} and eqns.~(3.10) -- (3.12) in Ref.~\cite{Nefedov:2024swu},  $L^{\mu}(\TT{q}{1},\TT{q}{2},k)$ is the universal Lipatov's central-emission ($R_+(\TT{q}{1})+R_-(\TT{q}{2})\to g(k)$) vertex~\cite{BFKL1,BFKL2,BFKL3}:
\begin{equation}
    L^\mu(\TT{q}{1},\TT{q}{2},k) = 2\bigg((q_{1T}-q_{2T})^\mu + \TT{q}{2}^2\frac{n^\mu_+}{k_+} - \TT{q}{1}^2\frac{n_-^\mu}{k_-} \bigg) + \big( k_+ n_-^\mu - k_-n_+^\mu \big),
\end{equation}
and $c_k$ is the colour index of an emitted gluon with momentum $k$.  The square of Lipatov's vertex is:
\begin{equation}
    \sum\limits_{\lambda=\pm 1} L^\mu L^\nu \varepsilon^*_\mu(k,\lambda) \varepsilon_\nu(k,\lambda) = \frac{16 \TT{q}{1}^2 \TT{q}{2}^2}{\T{k}^2}.\label{eq:L^2}
\end{equation}

While squaring the amplitude (\ref{eq:A-MRK}), one takes into account that the IFs summed over colours of initial and final-state particles are colour-singlet w.r.t. colour indices of Reggeons in the amplitude and complex-conjugate amplitude:
\begin{equation}
    \sum\limits_{\text{col., spins: } a,{\cal Q}} {\cal M}(a+R^{c}_- \to {\cal Q}) {\cal M}^*(a+R^{\bar{c}}_- \to {\cal Q}) = \frac{\delta_{c\bar{c}}}{N_c^2-1} |{\cal M}(a+R_- \to {\cal Q})|^2. 
\end{equation}
Thus, for the squared amplitude (\ref{eq:A-MRK}), one obtains:
\begin{equation}
    |{\cal A}_{\text{MRK}}|^2 = \frac{|{\cal M}(a+R_-(\TT{q}{1})\to {\cal Q}_1)|^2}{2\TT{q}{1}^2 \sqrt{N_c^2-1} } \frac{4C_A g_s^2}{\T{k}^2}   \frac{|{\cal M}(b+R_+(\TT{q}{2})\to {\cal Q}_2)|^2}{2\TT{q}{2}^2 \sqrt{N_c^2-1} }.
\end{equation}
Inserting this result into eqn.~(\ref{eq:sigma-MRK-der-01}), introducing the variables $z_{1}=p_1^+/q_1^+$, $z_2=p_2^-/q_2^-$ and taking into account approximations (\ref{eq:delta-MC-appr}), one can rewrite the partonic cross section in a form of eqn.~(\ref{eq:sighat-res})\footnote{Note the difference between the measures $d^2\T{p}$ and $d|\T{p}|d\phi$ w.r.t. eqn.~(\ref{eq:sighat-res}).}: 
\begin{eqnarray}
    && \frac{d\hat{\sigma}}{ d^{2-2\epsilon}\TT{p}{1} dy_1 d^{2-2\epsilon}\TT{p}{2} dy_2 } =  \int d^{2-2\epsilon}\TT{q}{1} \int d^{2-2\epsilon}\TT{q}{2}\,  V_a^{({\cal Q}_1,\text{LO})}(\TT{q}{1},\TT{p}{1},z_1) \nonumber \\
    && \times G_{O(\alpha_s)}^{\text{(real)}} (y_1-y_2,\TT{q}{1},\TT{q}{2}) V_b^{({\cal Q}_2,\text{LO})}(\TT{q}{2},\TT{p}{2},z_2). \label{eq:sigma-MRK-der-02}
\end{eqnarray}
Here, taking into account  NRQCD factorisation, the LO IFs are defined as:
\begin{equation}
    V_a^{({\cal Q},\text{LO})}(\T{q},\T{p},z) = \delta(1-z) \delta^{(2)}(\T{q}-\T{p}) \sum\limits_{m={}^{2S+1}L_J^{[1,8]}} h_m^{(0)}(\T{p}) \frac{\langle {\cal O}^{\cal Q}[m] \rangle}{M^{3+2L}} ,
\end{equation}
where
\begin{equation}
   h_m^{(0)}(\T{p}^2)=\frac{|{\cal M}(a+R_-(\T{q})\to Q\bar{Q}[m])|^2 M^{3+2L} }{4(2\pi)^{D/2-1}(q_+)^2 \T{q}^2 \sqrt{N_c^2-1} N_a^{\text{(col.)}} N_a^{\text{(pol.)}}  } .
\end{equation}
These results lead to eqns.~(\ref{eq:IF-as-expansion}) -- (\ref{eq:LO-IF-3S18}). 

In the eqn.~(\ref{eq:sigma-MRK-der-02}), the $O(\alpha_s)$ real-emission contribution to the BFKL Green's function appears:
\begin{equation}
    G_{O(\alpha_s)}^{\text{(real)}} (Y,\TT{q}{1},\TT{q}{2}) = \frac{\alpha_s C_A}{\pi^2 (2\pi)^{-2\epsilon}} \int\limits_0^Y  \frac{dy_k}{(\TT{q}{1}+\TT{q}{2})^2} , \label{eq:G-BFKL-as}
\end{equation}
which can be understood as a single iteration of the real-emission part of the LO BFKL kernel, see  eqn.~(\ref{eq:K-BFKL-DR}) below.

\subsection{Different forms of the LO BFKL kernel}
\label{append:LO-kernel}

Let us start with the form of the LO BFKL kernel, where the real and virtual parts are written explicitly in dimensional regularisation: 
\begin{equation}
  K^{\text{(dim. reg.)}}_{\text{BFKL}}(\T{q},\T{p})=  \frac{\alpha_s C_A}{\pi^2 (2\pi)^{-2\epsilon}} \frac{1}{(\T{p}-\T{q})^2} + 2\omega_g^{(1)}(\T{p}^2) \delta^{(2{-2\epsilon})}(\T{q}-\T{p}) , \label{eq:K-BFKL-DR}
\end{equation}
where the first term corresponds to the emission of a real gluon  and is equal to the square of the Lipatov's vertex (see eqn.~(\ref{eq:G-BFKL-as})), while the second term is the virtual one, with $\omega_g^{(1)}(\T{q})$ being the one-loop Regge trajectory of a gluon:
\begin{equation}
    \omega_g^{(1)}(\T{q}^2) = -\frac{\alpha_s C_A}{4\pi} \int \frac{d^{2-2\epsilon} \T{l}}{\pi (2\pi)^{-2\epsilon}} \frac{\T{q}^2}{\T{l}^2(\T{q}-\T{l})^2} = \tilde{C}_{\overline{MS}}\frac{\alpha_s C_A}{2\pi} \frac{1}{\epsilon} \left( \frac{\mu^2}{\T{q}^2} \right)^\epsilon ,\label{eq:1-loop-omega}
\end{equation}
and the factor 2 in front of the virtual term in eqn.~(\ref{eq:K-BFKL-DR}) takes into account contributions of an amplitude and complex conjugate amplitude in the interference of the one-loop and tree-level amplitudes. The factor $\tilde{C}_{\overline{MS}}=(4\pi)^{\epsilon}\Gamma^2(1-\epsilon)\Gamma(1+\epsilon)/\Gamma(1-2\epsilon)$ is different from $C_{\overline{MS}}$ defined after eqn.~(\ref{eq:V-NLO-sub2-q-01}) only at $O(\epsilon^3)$, so the difference between them is irrelevant at one loop.

To see the cancellation of IR-divergence between real-emission and virtual terms of eqn.~(\ref{eq:K-BFKL-DR}), let us convolute it with a smooth test function $G(\T{q})$ and do a few transformations:
\begin{eqnarray}
&& \int d^{2-2\epsilon}\T{q} K^{\text{(dim. reg.)}}_{\text{BFKL}}(\T{q},\T{p}) G(\T{q}) \nonumber \\
&& = \frac{\alpha_s C_A}{\pi} \bigg[ \tilde{C}_{\overline{MS}} \frac{G(\T{p})}{\epsilon}\bigg(\frac{\mu^2}{\T{p}^2} \bigg)^{\epsilon} + \int \frac{d^{2-2\epsilon} \T{q}}{\pi (2\pi)^{-2\epsilon}} \frac{G(\T{q})}{(\T{p}-\T{q})^2}  \bigg]  \nonumber \\
&& = \frac{\alpha_s C_A}{\pi} \bigg[ \int \frac{d^{2-2\epsilon} \T{q}}{\pi (2\pi)^{-2\epsilon}} \frac{G(\T{q})-G(\T{p})\Theta(\T{p},\T{q})}{(\T{p}-\T{q})^2} \nonumber \\
&& + G(\T{p})\bigg(  \frac{\tilde{C}_{\overline{MS}}}{\epsilon}\bigg(\frac{\mu^2}{\T{p}^2} \bigg)^{\epsilon} + \int \frac{d^{2-2\epsilon} \T{q}}{\pi (2\pi)^{-2\epsilon}} \frac{\Theta(\T{p},\T{q})}{(\T{p}-\T{q})^2} \bigg)\bigg]. \label{eq:KG-dim-reg-exp}
\end{eqnarray}
In order for the integral in the second line of eqn.~(\ref{eq:KG-dim-reg-exp}) to be explicitly IR and UV finite, the function $\Theta(\T{p},\T{q})$ has to satisfy two conditions: (i) it should be equal to $1$ for $\T{p}=\T{q}$ in order for IR-divergence to cancel, (ii) it should decrease at $\T{q}^2\gg \T{p}^2$ in order to avoid appearance of an artificial UV divergence at large $\T{q}$. Of course, there exist infinitely many such functions. We will examine the following examples:
\begin{eqnarray}
  &&  \Theta_1(\T{p},\T{q}) = \frac{\T{p}^2}{\T{p}^2 + (\T{p}-\T{q})^2}, \\
  &&  \Theta_2(\T{p},\T{q}) = \theta(\T{p}^2-(\T{p}-\T{q})^2), \\
  &&  \Theta_3(\T{p},\T{q}) = \frac{\T{p}^2}{\T{q}^2 + (\T{p}-\T{q})^2}\,.
\end{eqnarray}
Let us compute the integrals corresponding to the $\Theta_1$ and $\Theta_2$ functions:
\begin{eqnarray}
    && \int \frac{d^{2-2\epsilon} \T{q}}{\pi (2\pi)^{-2\epsilon}} \frac{\Theta_1(\T{p},\T{q})}{(\T{p}-\T{q})^2} = \frac{\Omega_{2-2\epsilon}}{(2\pi)^{1-2\epsilon}} \int\limits_0^{\infty} \frac{d\T{k}^2}{\T{k}^2} \left(\frac{\T{k}^2}{\mu^2}\right)^{-\epsilon} \frac{\T{p}^2}{\T{p}^2 + \T{k}^2}\nonumber \\
    &&= C_{\overline{MS}} \left(\frac{\mu^2}{\T{p}^2}\right)^{\epsilon} \Gamma(1+\epsilon)\Gamma(-\epsilon ) = C_{\overline{MS}} \left(\frac{\mu^2}{\T{p}^2}\right)^{\epsilon} \bigg( -\frac{1}{\epsilon} + O(\epsilon)\bigg), \\
    && \int \frac{d^{2-2\epsilon} \T{q}}{\pi (2\pi)^{-2\epsilon}} \frac{\Theta_2(\T{p},\T{q})}{(\T{p}-\T{q})^2} = \frac{\Omega_{2-2\epsilon}}{(2\pi)^{1-2\epsilon}} \int\limits_0^{\T{p}^2} \frac{d\T{k}^2}{\T{k}^2} \left(\frac{\T{k}^2}{\mu^2}\right)^{-\epsilon} = - \frac{C_{\overline{MS}}}{\epsilon} \left(\frac{\mu^2}{\T{p}^2}\right)^{\epsilon},
\end{eqnarray}
where we have put $\T{q}=\T{p}+\T{k}$. Doing so, one can see that, for both of these choices, the integral of $\Theta$ in the last line of eqn.~(\ref{eq:KG-dim-reg-exp}) cancels the virtual contribution up to $O(\epsilon)$ and for $\epsilon=0$ one has:
\begin{eqnarray}
    && \int d^{2-2\epsilon}\T{q} K^{\text{(dim. reg.)}}_{\text{BFKL}}(\T{q},\T{p}) G(\T{q}) \nonumber \\
    && = \frac{\alpha_s C_A}{\pi} \int \frac{d^{2-2\epsilon} \T{q}}{\pi (2\pi)^{-2\epsilon}} \frac{G(\T{q})-G(\T{p})\Theta(\T{p},\T{q})}{(\T{p}-\T{q})^2} + O(\epsilon^1).
\end{eqnarray}
Therefore, for the case of $\Theta_1$, one recovers eqn.~(\ref{eq:K-BFKL-IR-safety}) and it corresponds to our default representation of the LO BFKL kernel (\ref{eq:K-BFKL}). 

The computation of the integral with the $\Theta_3$ requires the introduction of a Feynman parameter:
\begin{eqnarray}
    && \int \frac{d^{2-2\epsilon} \T{q}}{\pi (2\pi)^{-2\epsilon}} \frac{\Theta_3(\T{p},\T{q})}{(\T{p}-\T{q})^2} =\int \frac{d^{2-2\epsilon} \T{k}}{\pi (2\pi)^{-2\epsilon}} \frac{\T{p}^2}{\T{k}^2 [(\T{p}+\T{k})^2 + \T{k}^2]} \nonumber \\
    && = \int\limits_0^{\infty} dx \int \frac{d^{2-2\epsilon} \T{k}}{\pi (2\pi)^{-2\epsilon}} \frac{\T{p}^2}{[(\T{p}+\T{k})^2 + \T{k}^2(1+x)]^2} \nonumber \\
    && = C_{\overline{MS}} \bigg(\frac{\mu^2}{\T{p}^2} \bigg)^{\epsilon} \pi\epsilon \csc(\pi\epsilon) \int\limits_0^{\infty} dx\, (1+x)^{-1-\epsilon} (2+x)^{2\epsilon} \nonumber \\
    && = C_{\overline{MS}} \bigg(\frac{\mu^2}{\T{p}^2} \bigg)^{\epsilon} \pi\epsilon \csc(\pi\epsilon) \bigg[ \int\limits_0^{\infty} dx\underbrace{\frac{(1+x)^{-\epsilon}(2+x)^{2\epsilon} - x^{\epsilon}}{1-x} }_{O(\epsilon)}  + \underbrace{\int\limits_0^{\infty} \frac{dx\, x^{\epsilon}}{1-x}}_{-\pi\csc(\pi\epsilon)}\bigg] \nonumber \\
    &&=  C_{\overline{MS}} \bigg(\frac{\mu^2}{\T{p}^2} \bigg)^{\epsilon} \bigg[-\frac{1}{\epsilon} + O(\epsilon) \bigg]. 
\end{eqnarray}
In conclusion, the integral of $\Theta_3$ also cancels the virtual part of the kernel up to $O(\epsilon)$. The corresponding explicitly IR-finite form of the kernel is used e.g. in~\cite{Forshaw:1997dc} (eqn. (4.18), Sec.~4.3).

\subsection{NLLA and Quasi-Multi-Regge Kinematics}\label{append:QMRK}

One of the contributions to the NLLA is when the MRK conditions (\ref{eq:MRK-1}) and (\ref{eq:MRK-2}) are violated and the emitted gluon carries away a significant fraction of momentum of the incoming parton $a$ in the process (\ref{eq:proc-MRK}):
\begin{eqnarray}
    p_1^+ \sim k^+ \sim q_1^+ \gg p_2^+, \label{eq:QMRK-1} \\
    p_1^- \sim k^- \ll p_2^-\sim q_2^-. \label{eq:QMRK-2}
\end{eqnarray}
This is one of the examples of Quasi-MRK (QMRK). In this kinematic limit, the squared tree-level QCD amplitude of the process (\ref{eq:proc-MRK}) factorises as:
\begin{equation}
    |{\cal A}_{\text{QMRK}}|^2 = \frac{|{\cal M}(a+R_-(\T{q})\to {\cal Q}_1+g(k))|^2}{2\T{q}^2 \sqrt{N_c^2-1} }  \frac{|{\cal M}(b+R_+(-\TT{q}{2})\to {\cal Q}_2)|^2}{2\T{q}^2 \sqrt{N_c^2-1} }, \label{eq:AQMRK^2}
\end{equation}
with the Reggeon transverse-momentum $\T{q}=-\TT{p}{2}=\TT{p}{1}+\T{k}$, see  fig.~\ref{fig:MRK-QMRK}(b). The corresponding approximation for momentum-conservation $\delta$ function is:
\begin{equation}
    \delta^{(D)}(q_1+q_2 - p_1-k-p_2) \simeq 2 \delta(q_1^+ - p_1^+-k^+) \delta(q_2^- - p_2^-) \delta(\TT{p}{1}+\T{k}+\TT{p}{2}),  \label{eq:delta-MC-appr-QMRK} 
\end{equation}
since we can no longer neglect the gluon $(+)$ momentum component. Substituting approximations (\ref{eq:AQMRK^2}) and (\ref{eq:delta-MC-appr-QMRK}) to  eqn.~(\ref{eq:sigma-MRK-der-01}) and rewriting it in a form of eqn.~(\ref{eq:sighat-res}) one obtains:
\begin{eqnarray}
    && \frac{d\hat{\sigma}}{ d^{2-2\epsilon}\TT{p}{1} dy_1 d^{2-2\epsilon}\TT{p}{2} dy_2 } =  \int d^{2-2\epsilon}\TT{q}{1} \int d^{2-2\epsilon}\TT{q}{2}\,  V_a^{({\cal Q}_1,\text{NLO, R})}(\TT{q}{1},\TT{p}{1},z_1) \nonumber \\
    && \times G_{O(\alpha_s^0)}(y_1-y_2,\TT{q}{1},\TT{q}{2}) V_b^{({\cal Q}_2,\text{LO})}(\TT{q}{2},\TT{p}{2},z_2), \label{eq:sigma-MRK-der-03}
\end{eqnarray}
where the LO in $\alpha_s$ BFKL Green's function is:
\begin{equation}
   G_{O(\alpha_s^0)}(Y,\TT{q}{1},\TT{q}{2}) = \delta(\TT{q}{1}+\TT{q}{2}), 
\end{equation}
while the NLO real-emission correction to the IF is:
\begin{eqnarray}
 &&   V_a^{({\cal Q},\text{NLO, R})}(\T{q},\T{p},z) = \int \frac{d^{2-2\epsilon}\T{k}}{ 2(2\pi)^{D-1} } \int\limits_0^\infty \frac{dk^+}{k^+} \delta(q_+ (1-z)-k_+) \delta(\T{q}-\T{p}-\T{k}) \nonumber \\
 &&   \times \frac{|{\cal M}(a(q)+R_-(\T{q})\to {\cal Q}(p)+g(k))|^2}{ 4(2\pi)^{D/2-1} q_+\T{q}^2 \sqrt{N_c^2-1} N_a^{\text{(col.)}} N_a^{\text{(pol.)}}  },
\end{eqnarray}
where the integrals over $\T{k}$ and $k^+$ can be calculated with the help of $\delta$ functions. We then obtain a rational function of $\T{q}$, $\T{p}$ and $z$.

For convenience, we express the NLO real-emission correction by factoring out the LO IF $h_m^{(0)}(\T{p})$ and the factor of $\alpha_s/(2\pi)$ (see eqns.~(\ref{eq:IF-as-expansion}) and (\ref{eq:IF-NLO-num})):
\begin{eqnarray}
    V_a^{({\cal Q},\text{NLO, R})}(\T{q},\T{p},z) = \frac{\alpha_s(\mu_R)}{2\pi}  \frac{z}{\pi(2\pi)^{-\epsilon}} h_m^{(0)}(\T{p}^2)  \Tilde{H}^{(m)}_{Ra}(\T{q},\T{p}, z), 
\end{eqnarray}
where the reduced matrix element is defined in terms of the squared matrix element in the EFT~\cite{Lipatov95} as:
\begin{eqnarray}
   \Tilde{H}^{(m)}_{Ra}(\T{q},\T{p}, z) =   \frac{(2\pi)^{2\epsilon}|{\cal M}(a(q)+R_-(\T{q})\to {\cal Q}(p)+g(k))|^2}{ (4\pi)^2 \alpha_s(\mu_R) z(1-z) (q_+)^2 \T{q}^2 \sqrt{N_c^2-1} N_a^{\text{(col.)}} h_m^{(0)}(\T{p}^2) }. \label{eq:Htil-M2-gen}
\end{eqnarray}


\section{Exact real-emission squared matrix elements}\label{append:exact-MEs}

In this appendix, we will collect the results for the exact squared matrix elements of the processes:
\begin{equation}
R_-(q_R)+a(q)\to Q\bar{Q}[m](p) + a(k), \label{eq:real-proc}
\end{equation}
where $R_-$ is the reggeised gluon carrying momentum $q^\mu_R=n_+^\mu q_R^-/2+q_T^\mu$, which is off-shell: $q_R^2=-\T{q}^2$, the parton $a=g/q$ is {\it on-shell} ($q^\mu=q^+n_-^\mu/2$, $q^2=0$) and the NRQCD states of the $Q\bar{Q}$ pair carrying the momentum $p$ ($p^2=M^2$) are $m={}^1S_0^{[1]}$, ${}^1S_0^{[8]}$ and ${}^3S_1^{[8]}$. The amplitude of the process (\ref{eq:real-proc}) is computed according to the Feynman rules of the gauge-invariant EFT for Multi-Regge processes in QCD~\cite{Lipatov95}. The relevant Reggeon-gluon couplings are summarised in Sec.~4 of ref.~\cite{Nefedov:2024swu}. 

The projection of the $Q\bar{Q}$ pair on the corresponding NRQCD intermediate state is done in a standard way. To project out the $L=0$ state, we simply put $p_Q=p_{\bar{Q}}=p/2$. The heavy-quark spinors are substituted by the corresponding spin and colour projectors as:
\begin{eqnarray}
    \bar{u}_i(p_Q) {\cal M}^{ij} v_j(p_{\bar{Q}}) = {\rm tr} \left[{\cal M}^{ij} \big( v_j(p_{\bar{Q}})\otimes \bar{u}_i(p_Q)\big) \right] \to {\rm tr} \left[{\cal M}^{ij} \Pi_S \right] \Pi^{\text{[1/8]}}_{ji}, 
\end{eqnarray}
with colour indices $i$ and $j$ and the well-known (see e.g.~\cite{Mangano:1996kg}) covariant projectors for the total spin $S=0$ and $S=1$ states given by:
\begin{eqnarray}
    \Pi_0&=&\frac{1}{\sqrt{M^3}} \left(\frac{p}{2} - \frac{M}{2} \right) \gamma_5 \left(\frac{p}{2} + \frac{M}{2}\right), \label{eq:SpinProj-0} \\
    \Pi_1&=& \frac{\varepsilon^*_\mu(p)}{\sqrt{M^3}} \left(\frac{p}{2} - \frac{M}{2} \right) \gamma^\mu \left(\frac{p}{2} + \frac{M}{2}\right), \label{eq:SpinProj-1}
\end{eqnarray}
where $\varepsilon^\mu(p)$ is the polarisation vector of the $S=1$ state.  The projectors for colour-singlet ($^{[1]}$) and colour-octet ($^{[8]}$) states are:
\begin{eqnarray}
    \Pi_{ij}^{[1]} = \frac{\delta_{ij}}{\sqrt{N_c}}, \\
    \Pi_{ij}^{[8,a]} = \sqrt{2}T^a_{ij},
\end{eqnarray}
with $T^a_{ij}$ being the $SU(N_c)$ generators in the fundamental representation. To obtain the hard-scattering coefficient in accordance with eqn.~(\ref{eq:NRQCD-exp}),  one has to multiply the projected squared amplitude by $M^3/(N_{\text{col.}}^{\text{(NRQCD)}} N_{\text{pol.}}^{\text{(NRQCD)}})$ with $N_{\text{pol.}}^{\text{(NRQCD)}} = 2J+1$ and $N_{\text{col.}}^{\text{(NRQCD)}}=2N_c$ for the CS state and $N_c^2-1$ for the CO state. 

Several useful quantities can be defined, given the EFT amplitude of the process (\ref{eq:real-proc}), which we  denote as ${\cal M}(a+R\to Q\bar{Q}[m]+a)$. The first one is the properly normalised squared matrix element, averaged over the quantum numbers of the initial state:
\begin{eqnarray}
    \overline{|{\cal M}(a+R\to Q\bar{Q}[m]+a)|^2} = \frac{1}{\T{q}^2} \bigg(\frac{q_R^-}{2} \bigg)^2 \frac{|{\cal M}(a+R\to Q\bar{Q}[m]+a)|^2}{N_a^{\text{(pol.)}} N_a^{\text{(col.)}} (N_c^2-1)}, \label{eq:|A|^2-aver}
\end{eqnarray}
where the factors $N_a^{\text{(pol.)}}$ and $N_a^{\text{(col.)}}$ comes from averaging over the helicities and the colour quantum numbers of the incoming on-shell parton and $(N_c^2-1)$ averages over colours of incoming Reggeon. The factor $(q_R^-)^2/(2\T{q}^2)$ in eqn.~\ref{eq:|A|^2-aver} ensures that the following {\it on-shell limit} property is satisfied by the squared matrix element:
\begin{eqnarray}
    \int\frac{d^2 \T{n}}{\Omega_{2-2\epsilon}} \lim\limits_{\T{q}^2\to 0} \overline{|{\cal M}(a+R\to Q\bar{Q}[m]+a)|^2} = \frac{|{\cal M}(a+g\to Q\bar{Q}[m]+a)|^2}{N_a^{\text{(pol.)}} N_a^{\text{(col.)}} (D-2)(N_c^2-1)},
\end{eqnarray}
where $\T{n}$ is the unit vector pointing in the direction of $\T{q}$ and the $|{\cal M}(a+g\to Q\bar{Q}[m]+a)|^2$ is the squared matrix element of the process (\ref{eq:real-proc}) with the Reggeon replaced by the on-shell gluon. We have checked that this property is indeed satisfied by the expressions for real-emission matrix elements, presented below.

The reduced matrix element $\Tilde{H}^{(m)}_{Ra}(\T{q},\T{p}, z)$ of eqn.~(\ref{eq:IF-NLO-num}) is defined in terms of the averaged matrix element of eqn.~(\ref{eq:|A|^2-aver}) as:
\begin{eqnarray}
  \Tilde{H}^{(m)}_{Ra}(\T{q},\T{p}, z) = \frac{\sqrt{N_c^2-1}}{(4\pi)^2 z(1-z) } \frac{\overline{|{\cal M}(a+R\to Q\bar{Q}[m]+a)|^2} }{ \alpha_s(\mu_R) \big( q_+ q_R^- \big)^2 h_0^{(m)}(\T{p})},  \label{eq:Htil-def-Mbar}
\end{eqnarray}
which is equivalent to eqn.~(\ref{eq:Htil-M2-gen}).

In the next two subsections, we give the expressions for the reduced matrix elements, written in terms of Mandelstam variables defined as:
\begin{eqnarray}
    s&=& (q+q_R)^2 = M^2y-\T{q}^2, \label{eq:s-def} \\
    t&=& (p-q_R)^2 = -\frac{1}{z}\left[ M^2(yz-1)-\T{p}^2 \right], \\
    u&=& (p-q)^2 = -\frac{1}{z}\left[ M^2(1-z)+\T{p}^2 \right],
\end{eqnarray}
where the variable $y=(q_R^- q_+)/M^2$ is related with $\T{q}$, $\T{p}$ and $z$ as:
\begin{equation}
y=\frac{(\T{q}-\T{p})^2}{M^2(1-z)} + \frac{M^2+\T{p}^2}{M^2 z}.  \label{eq:y-expr}
\end{equation}
Thanks to eqns.~(\ref{eq:s-def}) -- (\ref{eq:y-expr}), the reduced matrix element can be completely expressed as a function of $\T{q}$, $\T{p}$ and $z$. In addition, we have the following relation between Mandelstam invariants: 
\begin{equation}
s+t+u=M^2-\T{q}^2.    
\end{equation}

\subsection{Gluon channel}

\begin{figure}[hbt!]
    \centering
    \includegraphics[scale=1.5]{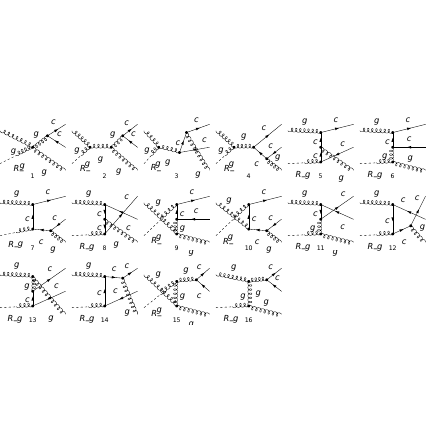} \vspace*{-3cm} 
    \caption{Feynman diagrams with $Rg$-transition vertex for the subprocess (\ref{eq:real-proc}). Heavy-quark ($Q$) lines are labelled by $c$. Dashed lines denote reggeised gluons. \vspace*{-3cm}}
    \label{fig:Rg-real-diags-Rg}
\end{figure}

\begin{figure}[hbt!]
    \centering
    \includegraphics[scale=1.1]{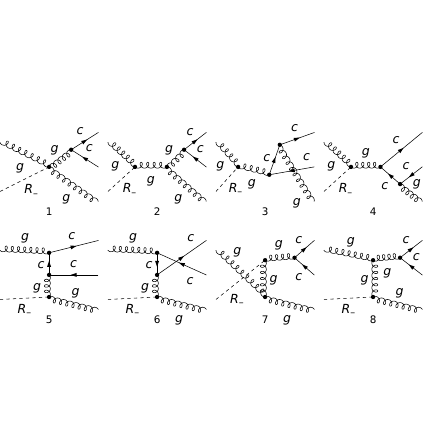} \vspace*{-2cm}
    \caption{Feynman diagrams with $Rgg$ and $Rggg$ induced vertices for the subprocess (\ref{eq:real-proc}). Heavy-quark ($Q$) lines are labelled by $c$. Dashed lines denote reggeised gluons. }
    \label{fig:Rg-real-diags-ind}
\end{figure}

The Feynman diagrams, contributing to the process (\ref{eq:real-proc}) for $a=g$, fall into two categories: diagrams with the $Rg$-transition vertex (fig.~\ref{fig:Rg-real-diags-Rg}) and with higher-order $Rgg$, $Rggg$ and $Rggg$ induced vertices (fig.~\ref{fig:Rg-real-diags-ind}). The diagrams of fig.~\ref{fig:Rg-real-diags-Rg} can be obtained from the usual QCD diagrams by the replacement of the polarisation vector of the reggeised gluon $\varepsilon_\mu(q_R)\to n^+_\mu$. However, this set of diagrams alone would not be gauge-invariant, because the reggeised gluon is off-shell ($q_R^2=-\T{q}^2<0$). Gauge invariance is restored by the addition of the diagrams with induced Reggeon-gluon couplings in fig.~\ref{fig:Rg-real-diags-ind}. In principle, all the diagrams in fig.~\ref{fig:Rg-real-diags-ind} will be nullified in the $A\cdot n^+=0$ gauge, so the alternative way to compute the impact factor would be to simply use this gauge for diagrams in fig.~\ref{fig:Rg-real-diags-Rg}. However no known gauge-choice exists for the computation of central-emission vertices ($R_+R_-\to X$), so the usage of the Lipatov EFT formalism of ref.~\cite{Lipatov95} is preferred in this case. 

On the technical side, we have used the \texttt{FeynArts}~\cite{FeynArts} package to generate the expressions for Feynman amplitudes in the EFT~\cite{Lipatov95} with the help of a in-house model file and the tools of the \texttt{FeynCalc}~\cite{FeynCalc,Shtabovenko:2020gxv,Shtabovenko:2023idz} framework to manipulate them together with the in-house interface to \texttt{FORM}~\cite{Vermaseren:2000nd,Ruijl:2017dtg}, which was used to perform Lorentz-index contractions. The expressions for the reduced squared matrix elements listed below are also provided in a plain-text format as an electronic appendix to the ArXiv submission of the present paper. 

Due to their size, we have put the reduced matrix elements for the NRQCD states $m={}^1S_0^{[1]}$ and ${}^1S_0^{[8]}$  with $a=g$ into the following form:
\begin{eqnarray}
    \Tilde{H}^{({}^1S_0^{[1]})}_{Rg}&=&\frac{1}{z(1-z) \T{q}^2}\frac{C_A \left(M^2+\T{p}^2\right)^2}{s^2 t^2 u^2 \left(M^2-s\right)^2 \left(M^2-t+\T{q}^2\right){}^2 \left(M^2-u\right)^2} \nonumber \\
    && \times \sum\limits_{n=-1}^3 (1-z)^n w^{({}^1S_0^{[1]})}_{g,n}(s,t,u), \label{eq:Htil-1S01} \\
    \Tilde{H}^{({}^1S_0^{[8]})}_{Rg}&=&\frac{-1}{2z(1-z) \T{q}^2}\frac{C_A \left(M^2+\T{p}^2\right)^2}{s^2 t^2 u^2 \left(M^2-s\right)^2 \left(M^2-t+\T{q}^2\right){}^2 \left(M^2-u\right)^2} \nonumber \\
    && \times \sum\limits_{n=-1}^3 (1-z)^n w^{({}^1S_0^{[8]})}_{g,n}(s,t,u), \label{eq:Htil-1S08}
\end{eqnarray}
while for $m={}^3S_1^{[8]}$ we directly substitute $N_c=3$ to reduce the size of the expressions:
\begin{eqnarray}
    \Tilde{H}^{({}^3S_1^{[8]})}_{Rg}&=&\frac{-1}{6z^2 (1-z) \T{q}^2 \T{p}^2}\frac{\left(M^2+\T{p}^2\right)^2}{ t  \left(M^2-s\right)^2 \left(M^2-t+\T{q}^2\right){}^2 \left(M^2-u\right)^2} \nonumber \\
    && \times \sum\limits_{n=-1}^4 (1-z)^n w^{({}^3S_1^{[8]})}_{g,n}(s,t,u), \label{eq:Htil-3S18}
\end{eqnarray}

The coefficients of $(1-z)$-expansion for the ${}^1S_0^{[1]}$-state are:
\begin{eqnarray}
&& w^{({}^1S_0^{[1]})}_{g,-1}= s t t_1 u \Bigl( t + t_1 + u\Bigr) \Bigl( s + 2 t_1 + u\Bigr) \
\Bigl( s^2 [ t_1 - u] [ t + t_1 \nonumber \\ 
 &&+ u] + t [ t + t_1] [ t_1 + u] [ 2 t_1 + u] + s [ t_1^3 - t_1 u^2 \
+ t^2 ( t_1 \nonumber \\ 
 &&+ u) + 2 t t_1 ( 2 t_1 + u)]\Bigr), 
 \end{eqnarray}
 
 \begin{eqnarray}
&& w_{g,0}^{({}^1S_0^{[1]})}=-u \Biggl\{ -s^5 t t_1 \Bigl( t + t_1 + u\Bigr)^2 - t^2 u \Bigl( t + \
t_1\Bigr)^2 \Bigl( t + t_1 + u\Bigr)^2 \Bigl( 2 t_1 \nonumber \\ 
 &&+ u\Bigr)^2 - s^4 \Bigl[ t + t_1 + u\Bigr] \Bigl[ -u ( t_1 + u)^3 + t ( t_1 + u)^2 [ 2 t_1 \nonumber \\ 
 &&- u] + t^3 [ 5 t_1 + u] + t^2 [ t_1 + u] [ 10 t_1 + u]\Bigr] - s t \
\Bigl[ t + t_1\Bigr] \Bigl[ t \nonumber \\ 
 &&+ t_1 + u\Bigr] \Bigl[ 2 t_1 + u\Bigr] \Bigl[ 2 t^3 u + t_1^2 u [ \
t_1 + u] + t u [ 2 t_1 + u] [ 5 t_1 + 2 u] \nonumber \\ 
 &&+ 2 t^2 [ t_1 + u] [ 4 t_1 + 3 u]\Bigr] + s^2 \Bigl[ t + t_1 + \
u\Bigr] \Bigl[ -t^5 u + t_1^2 u ( t_1 \nonumber \\ 
 &&+ u)^3 + t t_1 [ t_1 + u] [ 2 t_1 + u] [ t_1^2 - t_1 u + 2 u^2] - \
t^4 [ 16 t_1^2 + 21 t_1 u \nonumber \\ 
 &&+ 7 u^2] - t^3 [ 2 t_1 + u] [ 22 t_1^2 + 29 t_1 u + 8 u^2] - 2 t^2 \
t_1 [ 5 t_1^3 + 24 t_1^2 u + 19 t_1 u^2 \nonumber \\ 
 &&+ 4 u^3]\Bigr] + s^3 \Bigl[ 2 t_1 u ( t_1 + u)^4 - 2 t^5 [ 2 t_1 + \
u] - 2 t^2 t_1 [ t_1 + u] [ 10 t_1^2 + 22 t_1 u \nonumber \\ 
 &&+ 7 u^2] - t^4 [ 32 t_1^2 + 33 t_1 u + 8 u^2] + t ( t_1 + u)^2 [ \
t_1^3 + 2 t_1^2 u + 7 t_1 u^2 \nonumber \\ 
 &&+ 2 u^3] - t^3 [ 49 t_1^3 + 101 t_1^2 u + 54 t_1 u^2 + 8 \
u^3]\Bigr]\Biggr\} ,
 \end{eqnarray}

 \begin{eqnarray}
&&w_{g,1}^{({}^1S_0^{[1]})}=s u \Bigl[ s + t + t_1\Bigr] \Bigl[ t + t_1 + u\Bigr] \Bigl[ s^3 \
\Bigl( t + u\Bigr) \Bigl( 2 t^2 + t t_1 \nonumber \\ 
 &&- 2 ( t_1 + u)^2\Bigr) + t \Bigl( 2 t_1 + u\Bigr) \Bigl( t t_1 u [ \
-4 t_1 + u] + 2 t^3 [ 2 t_1 + u] \nonumber \\ 
 &&- t_1^2 u [ t_1 + 2 u] + 2 t^2 [ -4 t_1^2 + t_1 u + u^2]\Bigr) + \
s^2 \Bigl( 2 t^4 - 4 t_1 u ( t_1 \nonumber \\ 
 &&+ u)^2 + 2 t^3 [ 3 t_1 + 5 u] + t^2 [ -2 t_1^2 + t_1 u + 6 u^2] - \
t [ 5 t_1^3 \nonumber \\ 
 &&+ 14 t_1^2 u + 14 t_1 u^2 + 2 u^3]\Bigr) + s \Bigl( -2 t_1^2 u ( \
t_1 + u)^2 + 4 t^4 [ 2 t_1 + u] \nonumber \\ 
 &&+ t^3 [ -4 t_1^2 + 20 t_1 u + 10 u^2] + t^2 [ -8 t_1^3 - 18 t_1^2 \
u + t_1 u^2 \nonumber \\ 
 &&+ 2 u^3] - t t_1 [ 2 t_1^3 + 10 t_1^2 u + 14 t_1 u^2 + 3 \
u^3]\Bigr)\Bigr] ,
\end{eqnarray}

\begin{eqnarray}
 &&   w_{g,2}^{({}^1S_0^{[1]})} = s \Biggl\{ s^5 \Bigl( t + t_1 + u\Bigr)^2 \Bigl[ t - u\Bigr] \Bigl[ \
t + u\Bigr] + s^4 \Bigl[ t + t_1 \nonumber \\ 
 &&+ u\Bigr] \Bigl[ -2 u^3 [ t + 2 t_1] + 2 t^2 [ t + t_1] [ t + 3 \
t_1] + u^2 [ 2 t^2 - 5 t t_1 \nonumber \\ 
 &&- 4 t_1^2] + t u [ 6 t^2 + 9 t t_1 + t_1^2]\Bigr] + s^2 \Bigl[ t \
t_1 u^5 + 4 t^2 t_1 ( t + t_1)^3 [ t + 3 t_1] \nonumber \\ 
 &&+ 2 u^4 [ t^3 + 6 t^2 t_1 - 2 t_1^3] + 2 u^3 [ 4 t^4 + 27 t^3 t_1 \
+ 29 t^2 t_1^2 - 6 t t_1^3 - 4 t_1^4] \nonumber \\ 
 &&+ t u [ t + t_1] [ 2 t^4 + 32 t^3 t_1 + 90 t^2 t_1^2 + 51 t t_1^3 \
+ 5 t_1^4] + u^2 [ 8 t^5 + 73 t^4 t_1 \nonumber \\ 
 &&+ 163 t^3 t_1^2 + 84 t^2 t_1^3 - 6 t t_1^4 - 4 t_1^5]\Bigr] + s \
\Bigl[ 4 t^2 t_1^2 ( t + t_1)^4 + 2 t t_1 u^5 [ t + t_1] \nonumber \\ 
 &&+ u^4 [ t^4 + 17 t^3 t_1 + 24 t^2 t_1^2 + 4 t t_1^3 - t_1^4] + 2 t \
t_1 u [ t + t_1] [ 2 t^4 + 20 t^3 t_1 \nonumber \\ 
 &&+ 36 t^2 t_1^2 + 11 t t_1^3 + t_1^4] + u^3 [ 2 t^5 + 33 t^4 t_1 + \
101 t^3 t_1^2 + 64 t^2 t_1^3 - 4 t t_1^4 \nonumber \\ 
 &&- 2 t_1^5] + u^2 [ t^6 + 22 t^5 t_1 + 117 t^4 t_1^2 + 172 t^3 \
t_1^3 + 57 t^2 t_1^4 - 4 t t_1^5 - t_1^6]\Bigr] \nonumber \\ 
 &&+ t t_1 u \Bigl[ t + t_1\Bigr] \Bigl[ t + t_1 + u\Bigr] \Bigl[ 2 \
t_1 + u\Bigr] \Bigl[ t_1 u [ -t_1 + u] + 4 t^2 [ 2 t_1 \nonumber \\ 
 &&+ u] + t u [ 6 t_1 + u]\Bigr] + s^3 \Bigl[ t^6 - 6 t_1^2 u^2 ( t_1 \
+ u)^2 + 4 t^5 [ 3 t_1 + 2 u] \nonumber \\ 
 &&+ 2 t t_1 u [ t_1 + u] [ 2 t_1^2 - 7 t_1 u - 2 u^2] + t^4 [ 34 \
t_1^2 + 53 t_1 u + 15 u^2] + t^3 [ 36 t_1^3 \nonumber \\ 
 &&+ 87 t_1^2 u + 61 t_1 u^2 + 8 u^3] + t^2 t_1 [ 13 t_1^3 + 46 t_1^2 \
u + 41 t_1 u^2 + 14 u^3]\Bigr]\Biggr\}, \\
&&w_{g,3}^{({}^1S_0^{[1]})} = s t t_1 u \Bigl( s + t + t_1\Bigr) \Bigl( s + 2 t_1 + u\Bigr) \
\Bigl( s^2 [ t - u] [ t + t_1 \nonumber \\ 
 &&+ u] + t_1 [ t + t_1 + u] [ 2 t t_1 + u ( t + t_1)] + s t [ ( 3 \
t_1 - u) ( t_1 \nonumber \\ 
 &&+ u) + t ( 3 t_1 + u)]\Bigr),
\end{eqnarray}
where $t_1=\T{q}^2$. We collect the expressions for the coefficients of the $1-z$ expansion for the ${}^1S_0^{[1]}$ squared amplitude (\ref{eq:Htil-1S01}), in a form without explicit $t_1$ dependence, in the file \texttt{MEcoeffs\_Rg-1S01g.m} as a Mathematica array, the first element of which is the coefficient $w^{({}^1S_0^{[1]})}_{g,-1}$. In the file \texttt{MatrixElementsTest\_S-wave.nb}, the exact matrix elements are compared with the subtraction terms~(\ref{eq:sub-term-Rg}) and (\ref{eq:sub-term-Rq}) numerically in the collinear, soft and Regge limits. 

For the ${}^1S_0^{[8]}$ state, the coefficients of $(1-z)$ expansion of reduced matrix element are:
\begin{eqnarray}
&& w_{g,-1}^{({}^1S_0^{[8]})}=  s t u \Bigl( M^2 - s\Bigr) \Bigl( M^2 - s - t - u\Bigr) \Bigl( -2 \
M^2 + s + 2 t + u\Bigr) \Bigl( M^6 [ s \nonumber \\ 
 && + 2 t] - M^4 [ 3 u ( s + t) + 2 ( s^2 + 2 s t + 2 t^2)] - u [ s^3 + \
2 s^2 ( t + u) - s t ( t \nonumber \\ 
 && + u) + t^2 ( 2 t + u)] + M^2 [ s^3 + 2 s^2 ( t + 2 u) + t ( 2 t^2 + \
5 t u + u^2) + s ( 3 t^2 \nonumber \\ 
 && + 3 t u + 2 u^2)]\Bigr),
\end{eqnarray}

\begin{eqnarray}
&&w_{g,0}^{({}^1S_0^{[8]})}=  u \Bigl[ M^8 \Bigl( u^3 [ 2 s - 13 t] [ s + 2 t] + u^2 [ 16 s^3 + \
15 s^2 t - 76 s t^2 - 40 t^3] + s t [ 23 s^3 + 66 s^2 t \nonumber \\ 
 && + 82 s t^2   + 48 t^3] + u [ 12 s^4 + 59 s^3 t + 20 s^2 t^2 + 2 s t^3 - 8 \
t^4]\Bigr) + s^2 u \Bigl( s t u^4 + t u^3 [ 3 s^2 + s t + 4 t^2] \nonumber \\ 
 &&+ 2 t [ s + t] [ s + 2 t] [ s^3 + s^2 t + s t^2 + 4 t^3] + u^2 [ 2 \
s^4 + 11 s^3 t + 10 s^2 t^2 + 8 s t^3 \nonumber \\ 
 &&+ 18 t^4] + t u [ 7 s^4 + 13 s^3 t + 8 s^2 t^2 + 24 s t^3 + 28 \
t^4]\Bigr) + M^6 \Bigl( t u^4 [ 5 s + 12 t] + 2 u^3 [ -4 s^3 \nonumber \\ 
 &&- 2 s^2 t + 33 s t^2 + 16 t^3] - s t [ s + 2 t] [ 23 s^3 + 44 s^2 \
t + 55 s t^2 + 24 t^3] + u^2 [ -24 s^4 \nonumber \\ 
 &&- 82 s^3 t + 14 s^2 t^2 + 39 s t^3 + 16 t^4] - 2 s u [ 4 s^4 + 44 \
s^3 t + 67 s^2 t^2 + 70 s t^3 + 45 t^4]\Bigr) \nonumber \\ 
 && + M^2 s \Bigl( 2 t^2 \
u^5  + t u^4 [ -7 s^2 + 5 s t + 4 t^2] - 2 s t [ s + t] [ s + 2 t] [ \
s^3 + 3 s^2 t + 3 s t^2 + 4 t^3] \nonumber \\ 
 &&- u^3 [ 8 s^4 + 38 s^3 t + 21 s^2 t^2 + 23 s t^3 + 10 t^4] - t u [ \
18 s^5 + 59 s^4 t + 89 s^3 t^2 + 132 s^2 t^3 + 112 s t^4 \nonumber \\ \
 &&+ 16 t^5] - u^2 [ 4 s^5 + 47 s^4 t + 67 s^3 t^2 + 68 s^2 t^3 + 108 \
s t^4 + 28 t^5]\Bigr) + 2 M^{12} \Bigl( -(s t u) - 4 t^2 u \nonumber \\ \
 &&+ s^2 [ t + u]\Bigr) - M^{10} \Bigl( -8 t^2 u [ 2 t + 3 u] + s^3 [ \
11 t + 8 u] + s t [ 16 t^2 - 30 t u - 7 u^2] \nonumber \\ 
 &&+ 2 s^2 [ 9 t^2 + 6 t u + 2 u^2]\Bigr) + M^4 \Bigl( -2 t^2 u^3 ( 2 \
t + u)^2 + s^6 [ 11 t + 2 u] + s^5 [ 2 t + u] [ 27 t \nonumber \\ 
 &&+ 16 u] + s^4 [ 105 t^3 + 143 t^2 u + 100 t u^2 + 12 u^3] + s^3 t [ \
124 t^3 + 195 t^2 u + 88 t u^2 + 40 u^3] + s t [ t \nonumber \\ 
 &&+ u] [ 16 t^4 + 60 t^3 u - 8 t^2 u^2 - 21 t u^3 - u^4] + s^2 t [ \
84 t^4 + 220 t^3 u + 77 t^2 u^2 - 21 t u^3 - u^4]\Bigr)\Bigr], \nonumber \\
\end{eqnarray}

\begin{eqnarray}
&& w_{g,1}^{({}^1S_0^{[8]})}=  s u \Bigl( M^2 - s\Bigr) \Bigl( M^2 - u\Bigr) \Bigl( -(s^5 t) - \
s^4 t [ 5 t + 2 u] - s^3 [ 12 t^3 + 5 t^2 u + t u^2 \nonumber \\ 
 &&- 4 u^3] - t [ t + u] [ 2 t + u] [ 12 t^3 + 4 t^2 u + 2 t u^2 + \
u^3] - s^2 t [ 28 t^3 + 8 t^2 u + 8 t u^2 \nonumber \\ 
 &&+ u^3] - s t [ 44 t^4 + 36 t^3 u + 8 t^2 u^2 + 5 t u^3 + 2 u^4] + \
2 M^8 [ t u + s ( t + 2 u)] - M^6 [ 2 s ( t \nonumber \\ 
 &&+ u) ( 3 t + 4 u) + s^2 ( 3 t + 8 u) + t ( -16 t^2 + 6 t u + 3 u^2)] \
+ M^4 [ -(s^3 ( t \nonumber \\ 
 &&- 4 u)) + s^2 ( 2 t^2 + 11 t u + 16 u^2) - t ( 56 t^3 + 34 t^2 u - \
2 t u^2 + u^3) + s ( -34 t^3 \nonumber \\ 
 &&+ 16 t^2 u + 11 t u^2 + 4 u^3)] + M^2 [ 3 s^4 t + s^3 ( 9 t^2 + 3 \
t u - 8 u^2) + s^2 ( 29 t^3 + t^2 u - 8 t u^2 \nonumber \\ 
 &&- 8 u^3) + s t ( 82 t^3 + 34 t^2 u + t u^2 + 3 u^3) + t ( 64 t^4 + \
82 t^3 u + 29 t^2 u^2 + 9 t u^3 + 3 u^4)]\Bigr), \nonumber \\
\end{eqnarray}

\begin{eqnarray}
&& w_{g,2}^{({}^1S_0^{[8]})}= - s \Bigl[ 2 M^{12} \Bigl( 4 s t^2 + s t u - u^2 [ s + t]\Bigr) + M^{10} \
\Bigl( -8 s t^2 [ 3 s + 2 t] + u^3 [ 8 s + 11 t] \nonumber \\ 
 &&+ 2 u^2 [ 2 s^2 + 6 s t + 9 t^2] + t u [ -7 s^2 - 30 s t + 16 \
t^2]\Bigr) + M^4 \Bigl( 2 s^3 t^2 ( s + 2 t)^2 - u^6 [ 2 s \nonumber \
\\ 
 &&+ 11 t] - u^5 [ s + 2 t] [ 16 s + 27 t] - u^4 [ 12 s^3 + 100 s^2 t \
+ 143 s t^2 + 105 t^3] - t u^3 [ 40 s^3 + 88 s^2 t \nonumber \\ 
 &&+ 195 s t^2 + 124 t^3] + t u^2 [ s^4 + 21 s^3 t - 77 s^2 t^2 - 220 \
s t^3 - 84 t^4] - t u [ s + t] [ -s^4 \nonumber \\ 
 &&- 21 s^3 t - 8 s^2 t^2 + 60 s t^3 + 16 t^4]\Bigr) + M^6 \Bigl( -4 \
s^2 t^2 [ s + 2 t] [ 3 s + 2 t] + u^5 [ 8 s + 23 t] \nonumber \\ 
 &&+ 2 u^4 [ 12 s^2 + 44 s t + 45 t^2] + 2 t u^2 [ 2 s^3 - 7 s^2 t + \
70 s t^2 + 67 t^3] + u^3 [ 8 s^3 + 82 s^2 t + 134 s t^2 \nonumber \\ 
 &&+ 143 t^3] + t u [ -5 s^4 - 66 s^3 t - 39 s^2 t^2 + 90 s t^3 + 48 \
t^4]\Bigr) + M^8 \Bigl( s^3 [ 13 t - 2 u] [ 2 t + u] \nonumber \\ 
 &&+ s^2 [ 40 t^3 + 76 t^2 u - 15 t u^2 - 16 u^3] - t u [ 48 t^3 + 82 \
t^2 u + 66 t u^2 + 23 u^3] + s [ 8 t^4 - 2 t^3 u \nonumber \\ 
 &&- 20 t^2 u^2 - 59 t u^3 - 12 u^4]\Bigr) - s u^2 \Bigl( s^4 t u + \
s^3 t [ 4 t^2 + t u + 3 u^2] + 2 t [ t + u] [ 2 t + u] [ 4 t^3 \nonumber \\ 
 &&+ t^2 u + t u^2 + u^3] + s^2 [ 18 t^4 + 8 t^3 u + 10 t^2 u^2 + 11 \
t u^3 + 2 u^4] + s t [ 28 t^4 + 24 t^3 u \nonumber \\ 
 &&+ 8 t^2 u^2 + 13 t u^3 + 7 u^4]\Bigr) + M^2 u \Bigl( -2 s^5 t^2 + \
s^4 t [ -4 t^2 - 5 t u + 7 u^2] + 2 t u [ t \nonumber \\ 
 &&+ u] [ 2 t + u] [ 4 t^3 + 3 t^2 u + 3 t u^2 + u^3] + s^3 [ 10 t^4 + \
23 t^3 u + 21 t^2 u^2 + 38 t u^3 + 8 u^4] \nonumber \\ 
 &&+ s^2 [ 28 t^5 + 108 t^4 u + 68 t^3 u^2 + 67 t^2 u^3 + 47 t u^4 + \
4 u^5] + s t [ 16 t^5 + 112 t^4 u + 132 t^3 u^2  \nonumber \\ 
 && + 89 t^2 u^3 + 59 t u^4 + 18 u^5]\Bigr)\Bigr] , \nonumber \\
\end{eqnarray}
and
\begin{eqnarray}
  &&  w_{g,3}^{({}^1S_0^{[8]})}=  s t u \Bigl( M^2 - u\Bigr) \Bigl( M^2 - s - t - u\Bigr) \Bigl( -2 \
M^2 + s + 2 t + u\Bigr) \Bigl( u^3 [ M^2 \nonumber \\ 
 &&- s] + t [ M^2 - s] [ 2 M^2 - s - 2 t] [ M^2 - t] - 2 u^2 [ M^2 - \
s] [ M^2 - s - t] \nonumber \\ 
 &&+ u [ M^6 + s t ( s + t) - M^4 ( 3 s + 4 t) + M^2 ( 2 s^2 + 3 s t + \
3 t^2)]\Bigr).
\end{eqnarray}
We collect the expressions for the coefficients of the $1-z$ expansion for the ${}^1S_0^{[8]}$ squared amplitude (\ref{eq:Htil-1S08}), in the file \texttt{MEcoeffs\_Rg-1S08g.m} as a Mathematica array, the first element of which is the coefficient $w^{({}^1S_0^{[8]})}_{g,-1}$.

For the ${}^3S_1^{[8]}$ state the coefficients in eqn.~(\ref{eq:Htil-3S18}) are: 

\begin{eqnarray}
&& w_{g,-1}^{({}^3S_1^{[8]})}= -54 \Bigl( M^2 - s\Bigr) \Bigl( M^2 - s - t - u\Bigr) \Bigl( -2 M^2 + \
s + 2 t + u\Bigr) \Bigl( M^6 - u [ s \nonumber \\ 
 &&- t] [ s + 2 t + u] - M^4 [ 2 s + t + 2 u] + M^2 [ 3 s u + u^2 + s \
( s + t)]\Bigr),
\end{eqnarray}

\begin{eqnarray}
&&w_{g,0}^{({}^3S_1^{[8]})}= - \Bigl[ 324 M^{12} + 54 s u [ s + 2 t + u]^2 \Bigl( 2 s - 3 t + \
u\Bigr) \Bigl( s + t \nonumber \\ 
 &&+ u\Bigr) - M^{10} \Bigl( 1323 s + 1102 t + 1269 u\Bigr) + M^8 \Bigl( \
2133 s^2 + 3538 s t + 1108 t^2 + 4401 s u \nonumber \\ 
 &&+ 3170 t u + 1890 u^2\Bigr) + M^4 \Bigl( 675 s^4 + 2306 s^3 t + \
1977 s^2 t^2 - 16 s t^3 - 200 t^4 + 432 u^4 \nonumber \\ 
 &&+ 27 u^3 ( 110 s + 53 t) + 27 u^2 ( 206 s^2 + 240 s t + 27 t^2) + \
u ( 3645 s^3 + 7571 s^2 t + 2074 s t^2 \nonumber \\ 
 &&- 848 t^3)\Bigr) - 27 M^2 \Bigl( s + 2 t + u\Bigr) \Bigl( 2 u^4 + \
4 u^3 ( 6 s + t) + 2 s ( 2 s - t) ( s \nonumber \\ 
 &&+ t) ( s + 2 t) + u^2 ( 55 s^2 + 15 s t - 6 t^2) + u ( 35 s^3 + 28 \
s^2 t - 33 s t^2 \nonumber \\ 
 &&- 10 t^3)\Bigr) - M^6 \Bigl( 1701 s^3 + 130 t^3 + 1669 t^2 u + \
3321 t u^2 + 1323 u^3 + 8 s^2 ( 532 t + 729 u) \nonumber \\ 
 &&+ s ( 2491 t^2 + 8203 t u + 5400 u^2)\Bigr)\Bigr],
\end{eqnarray}

\begin{eqnarray}
&&w_{g,1}^{({}^3S_1^{[8]})}=  540 M^{12} + 54 s u [ s + 2 t + u]^2 \Bigl( s + t + u\Bigr) \
\Bigl( 3 s - 5 t \nonumber \\ 
 &&+ 2 u\Bigr) - 3 M^{10} \Bigl( 747 s + 482 t + 693 u\Bigr) + M^8 \
\Bigl( 3618 s^2 + 4874 s t + 868 t^2 + 7317 s u \nonumber \\ 
 &&+ 4290 t u + 3105 u^2\Bigr) - M^2 \Bigl( 108 u^5 + 27 u^4 ( 47 s + \
14 t) + 2 s u^3 ( 1809 s + 1501 t) + 54 s ( s \nonumber \\ 
 &&+ t) ( s + 2 t) ( 3 s^2 + 3 s t - 7 t^2) + u^2 ( 3969 s^3 + 6463 \
s^2 t - 1043 s t^2 - 1080 t^3) \nonumber \\ 
 &&+ u ( 1674 s^4 + 4271 s^3 t - 125 s^2 t^2 - 4504 s t^3 - 864 \
t^4)\Bigr) - M^6 \Bigl( 2835 s^3 - 670 t^3 + 1191 t^2 u \nonumber \\ 
 &&+ 4642 t u^2 + 2241 u^3 + 15 s^2 ( 397 t + 648 u) + s ( 1905 t^2 + \
11675 t u + 8964 u^2)\Bigr) + M^4 \Bigl( 1080 s^4 \nonumber \\ 
 &&- 632 t^4 - 2112 t^3 u + 361 t^2 u^2 + 2138 t u^3 + 783 u^4 + s^3 ( \
3175 t + 5994 u) + s^2 ( 1393 t^2 \nonumber \\ 
 &&+ 11116 t u + 9234 u^2) + s ( -1712 t^3 + 818 t^2 u + 9647 t u^2 + \
5049 u^3)\Bigr),
\end{eqnarray}

\begin{eqnarray}
&&w_{g,2}^{({}^3S_1^{[8]})}= - \Bigl[ 540 M^{12} + 54 s u [ s + 2 t + u]^2 \Bigl( s + t + u\Bigr) \
\Bigl( 2 s - 5 t \nonumber \\ 
 &&+ 3 u\Bigr) - 3 M^{10} \Bigl( 693 s + 482 t + 747 u\Bigr) + M^8 \
\Bigl( 3105 s^2 + 4290 s t + 868 t^2 + 7317 s u \nonumber \\ 
 &&+ 4874 t u + 3618 u^2\Bigr) - M^6 \Bigl( 2241 s^3 - 670 t^3 + 1905 \
t^2 u + 5955 t u^2 + 2835 u^3 + s^2 ( 4642 t \nonumber \\ 
 &&+ 8964 u) + s ( 1191 t^2 + 11675 t u + 9720 u^2)\Bigr) + M^4 \Bigl( \
783 s^4 - 632 t^4 - 1712 t^3 u + 1393 t^2 u^2 \nonumber \\ 
 &&+ 3175 t u^3 + 1080 u^4 + s^3 ( 2138 t + 5049 u) + s^2 ( 361 t^2 + \
9647 t u + 9234 u^2) + s ( \nonumber \\ 
 &&-2112 t^3 + 818 t^2 u + 11116 t u^2 + 5994 u^3)\Bigr) - M^2 \Bigl( \
108 s^5 + 27 s^4 ( 14 t + 47 u) + 2 s^3 u ( 1501 t \nonumber \\ 
 &&+ 1809 u) + 54 u ( t + u) ( 2 t + u) ( -7 t^2 + 3 t u + 3 u^2) + \
s^2 ( \nonumber \\ 
 &&-1080 t^3 - 1043 t^2 u + 6463 t u^2 + 3969 u^3) + s ( -864 t^4 - \
4504 t^3 u - 125 t^2 u^2 \nonumber \\ 
 &&+ 4271 t u^3 + 1674 u^4)\Bigr)\Bigr],
\end{eqnarray}

\begin{eqnarray}
&& w_{g,3}^{({}^3S_1^{[8]})}=  324 M^{12} + 54 s u [ s + 2 t + u]^2 \Bigl( s + t + u\Bigr) \
\Bigl( s - 3 t \nonumber \\ 
 &&+ 2 u\Bigr) - M^{10} \Bigl( 1269 s + 1102 t + 1323 u\Bigr) + M^8 \
\Bigl( 1890 s^2 + 3170 s t + 1108 t^2 + 4401 s u \nonumber \\ 
 &&+ 3538 t u + 2133 u^2\Bigr) - M^6 \Bigl( 1323 s^3 + 3321 s^2 t + \
1669 s t^2 + 130 t^3 + 1701 u^3 + 8 u^2 ( 729 s \nonumber \\ 
 &&+ 532 t) + u ( 5400 s^2 + 8203 s t + 2491 t^2)\Bigr) - 27 M^2 \
\Bigl( s + 2 t + u\Bigr) \Bigl( 2 s^4 - 2 u ( t \nonumber \\ 
 &&- 2 u) ( t + u) ( 2 t + u) + 4 s^3 ( t + 6 u) + s^2 ( -6 t^2 + 15 \
t u \nonumber \\ 
 &&+ 55 u^2) + s ( -10 t^3 - 33 t^2 u + 28 t u^2 + 35 u^3)\Bigr) + \
M^4 \Bigl( 432 s^4 - 200 t^4 \nonumber \\ 
 &&- 16 t^3 u + 1977 t^2 u^2 + 2306 t u^3 + 675 u^4 + 27 s^3 ( 53 t + \
110 u) + 27 s^2 ( 27 t^2 + 240 t u \nonumber \\ 
 &&+ 206 u^2) + s ( -848 t^3 + 2074 t^2 u + 7571 t u^2 + 3645 \
u^3)\Bigr),
\end{eqnarray}

\begin{eqnarray}
&&w_{g,4}^{({}^3S_1^{[8]})}= 54 \Bigl( M^2 - u\Bigr) \Bigl( M^2 - s - t - u\Bigr) \Bigl( -2 M^2 + \
s + 2 t + u\Bigr) \Bigl( M^6 + s [ t \nonumber \\ 
 &&- u] [ s + 2 t + u] - M^4 [ 2 s + t + 2 u] + M^2 [ s^2 + 3 s u + u \
( t + u)]\Bigr).
\end{eqnarray}

We collect the expressions for the coefficients of the $1-z$ expansion for the ${}^3S_1^{[8]}$ squared amplitude (\ref{eq:Htil-3S18}), in the file \texttt{MEcoeffs\_Rg-3S18g.m} as a Mathematica array, the first element of which is the coefficient $w^{({}^3S_1^{[8]})}_{g,-1}$.

\subsection{Quark channel}

\begin{figure}
    \centering
\begin{tabular}{cc}
\parbox{0.5\linewidth}{\includegraphics[width=\linewidth]{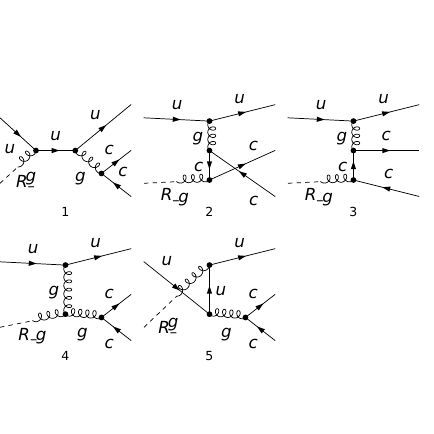}} & \parbox{0.2\linewidth}{\includegraphics[width=\linewidth]{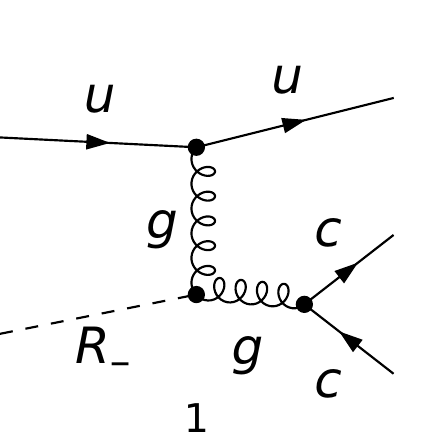}} \vspace*{-1cm} \\ 
(a) & (b)
\end{tabular}
    \caption{Direct $Rg$-coupling (a) and induced $Rgg$-coupling (b) high-energy EFT~\cite{Lipatov95} diagrams, contributing to the process (\ref{eq:real-proc}) with $a=q$}
    \label{fig:real-diags-quark}
\end{figure}

The reduced real-emission matrix element for $a=q$ and $m={}^1S_0^{[1]}$ gives a relatively compact expression:
\begin{eqnarray}
&& \Tilde{H}_{Rq}^{({}^1S_0^{[1]})}=-\frac{1}{z(1-z) \T{q}^2}\frac{C_F \left(\T{p}^2+M^2\right)^2}{t^2 \left(M^2-t+\T{q}^2\right)^2}\bigg( M^4-2 M^2 \left(z (s+\T{q}^2)+t-\T{q}^2\right) \nonumber \\
&&  +\left(\T{q}^2-z (s+\T{q}^2)\right)^2+2 s t z+t^2+t \T{q}^2
   ((z-2) z+2)\bigg), \label{eq:HRq-1S01}
\end{eqnarray}
because only diagrams \#2 and \#3 from the fig.~\ref{fig:real-diags-quark}(a) give a nonzero contribution for the ${}^1S_0^{[1/8]}$ states. The result for the reduced matrix element of production of ${}^1S_0^{[8]}$ state turns out to be the same as in the CS case:
\begin{equation}
    \Tilde{H}_{Rq}^{({}^1S_0^{[8]})} = \Tilde{H}_{Rq}^{({}^1S_0^{[1]})},
\end{equation}
because in both cases only the diagrams\#2 and \#3 in the fig.~\ref{fig:real-diags-quark}(a) contribute so the whole difference between CS and CO cases reduces to an overall colour factor, which is cancelled in the definition of the reduced squared matrix element~(\ref{eq:Htil-def-Mbar}). 

On the other hand, for the ${}^3S_1^{[8]}$ state, all the diagrams in fig.~\ref{fig:real-diags-quark}(a) and (b) contribute and we have:
\begin{eqnarray}
 \Tilde{H}_{Rq}^{({}^3S_1^{[8]})} = \frac{-C_F(M^2+\T{p}^2)^2}{12N_c\T{p}^2 \T{q}^2 z^2 (1-z) s^2 t u^2 (M^2-t+\T{q}^2)^2}\sum\limits_{n=0}^3 (1-z)^n w^{({}^3S_1^{[8]})}_{q,n}(s,t,u), \label{eq:Hq-3S18}
\end{eqnarray}
with
\begin{eqnarray}
&& w_{q,0}^{({}^3S_1^{[8]})}= 2 u \Biggl \{ t \Bigl( -2 M^2 + s + 2 t + u\Bigr)^2 \Bigl( -(M^2 ( s \
+ u)) + s ( s \nonumber \\ 
 &&+ t + u) \Bigr) + C_A^2 \Biggl( s^2 ( s + 2 t + u)^2 \Bigl( s \
+ t + u\Bigr) + M^6 \Bigl( -2 s^2 \nonumber \\ 
 && + 2 s u + 4 t u\Bigr) + M^4 \Bigl( 5 s^3 + 8 s^2 t - 4 t u \
( 2 t + u) - 3 s u ( 4 t \nonumber \\ 
 &&+ u)\Bigr) - M^2 \Bigl( s + 2 t + u\Bigr) \Bigl( 4 s^3 - t u ( 2 t \
+ u) - s u ( 4 t + u) \nonumber \\ 
 &&+ s^2 ( 5 t + u)\Bigr)\Biggr)\Biggr \} \; ,
\end{eqnarray}

\begin{eqnarray}
&& w_{q,1}^{({}^3S_1^{[8]})}= -2 u \Biggl\{ t \Bigl( -2 M^2 + s + 2 t + u\Bigr)^2 \Biggl( M^2 \
\Bigl( s - u\Bigr) + s \Bigl( s \nonumber \\ 
 &&+ t + u\Bigr) \Biggr) + C_A^2 \Biggl( s u \Bigl( s + 2 t + u\Bigr)^2 \
\Bigl( s + t + u\Bigr) + 2 M^6 \Bigl( s^2 \nonumber \\ 
 &&- s u + 2 t u\Bigr) - M^4 \Bigl( 3 s^3 + 8 s^2 t + s u ( 4 t - 5 \
u) + 4 t u ( 2 t + u ) \Bigr) \nonumber \\ 
 &&+ M^2 \Bigl( s^4 + 5 s^3 t + t u ( 2 t + u)^2 + s u ( t - 4 u ) ( 2 \
t + u) + s^2 ( 6 t^2 \nonumber \\ 
 &&+ 5 t u - 5 u^2) \Bigr) \Biggr) \Biggr\} \; ,
\end{eqnarray}

\begin{eqnarray}
&& w_{q,2}^{({}^3S_1^{[8]})}= 2 s \Biggl\{ t \Bigl( -2 M^2 + s + 2 t + u\Bigr)^2 \Biggl( M^2 \
\Bigl( -s + u\Bigr) + u \Bigl( s \nonumber \\ 
 &&+ t + u\Bigr) \Biggr) + C_A^2 \Biggl( s u \Bigl( s + 2 t + u\Bigr)^2 \
\Bigl( s + t + u\Bigr) + 2 M^6 \Bigl( 2 s t \nonumber \\ 
 &&- s u + u^2\Bigr) - M^4 \Bigl( 8 t u^2 + 3 u^3 + 4 s t ( s + 2 t) \
+ s u ( -5 s \nonumber \\ 
 &&+ 4 t) \Bigr) + M^2 \Bigl( 5 t u^3 + u^4 + s t ( s + 2 t)^2 - s u ( \
4 s - t) ( s + 2 t) \nonumber \\ 
 &&+ u^2 ( -5 s^2 + 5 s t + 6 t^2) \Bigr) \Biggr) \Biggr\} \; ,
\end{eqnarray}

\begin{eqnarray}
&& w_{q,3}^{({}^3S_1^{[8]})}= - \Biggl(t \bigl( -2 M^2 + s + 2 t + u\bigr)^2 \biggl( -2 M^2 s \bigl( s + u 
\bigr) \nonumber \\ 
 &&+ 2 s u \bigl( s + t + u\bigr) \biggr) \Biggr) - 2 C_A^2 s \Biggl( u^2 ( s \
+ 2 t + u )^2 \bigl( s + t + u \bigr) \nonumber \\ 
 &&- M^4 \bigl( -8 t u^2 - 5 u^3 + 4 s t ( s + 2 t) + 3 s u ( s + 4 \
t)\bigr) + M^2 \Bigl( s \nonumber \\ 
 &&+ 2 t + u\Bigr) \Bigl( -4 u^3 + s t ( s + 2 t) + s u ( s + 4 t) - \
u^2 ( s \nonumber \\ 
 &&+ 5 t)\Bigr) + 2 M^6 \Bigl( -u^2 + s ( 2 t + u)\Bigr)\Biggr) \; .
\end{eqnarray}
We collect the expressions for the coefficients of the $1-z$-expansion for the ${}^3S_1^{[8]}$ squared amplitude (\ref{eq:Hq-3S18}), in the file \texttt{MEcoeffs\_Rq-3S18q.m} as a Mathematica array, the first element of which is the coefficient $w^{({}^3S_1^{[8]})}_{q,0}$.



\bibliographystyle{JHEP}
\bibliography{mybibfile}

\end{document}